\documentstyle[aps,epsf,multicol,epsfig]{revtex}

\newcommand{\bleq}{\ifpreprintsty
                   \else
                   \end{multicols}\vspace*{-3.5ex}\widetext{\tiny
                   \noindent\begin{tabular}[t]{c|}
                   \parbox{0.493\hsize}{~} \\ \hline \end{tabular}}
                   \fi}
\newcommand{\eleq}{\ifpreprintsty
                   \else
                   {\tiny\hspace*{\fill}\begin{tabular}[t]{|c}\hline
                    \parbox{0.49\hsize}{~} \\
                    \end{tabular}}\vspace*{-2.5ex}\begin{multicols}{2}
                    \narrowtext \fi}
\begin{document}
\draft
\title{Morphology of Ion-Sputtered Surfaces}

\author{Maxim Makeev$^{\dag}$, Rodolfo Cuerno$^{\ddag}$, 
and Albert-L\'aszl\'o Barab\'asi$^{\dag}$}

\address{$^{\dag}$ Department of Physics, University of Notre Dame, Notre
Dame, IN 46556, USA
\\
$^{\ddag}$ 
Departamento de Matem\'{a}ticas and Grupo Interdisciplinar de Sistemas
Complicados, Universidad Carlos III de Madrid, 
Avenida de la Universidad 30, 28911 Legan\'{e}s, SPAIN}

\date{\today}

\maketitle

\begin{abstract}
 We derive a stochastic nonlinear continuum theory to describe the
morphological evolution of amorphous surfaces eroded by ion bombardment.
Starting from Sigmund's theory of sputter erosion, we calculate the 
coefficients appearing in the continuum equation in terms of the physical
parameters characterizing the sputtering process. We analyze the 
morphological features predicted by the continuum theory, comparing them 
with the experimentally reported morphologies. We show that for short time
scales, where the effect of nonlinear terms is negligible, the continuum 
theory predicts ripple formation. We demonstrate that in addition to 
relaxation by thermal surface diffusion, the sputtering process can also
contribute to the smoothing mechanisms shaping the surface morphology. 
We explicitly calculate an effective surface diffusion constant 
characterizing this smoothing effect, and show that it is responsible 
for the low 
temperature ripple formation observed in various experiments. At long 
time scales the nonlinear terms dominate the evolution of the surface
morphology. The nonlinear terms lead to the stabilization of the ripple
wavelength and we show that, depending on the experimental parameters 
such as angle of incidence and ion energy, different morphologies can be 
observed: asymptotically, sputter eroded surfaces could undergo kinetic
roughening, or can display novel ordered structures with rotated ripples.
Finally, we discuss in detail the existing experimental support for the 
proposed theory, and uncover novel features of the surface morphology and
evolution, that could be directly tested experimentally.
\end{abstract}


% 79.20.Rf = Impact phenomena (including electron spectra and
%sputtering):
%            Atomic, molecular, and ion beam interaction with
%surfaces
% 64.60.Ht = General Studies of phase transitions:
%            Dynamic critical phenomena
% 68.35.Rh = Solid surfaces and solid-solid interfaces:
%            Phase trans. & crit. phenomena

%\newpage

\begin{multicols}{2}

\narrowtext

%\tableofcontents

\section{Introduction}
\label{INT}
\addtocontents{top:}{Introduction}

  Sputtering is the removal of material from the surface of solids 
through the impact of energetic particles \cite{blue,revsput,book}. It 
is a widespread experimental technique, used in a large number of 
applications with a remarkable level of sophistication. It is a basic 
tool in surface analysis, depth profiling, sputter cleaning, micromachining, 
and sputter deposition. Perhaps the largest community of users is in the 
thin film and semiconductor fabrication areas, sputter erosion being 
routinely used for etching patterns  important to the production of 
integrated circuits and device packaging. 

  To have a better control over this important tool, we need to understand
the effect of the sputtering process on the surface morphology. In many 
cases sputtering is routinely used to smooth out surface features. On the 
other hand, some investigations indicate that sputtering can also roughen 
the surface. Consequently, sputter erosion may have different effects on 
the surface, depending on many factors, such as incident ion energy, mass,
angle of incidence, sputtered substrate temperature and material composition.
The experimental results on the effect of sputter erosion on the surface
morphology can be classified in two main classes. There exists ample
experimental evidence that ion sputtering can lead to the development of
periodic ripples on the surface 
\cite{barber,vasiliu,carter0,stevie,ishitani,karen1,karen2,witt,maclaren,umbach,carter2,chason94,chason94a,ChasonMRS,vajo,erleb,aziz,elst,elst1,shichi,rusponi,rusponi2,cirlin1,cirlin2}.
Depending on the sputtered substrate and the sputtering conditions these 
ripples can be surprisingly straight and ordered. However, a number of recent
investigations \cite{eklund,eklund2,krim,yang,wang,smilgies,chan1,konarski,csahok}
have provided rather detailed and convincing experimental evidence that 
under certain experimental conditions ion eroded surfaces become rough, and 
the roughness follows the predictions of various scaling theories
\cite{revrough}. Moreover, these investigations did not find any evidence 
of ripple formation on the surface. Up to recently these two morphological
features were treated separately and  no unified theoretical framework
describing these morphologies was available.

  The first widely accepted theoretical approach describing the process of
ripple formation on amorphous substrates was developed by Bradley and Harper
(BH) \cite{bh}. This theory is rather successful in predicting the ripple
wavelength and orientation in agreement with numerous experimental 
observations. However, a number of experimental results have systematically
eluded this theory. For example, the BH theory predicts an unlimited 
exponential increase in ripple amplitude in contrast with the observed
saturation of the surface width. Similarly, it cannot account for surface
roughening, or for ripple orientations different from those defined by the
incoming ion direction or perpendicular to it. Finally, recent experiments
\cite{maclaren,umbach} have observed ripples whose wavelength is independent
of the substrate temperature, and linear in the ion energy, in contrast with 
the BH prediction of a ripple wavelength which depends exponentially with
temperature and decreases with ion energy.
%------------------------------------------------------------------------

  In the light of the accumulated experimental results, it is clear that
a theory going beyond the BH approach is required, motivating the results
described in this paper. Thus here we investigate the morphology of
ion-sputtered amorphous
surfaces aiming to describe in {\it an unified framework} the
dynamic and scaling behavior of the experimentally observed surface
morphologies. For this we derive a  nonlinear theory that describes the time
evolution of the surface morphology. 
At short time scales the nonlinear theory predicts the development of a
periodic ripple structure, while at large time scales the surface morphology 
may be either rough or dominated by new ripples, that are different from 
those existing at short time scales. We find that transitions may take place
between various surface morphologies as the experimental parameters (e.g.\ 
angle of incidence, penetration depth) are varied. Usually stochastic 
equations describing growth and erosion models are constructed using symmetry
arguments and conservation laws. In contrast, here we show that for sputter
eroded surfaces the growth equation can be {\it derived} directly from a
microscopic model of the elementary processes taking place in the system. 
A particular case of our theory was presented in \cite{us}.
  In addition, we show that the presented theory can be extended to 
describe low temperature ripple formation as well. We demonstrate that
under certain conditions ion-sputtering can lead to preferential erosion 
that appears as a surface diffusion term in the equation of motion even 
though no mass transport along the surface takes place in the system. To
distinguish it from ordinary surface diffusion, in the following we refer
to this phenomenon as effective smoothing (ES). We calculate
analytically an effective surface diffusion constant accounting for the 
ES effect, and study its dependence on the ion energy, flux, 
angle of incidence, and penetration depth. The effect of ES on
the morphology of ion-sputtered surfaces is summarized in a morphological 
phase diagram, allowing for direct experimental verification of our 
predictions. A restricted study along these lines appeared in \cite{mb}.
%-------------------------------------------------------------------------
%-------------------------------------------------------------------------

  The paper is organized as follows. In Section II we review the recent
advances in the scaling theory of rough (self-affine) interfaces. Section 
III is dedicated to a brief overview of the experimental results on surface
morphology development under ion sputtering. A short summary of the 
theoretical approaches developed to describe the morphology of ion 
sputtered surfaces is presented in  Section IV. This section also contains
a short description of Sigmund's theory of sputtering, that is the basis 
for our calculations. In Section V we derive the nonlinear stochastic 
equation describing sputter erosion. Analysis of this equation is presented 
in Section VI, discussing separately both the high and low temperature 
ripple formation. We compare the predictions of our theory with 
experimental results on surface roughening and ripple formation in 
Section VII, followed by Section VIII, that summarizes our findings.

\section{Scaling theory \index {anisotropic KPZ equation}}
\label{st:2}

  In the last decade we witnessed the development of an array of 
theoretical tools and techniques intended to describe and characterize the
roughening of nonequilibrium surfaces and interfaces \cite{revrough}. 
Initiated by advances in understanding the statistical mechanics of various
nonequilibrium systems, it has been observed that the roughness of many
natural surfaces follows rather simple scaling laws, which can be 
quantified using scaling exponents. Since kinetic roughening is a common 
feature of ion-bombarded surfaces, before we discuss the experimental results 
on sputtering, we need to introduce the main quantities characterizing the
surface morphology. 

Let us consider a two-dimensional surface that is characterized by the height
function $h(x,y,t)$. The morphology and dynamics of a rough surface can be
quantified by the {\it interface width\/}, defined by the rms fluctuations 
in the height $h(x,y,t)$,
\begin{equation}
w(L,t)\equiv \sqrt{{1 \over {L^2}} \sum_{x,y=1,L} [h({x,y})-\bar h]^2},
\label{eq1}
\end{equation}
where $L$ is the linear size of the sample and $\bar h$ is the {\it mean
surface height\/} of the surface
\begin{equation}
\bar h(t) \equiv {1 \over {L^2}} \sum_{x,y=1,L} h({x,y,t}).
\label{eq2}
\end{equation}

  Instead of measuring the roughness of a surface over the
whole sample size $L \times L$, we can choose a window of size $\ell
\times \ell$, and measure the local width, $w(\ell)$. A general 
property of many rough surfaces is that the roughness depends on the 
length scale of observation. This can be quantified by plotting $w(\ell)$ 
as a function of $\ell$. There are two characteristic regimes one can
distinguish:

(i) For length scales smaller than $\ell_\times$, the local width 
increases as
\begin{equation}
w(\ell) \sim \ell^\alpha,
\label{eq3}
\end{equation}
where $\alpha$ is  the {\it roughness exponent}. If we are interested in 
surface phenomena that take place at length scales shorter than $\ell_\times$ 
then we cannot neglect the roughness of the surface. In this regime 
the roughness is not simply a number, but it depends on the length
scale accessible to the method probing the surface.

(ii) For $\ell \gg \ell_\times$, $w(\ell)$ is independent of $\ell$, thus, at
length scales larger than $\ell_\times$, the surface is {\it smooth}. In this
regime we can characterize the surface roughness with the saturation width
$w_{sat}=w(\ell_\times)$.

  The {\it dynamics} of the roughening process can be best characterized
by the time dependent total width (\ref{eq1}). At early times the total 
width increases as $w(L,t) \sim t^\beta$, where $\beta$ is the {\it 
growth exponent}. However, for finite systems, after a crossover time
$t_\times$, the width saturates, following the Family-Vicsek scaling 
function \cite{Family}

\begin{eqnarray}
w(L,t) \equiv t^{\beta} g\left( \frac{t}{L^{z}}\right),
\label{eq4}
\end{eqnarray} 
where $z=\alpha/\beta$ is the dynamic exponent and  $g(u \ll 1) 
\sim 1$, while  $g(u \gg 1 ) \sim u^{-\beta}$.

  Scaling relations such as Eq.\ (\ref{eq4}) allow us to define {\it
universality\/} classes. The universality class concept is a product of 
modern statistical mechanics, and encodes the fact that there are but a few
essential factors that determine the exponents characterizing the scaling 
behavior. Thus different systems, which at first sight may appear to have no
connection between them, behave in a remarkably similar fashion. The values of
the exponents $\alpha$ and $\beta$ are independent of many ``details'' of the
system, and they are uniquely defined for a given universality class. In
contrast, other quantities, such as $A$, $\ell_\times$, or $w_{sat}$, are
non-universal, i.e. they depend on almost every detail of the system.

\section{Experimental results}
\label{er:M}

  The morphology of surfaces bombarded by energetic ions has long
fascinated the experimental community.  Lately, with the development
of high resolution observation techniques such as atomic force (AFM)
and scanning tunneling (STM) microscopies, this 
problem is living a new life.  The various experimental investigations 
can be classified into two main classes. First, early investigations,
corroborated by numerous recent studies, have found 
that sputter eroded surfaces develop a ripple morphology with a rather
characteristic wavelength of the order of a few micrometers 
\cite{barber,vasiliu,carter0,stevie,ishitani,karen1,karen2,witt,maclaren,umbach,carter2,chason94,chason94a,ChasonMRS,vajo,erleb,aziz,elst,elst1,shichi,rusponi,rusponi2,cirlin1,cirlin2}. However, a number of research groups have found no evidence of
ripples, but observed the development of apparently random, rough surfaces
\cite{eklund,eklund2,krim,yang,wang,smilgies,chan1,konarski,csahok}, that
were characterized using scaling theories. In the following we summarize the
key experimental observations for both ripple development and kinetic
roughening.

\subsection{\bf Ripple  formation}
\label{er:A}     

  The ripple morphology of ion bombarded surfaces has been initially  
discovered in the mid 1970's \cite{barber,vasiliu,carter0}. Since then, a 
number of research groups have provided detailed quantitative results 
regarding the ripple characteristics and dynamics of ripple formation. It 
is beyond the scope of this paper to offer a comprehensive review of the vast 
body of the experimental literature on the subject. Thus, we selected a few
experiments that offer a representative picture of the experimental features
that appear to be common to different materials. 

{\it Angle of incidence:}
 An experimental parameter which is rather easy to change in sputtering is
the angle of incidence $\theta$ of the incoming ions relative to the normal
to the average surface configuration. Consequently, numerous research 
groups have investigated the effect of $\theta$ on the ripples. These 
results indicate that ripples appear only for a limited 
range of incidence angles, which, depending on materials and ions involved,
typically vary between  30$^\circ$ and 60$^\circ$. 

  Support for a well defined window in $\theta$ for ripple formation was 
offered by Stevie {\it et al.} \cite{stevie}, who observed abrupt secondary 
ion yield changes (correlated with the onset of ripple morphology development) 
in experiments on $6$ and $8$ keV O$^{+}_2$ sputtering of Si and 
$8$, $5.5$, and $2.5$ keV O$^{+}_2$ sputtering of GaAs 
at incidence angles between 39$^\circ$ and 52$^\circ$. These results were
corroborated by Karen {\it et al.} \cite{ishitani,karen1,karen2}, who 
investigated ripple formation on GaAs surfaces under bombardment with
$10.5$ keV O$_{2}^{+}$ ions. They found that ripple formation takes place 
for angles of incidence between 30$^\circ$ and 60$^\circ$ (see 
Table I of Ref. \cite{karen2}). Similarly, Wittmaack \cite{witt} found 
that ripple formation occurs at incidence angles between 32$^\circ$ and
58$^\circ$ during $10$ keV O$^{+}_2$-ion bombardment of a Si target.

{\it Temperature dependence:}
Another parameter that has been found to influence the ripple structure, 
and in particular the ripple wavelength, is the temperature of the substrate,
$T$. Two different behaviors have been 
documented: exponential dependence of the ripple wavelength on $T$ has been
observed at high temperatures, while the wavelength was found to be constant
at low temperatures.

  A series of experiments on the temperature dependence of ripple formation 
were reported by MacLaren {\it et al.} \cite{maclaren}. They studied InP and
GaAs surfaces bombarded with 17.5 KeV Cs$^+$ and 5.5 keV O$_2^+$ ion
beams in the
temperature range from $-50^\circ$ C to 200$^\circ$ C. For GaAs bombarded by
Cs$^+$ ions the ripple wavelength increased from 0.89 $\mu$m to 2.1 $\mu$m 
as the temperature increased from 0$^\circ$ C to 100$^\circ$ C. 
Probably the most
interesting finding of their study was that lowering the temperature, the
ripple wavelength did not go continuously to zero as one would expect, 
since the surface diffusion constant decreases exponentially with  
temperature (see Sect.\ \ref{ta:E}), but rather at approximately 60$^\circ$
C it stabilized at a constant value. MacLaren {\it et al.} interpreted this 
as the emergence of radiation enhanced diffusion, that gives a constant
(temperature independent) contribution to the diffusion constant. Recently,
Umbach {\it et al.} \cite{umbach} have studied the sputter-induced ripple
formation on SiO$_{2}$ surfaces using $0.5-2.0$ keV Ar ion beams. The
temperature dependence of the ripple wavelength $\ell$ was investigated for
temperatures ranging from room temperature to 800$^\circ$ C. It was found 
that for high temperatures, $T \ge 400^\circ$ C, the ripple wavelength 
follows the Arrhenius law $(1/T^{1/2}) \exp{(-\Delta E/2 k_{B}T)}$, 
indicating the thermally activated character of the relaxation processes.
However, at low temperatures the ripple wavelength was independent of
temperature, indicating the presence of a temperature independent 
relaxation mechanism.    

  Results indicating temperature independent non-diffusive relaxation have 
been reported for crystalline materials as well by Carter {\it et al.}
\cite{carter2}. In these experiments Si bombarded with highly energetic 
$10-40$ keV Xe$^{+}$-ions led to ripple formation with wavelength $\ell 
\simeq$ 0.4 $\mu$m for angles of incidence close to 45$^\circ$. Changing 
the surface temperature from 100 K to 300 K the ripple wavelength and
orientation did not change. This observation led the authors to conclude 
that the smoothing mechanism is not of thermal origin. They also
found that the ripple wavevector is relatively insensitive to the primary
ion energy.

{\it Flux and fluence dependence:}
The effect of the flux on the surface dynamics was studied by Chason
{\it et al.} \cite{chason94,chason94a}. In these experiments a 
$1$ keV Xe ion beam was directed towards a SiO$_{2}$ sample with an 
angle of incidence of 55$^\circ$. The typical incoming flux  was 10$^{13}$
cm$^{-2}$s$^{-1}$ and fluence (the number of incoming atoms per surface 
area, playing the role of time) was up to $10 \times 10^{15}$ cm$^{-2}$. 
The surface was analyzed using {\it in situ\/} 
energy dispersive X-ray reflectivity and {\it ex situ} 
AFM. It was found that the interface roughness, which is
proportional to the ripple amplitude, increases linearly with the
fluence. Similar experiments were performed on Ge(001) surfaces 
\cite{ChasonMRS} using $0.3$, $0.5$, and $1$ keV Xe ion beams for 
$T=350^\circ$ C. For flux values up to $3$ $\mu$A/cm$^2$ and fluences 
up to $6 \times 10^{16}$ cm$^{-2}$, the roughness 
is seen to increase as the square of the flux. 

{\it Ion energy:} The ripple wavelength dependence on the incident ion energy
and the angle of incidence was reported in Refs. \cite{ishitani,karen1,karen2}.
The experiments indicate that the ripple wavelength is linear in the energy,
following $\ell \sim \epsilon \cos \theta$. Similar results were obtained 
in Ref. \cite{vajo}, providing an extensive study of ripple formation by
secondary ion spectrometry and scanning electron microscopy. The ripple
topography was observed during O$_2^{+}$ primary ion bombardment of a Si(100)
substrate with ion energies ranging between $1.5$ keV and $9$ keV. No
ripples were observed for energies less than $1.5$ keV or for high energies,
such as $1.5$ keV and $7$ keV,  when Ar$^{+}$ primary ions were used. 
The experiments indicate that the ripple wavelength increases linearly 
from 100 to 400 nm when the energy of the primary ion changes from $1$ to 
$9$ keV. Furthermore, the experimental data indicated that the primary ion
penetration depth $a$ and the ripple wavelength $\ell$ are related by the 
empirical relation $\ell = 40 a$. The wavelength of the ripples is found to
be independent of the primary ion flux and dependent merely on fluence, i.e.\
sputtered depth. The recent results by Umbach {\it et al.} \cite{umbach}
provided further strong evidence for the linear relationship between the ion
energy and the ripple wavelength for SiO$_{2}$ substrates (see below).

{\it Ripple amplitude:}
Indirect results on the ripple amplitude were presented by Vajo {\it et al.}
\cite{vajo} in their study of the surface topography induced secondary ion 
yield changes on SiO$_{2}$ surfaces bombarded by O$^{+}_{2}$ ions. The authors
have found that the yield changes exponentially in the first stages of ripple
development and saturates for large sputtered depth. Direct evidence on ripple
amplitude saturation was obtained by Erlebacher {\it et al.} \cite{aziz}, who
measured the time evolution of the ripple amplitude in experiments 
bombarding Si(100) surfaces with $0.75$ keV Ar$^{+}$ ions. They found that,
while at short times the ripple amplitude increases exponentially, it 
saturates  
%----------------------------------------------------------------------------%
\vbox{
%\widetext
\narrowtext
\begin{table}%[h]
\vspace{-0.3cm}
\hspace{-.15in}
\begin{center}
\begin{tabular}{|c|c|c|c|c|c|}
\hline\hline
Material  &Ion     &Angle       &Ion energy   &Ripple 
& Ref. \\
          &type    &of          &(keV)        &wavelength 
&       \\

          &        &incidence   &             &($\mu$m)
&       \\  
\hline \hline
GaAs(100)  &O$^+_2$   &$39^\circ$ &8    &0.2   &\cite{stevie} 
\\   \hline
GaAs(100)  &O$^+_2$   &$42^\circ$ &5.5  &0.1   &\cite{stevie}
\\   \hline
GaAs(100)  &O$^+_2$   &$37^\circ$  &10.5 &0.23 &\cite{karen1} 
\\   \hline
GaAs(100)  &O$^+_2$   &$42^\circ$  &5.5  &0.13 &\cite{karen2} 
\\   \hline
GaAs(100)  &O$^+_2$   &$39^\circ$  &8.0  &0.21 &\cite{karen2} 
\\   \hline
GaAs(100)  &O$^+_2$   &$37^\circ$  &10.5 &0.27 &\cite{karen2} 
\\   \hline
GaAs(100)  &O$^+_2$   &$57^\circ$  &13   &0.33 &\cite{karen2}
\\  \hline
GaAs       &O$^+_2$ &$40^\circ$   &3.0  &0.075 &\cite{cirlin1}
\\ \hline
GaAs       &O$^+_2$ &$40^\circ$   &7.0  &0.130 &\cite{cirlin1}
\\ \hline
Ge(001)    &Xe$^+$  &$55^\circ$   &1   &0.2   &\cite{chason94}
\\ \hline
Si(001)    &O$_2^+$   &$41^\circ$ &6    &0.4   &\cite{stevie}
\\    \hline
Si(001)    &O$_2^+$   &$42^\circ$ &5.5  &0.5   &\cite{stevie}
\\    \hline
Si(100)    &O$_2^+$   &$39^\circ$ &8    &0.5   &\cite{stevie}
\\   \hline
Si(100)    &O$^+_2$ &$40^\circ$  &3    &0.198  &\cite{vajo}
\\ \hline
Si(100)    &O$^+_2$ &$40^\circ$  &5    &0.302  &\cite{vajo}
\\ \hline
Si(100)    &O$^+_2$ &$40^\circ$  &9    &0.408  &\cite{vajo}
\\ \hline
Si(100)    &Ar$^+$  &$67.5^\circ$ &0.75  &0.57 &\cite{erleb}
\\ \hline
Si         &Xe$^+$  &$45^\circ$  &40  &0.4    &\cite{carter2}
\\ \hline
Si         &O$_2^+$ &$37^\circ$  &12.5 &0.35  &\cite{elst}
\\ \hline 
SiO$_2$    &Ar$^+$  &$45^\circ$  &0.5-2 &0.2-0.55 &\cite{umbach} 
\\   \hline
SiO$_2$      &Xe$^+$  &$55^\circ$   &1   &0.03  &\cite{chason94a} 
\\  
\hline \hline
\end{tabular}
%\vspace{.15in}
\end{center}
\caption{Summary of the ripple characteristics reported in sputter erosion
experiments of non-metallic substrates. In all cases shown, 
the ripple wave vector is parallel to the ion beam
direction. Note that a number of experiments have obtained indirect
information on ripple formation from secondary ion yield changes.
These have not been included in the table.}
\label{table:1}
\end{table}
}
%\narrowtext
%--------------------------------------------------------------------------
after a crossover time has been reached. Furthermore, the
experiments indicate that the crossover time scales with the temperature
induced surface diffusion constant.

{\it Surface chemistry and other morphological features:}
While a number of attempts have been made to explain ripple formation based
on chemical effects, such as O$^+_2$ variations \cite{vajo,elst,elst1}, most
of these studies were contradicted by subsequent investigations  \cite{shichi}
where such chemical component were not present. Furthermore, in Refs.\ 
\cite{ishitani,karen1,karen2} it was unambiguously shown that the process
of ripple formation is not caused by defects or inherited irregularities on 
the surface, but is determined merely by the primary ion characteristics. 
These results indicate that ripple formation is independent of microscopic
details and the surface chemistry. 

{\it Ripple formation on crystalline and metallic surfaces:}
As the discussed experimental results have indicated, ripple formation takes
place under a variety of conditions and on surfaces of different materials,
including both crystalline and amorphous materials. Despite the fact that 
Sigmund's theory, the basis of all theories of ripple formation, has been
developed for amorphous targets, it is worth noting that these approaches
describe many features of ripple formation on crystalline surfaces as well.
However, when discussing ripple formation on crystalline materials, we always
have to be aware that additional effects, induced by the crystalline 
anisotropy, could be present.
  An example of ripple development on crystalline materials has been obtained
for Ag(110) surfaces under low energy ($\epsilon \ge 800$ eV) Ar$^{+}$ primary
beam bombardment by Rusponi {\it et al.} \cite{rusponi}. Ripples with 
wavelength of approximately 600 \AA, oriented along the $\langle 110 \rangle$ 
crystallographic direction, appeared in a temperature range 270$^\circ$ K 
$\le T$ $\le$ 320$^\circ$ K {\it at normal ion incidence}. The ripple 
structure was found to be unstable at room temperature, i.e. substantial
smoothing of the surface with time takes place. The structure depends on the
ion flux and ion energy.
Similar results are available for ion-sputtered Cu$(110)$ monocrystals using 
a $1$ keV Ar$^{+}$ ion beam \cite{rusponi2}. For normal incidence a well 
defined ripple structure was observed with wave vectors whose direction 
changes from $\langle 001 \rangle$ to $\langle 110\rangle $ as the 
temperature of the substrate is raised. Off-normal
sputtering generated ripples whose orientation depends both on the ion 
direction and the surface orientation. The authors suggested that this
phenomenon can be explained in terms of anisotropic surface diffusion. 
%----------------------------------------------------------------------------

 {\it Summary:}
As the presented results indicate, ripple formation on ion-sputtered 
surfaces has been observed by many groups in various systems (for a 
partial summary see Table \ref{table:1}). The main experimental results, 
common to most studied materials, can be summarized as follows:
  
$\bullet$
Off-normal ion bombardment often produces  periodically modulated 
structures (ripples) on the surfaces of amorphous and crystalline 
materials. The wavelength of the ripples $\ell$ is usually of the order 
of tenths of micrometers.

$\bullet$
For non metallic substrates, the orientation of the ripples depends on 
the angle of incidence $\theta$, and in most cases is either parallel or 
perpendicular to the direction of the incoming ions.

$\bullet$
 At low temperatures the ripple wavelength is independent of $T$, while it
follows the Arrhenius law $\ell \sim (1/T^{1/2}) \exp{(-\Delta E/k_{B}T)}$ 
at higher temperatures.

$\bullet$
Numerous experiments find that the ripple wavelength is proportional to the 
ion range, and thus to the ion energy for intermediate energies. 

$\bullet$
The ripple wavelength in many cases is independent of the ion flux, but
systematic flux dependence has also been reported.

$\bullet$
The amplitude of the periodic modulations grows exponentially for early 
times, but saturates after a typical crossover time has been reached. 
In many instances, the ripple wavelength $\ell$ is found to be independent 
of time.

$\bullet$
Evidence for ripple formation was obtained for different materials and
different primary ions, suggesting that the mechanism responsible for ripple
formation is largely independent of surface chemistry, chemical reactions on
the surface, or defects.         
    
\subsection{\bf Kinetic roughening}
\label{er:B}

  Motivated by the advances in characterizing the morphology of rough 
surfaces, recently a number of experimental studies have focused on the 
scaling properties of surfaces eroded by ion bombardment
\cite{eklund,eklund2,krim,yang,wang,smilgies,chan1,konarski,csahok}. 
These experiments have demonstrated that under certain ion bombardment
conditions ripples do not form, and the surface undergoes kinetic roughening.
The goal of the present section is to review the pertinent experimental
results, aiming to summarize the key features that a comprehensive theory
should address.     

{\it Surface roughness and dynamical exponents:}
In the experiments of Eklund {\it et al.} \cite{eklund,eklund2} 
pyrolytic graphite
was bombarded by 5 keV Ar ions, at an angle of incidence of
60$^\circ$. The experiments were carried out for two flux values, 
$6.9\times 10^{13}$ and $3.5 \times 10^{14}$ ions s$^{-1}$ cm$^{-2}$, 
the total fluences being $10^{16}$, $10^{17}$ and $10^{18}$ ions $\cdot$
cm$^{-2}$. STM micrographs indicated that large scale features develop with
continuous bombardment, the interface becoming highly correlated and rough. 
The scaling properties have also been probed using the Fourier transform of
the height-height correlation function, obtaining a dynamic exponent  $z$
in the range $1.6-1.8$, and a roughness exponent  in the range $0.2-0.4$. 
These exponents are consistent with the predictions of the continuum theory,
describing kinetic roughening, proposed by Kardar, Parisi and Zhang (KPZ)
\cite{kpz}, that predicts $z \simeq 1.6$ and $\alpha \simeq 0.38$ (see
section \ref{ta:A:1}).

  A somewhat larger roughness exponent has been measured for samples of iron
bombarded with 5 keV Ar ions at an angle of incidence of 25$^\circ$
\cite{krim}. The interface morphology was observed using STM, and the
height--height correlation function indicated a roughness exponent $\alpha
= 0.53 \pm 0.02$ \cite{krim}. The mechanism leading to such a roughness
exponent is not yet understood in terms of continuum theories, since for
two dimensional surfaces the existing continuum theories predict $\alpha$ 
values of 0.38, 2/3 and 1 \cite{revrough}, far from the observed 
roughness exponent.

  Anomalous dynamic-scaling behavior of sputtered surfaces was reported by
Yang {\it et al.}\cite{yang}. The experiments performed on Si(111)
surfaces with $0.5$ keV Ar$^{+}$ ions with flux  $0.2$ $\mu$A/mm$^2$ in a
wide range of substrate temperatures have provided evidence of 
scaling behavior in the limit of small distances $r$. The height--height
correlation function has been found to follow $C(r) = \langle
(h(r_{o})-h(r+r_{o}))^2 \rangle 
\sim r^{2\alpha} \log t $, with $\alpha \simeq 1.15 \pm 0.08$ for temperatures
lower than 530$^\circ$ C. No roughening was observed for higher 
temperatures, demonstrating the temperature dependence of kinetic roughening. 

{\it  Temperature dependence:}
The effect of surface relaxation due to surface diffusion on roughening of
GaAs(110) surfaces eroded by $2$ keV 
Ar$^{+}$ and Xe$^{+}$ was reported by Wang {et al.} \cite{wang}. They found 
that both the height-height correlation function and the small scale roughness
increase significantly faster during erosion at higher temperatures than at
lower ones. The surface width in these experiments increased 
with $\beta =0.3$ at $T=725$ K and there was no evidence of
scaling for lower temperatures, such as $T = 625$ K. The roughness exponent
has been determined as $\alpha=0.38\pm 0.03$. In general, Ref. \cite{yang} 
concludes that on large scales the surfaces are rougher at higher temperatures,
contrary to the expectation of smaller roughness due to increased
relaxation by surface diffusion. Similar conclusions on the temperature 
dependence of the scaling properties were drawn in Ref. \cite{smilgies}. A
sharp transition between scaling regimes in ion-bombardment of Ge(001)
surfaces with $1$ keV Xe ions was observed at $T_c=488$ K. The regimes above 
and below $T_{c}$ are characterized by dynamic scaling exponents $\beta$ with
values 0.4 and 0.1, respectively. The surface roughness of Si(111)
during low-energy (500 eV) ion bombardment at $T= 610$ K was studied in Ref.
\cite{chan1} using STM. It was found that the rough morphology is consistent
with the early time behavior of the noisy Kuramoto-Sivashinsky (KS) equation 
(see Sect.\ \ref{ta:A:3}). The measured roughness exponent was  
$\alpha=0.7$ and the dynamic exponent was $\beta=0.25$.

{\it Low energy ion bombardment:}
 Recently a number of experiments and simulations have focused on low energy
ion bombardment (i.e., at energies $50$-$500$ eV), for which the 
secondary ion yields are smaller than one 
\cite{chey1,smil1,smil2,chasonE,watan1,poel,bedro}. In this systems, 
the effect of the ions is limited to the surface of the material, the 
collective effect created by the collision cascade being less relevant. 
Often, such low energy sputtering leads to layer-by-layer erosion, almost
mirroring layer-by-layer growth phenomena. The effect of vacancy 
diffusion and Schwoebel barriers can be rather well studied in these systems,
that include Ge(001) surface etching, by $240$ eV Xe ions \cite{chey1,chasonE},
and Si(111) surfaces etched by $100$ eV Ar ions \cite{watan1}. In the 
absence of the collision cascade, ripple formation and kinetic roughening 
seen at higher energies, the subject of this paper, do not appear.   

  Various experimental results on ion-bombardment induced surface 
roughening are summarized in Table \ref{table:2}. These experiments 
demonstrate that kinetic roughening is one of the major experimental
morphologies generated by ion bombardment. However, as Table \ref{table:2}
indicates, there is a considerable scattering in the scaling exponents.
This scattering is not too disturbing at this point: accurate determination
of the scaling exponents from experimental data is rather difficult, since 
often the scaling regime is masked by strong crossover effects. As we 
demonstrate later, due to the separation of the linear and nonlinear
regimes, such crossovers are, indeed, expected in sputter erosion. Thus the 
main conclusion we would like to extract from this section is that numerous
experiments do observe kinetic roughening, and find that scaling concepts
can successfully characterize the surface morphology. It will be a major 
aim of the theory proposed here to account for the origin of kinetic 
roughening and predict the scaling exponents.
 
%----------------------------------------------------------------------------
\vbox{
%\widetext
\begin{table}[hp]
\vspace{-.3cm}
\hspace{-.1in}
\begin{center}
\begin{tabular}{|c|c|c|c|c|c|c|}
\hline\hline
Surface     &Ion        &Ion           &Angle
        &$\alpha$   &$\beta$    &Ref. \\

material    &type       &energy   &of 

         &           &           &       \\
              &           & (keV)       &incidence
         &            &          &       \\
\hline\hline
Graphite   &Ar$^+$ &5      &$60^\circ$  &0.2-0.4 &2.5-2.9   &\cite{eklund}
\\ \hline
Iron   &Ar$^+$ &5      &$25^\circ$  &0.53    &$--$          &\cite{krim}
\\ \hline
Si(111)   &Ar   &0.5    &$0^\circ$   &1.15$^*$ &$--$      &\cite{yang}
\\ \hline
Si(111)    &Ar$^+$ &0.5    &$0^\circ$   &0.7     &0.25      &\cite{chan1}
\\ \hline
GaAs(110) &Ar$^+$ &2      &$0^\circ$    &0.38(3)   &0.3     &\cite{wang}
\\ \hline
Ge(001)    &Xe   &1      &$30^\circ$  &$--$    &0.1, 0.4  &\cite{smilgies}
\\ \hline
Ni, Cr, Cu &Ar$^+$  &1      &$86^\circ$  &$0.49$  &$--$      &\cite{csahok}
\\
\hline\hline
\end{tabular}
\vspace{.15in}
\end{center}
\caption{Summary of the scaling exponents, characterizing the surface
morphology, reported in various experiments on sputter eroded surfaces.
$^*$ - anomalous logarithmic scaling was reported in this experiment.}
\label{table:2}
\end{table}
}
%\begin{multicols}{2}
\narrowtext
%----------------------------------------------------------------------------

\section{\bf Theoretical approaches}
\label{ta:M}

  The recent theoretical studies focusing on the characterization of
various surface morphologies and their time evolution have revolutionized
our understanding of growth and erosion (for reviews, see \cite{revrough}). 
The physical understanding of the processes associated with interface  
roughening require the use of the modern concepts of fractal geometry,
universality and scaling. In Sect.\ \ref{ta:A} we review the major theoretical
contributions to this area, necessary to describe the morphology of 
ion-eroded surfaces. In Sects.\ \ref{ta:B} to \ref{ta:G} we then review 
the available theoretical approaches (whether through continuum equations 
or by the use of discrete atomistic models) that specifically
describe surfaces eroded by ion-bombardment, emphasizing the procedures 
which allow to describe within a continuum approach some of the relevant
physical processes taking place at the surface, such as surface diffusion
and beam fluctuations. 

\subsection{\bf Continuum theories of kinetic roughening }
\label{ta:A}

The full strength of the continuum theories comes from the prediction of 
the asymptotic behavior of the growth process valid in the long time and 
large length scale limits. These limits are often beyond the experimentally
or practically interesting time and length scales. A notable exception 
is sputter erosion, where both the short time ripple development and 
the asymptotic kinetic roughening have been observed experimentally. 
Consequently, next we discuss separately the continuum theories needed
to understand sputter erosion.

\subsubsection{\bf Kardar-Parisi-Zhang (KPZ) equation \index
{anisotropic KPZ equation}}
\label{ta:A:1}

 The time evolution of a nonequilibrium interface can be described 
by the Kardar-Parisi-Zhang (KPZ) equation \cite{kpz}
\begin{equation}
\frac{\partial h}{\partial t} =  \nu \nabla^2 h + \frac{\lambda}{2} (\nabla
h)^2 + \eta .
%~~~~~[KPZ].
\label{eq5} 
\end{equation}
The first term on the rhs describes the relaxation of the interface due to 
the surface tension ($\nu$ is here a positive constant)
and the second is a generic nonlinear term incorporating
lateral growth or erosion. The noise, $\eta({ x},y,t)$, reflects the random
fluctuations in the growth process and is a set of uncorrelated random 
numbers with zero configurational average. 
For one dimensional interfaces the scaling
exponents of the KPZ equation are known exactly, as $\alpha=1/2$, $\beta=1/3$,
and $z=3/2$. However, for higher dimensions they are known only from 
numerical simulations. For the physically most relevant two dimensional
surfaces we have $\alpha \simeq 0.38$ and $\beta \simeq 0.25$ \cite{numer}.

If $\lambda=0$ in Eq.\ (\ref{eq5}), the remaining equation describes the 
equilibrium fluctuations of an interface which tries to minimize its area 
under the influence of the external noise. This equation, first introduced
by Edwards and Wilkinson (EW) \cite{ew}, can be solved exactly due to its
linear character, giving the scaling exponents $\alpha=(2-d)/2$ and
$\beta=(2-d)/4$. For two dimensional interfaces ($d=2$)
we have $\alpha=\beta=0$,
meaning logarithmic roughening of the interface, i.e., $w(L) \sim 
\log L $ for saturated interfaces, and $w(t) \sim \log t $ for early times.

\subsubsection{\bf Anisotropic KPZ equation\index{anisotropic KPZ equation}}
\label{ta:A:2}

 The presence of anisotropy along the substrate \index{anisotropy} 
may drastically change the scaling properties of the KPZ equation. 
As a physical example consider an ion bombarded surface, where 
the ions arrive under oblique incidence in the $x-h$ plane. As a result,
the $x$ and $y$ directions along the substrate will not be equivalent. 
This anisotropy is expected to appear in the erosion equation, leading 
to an {\it anisotropic} equation \index{anisotropic growth equation} of 
the form ($d=2$) 

\begin{eqnarray}
\label{akpz}
{\partial h \over \partial t} & = & \nu_x \partial^2_x h
+\nu_y \partial^2_y h + {\lambda_x \over 2} (\partial_x h)^2 
\nonumber\\ & &+ {\lambda_y \over 2} (\partial_y h)^2
+ \eta(x,y,t), \qquad~~ [{\rm AKPZ}]
\label{eq6}
\end{eqnarray}
where $\partial_x h \equiv \partial h/\partial x$ and 
$\partial_y h \equiv \partial h/\partial y$.
The anisotropy leads to surface tension\index{surface tension} and nonlinear
terms\index{nonlinear terms} that are different in the two directions, which
have been incorporated in the growth equation by considering different values
for the coefficients $\nu$ and $\lambda$ (in Eq.\ (\ref{eq6}), $\nu_x$ 
and $\nu_y$ are positive constants). Equation (\ref{eq6}) is called
the anisotropic KPZ \index{anisotropic $KPZ$ equation} (AKPZ) equation. It
was introduced by Villain\cite{Villain91a}, and its nontrivial properties were
studied by Wolf \cite{wolf,bruins}. 
We note that if $\nu_x=\nu_y$ and $\lambda_x=
\lambda_y$, Eq.\ (\ref{eq6}) reduces to the KPZ equation \index{KPZ equation} 
(\ref{eq5}). The AKPZ equation has different scaling properties depending
on the signs of the coefficients $\lambda_x$ and $\lambda_y$. When $\lambda_x
\cdot \lambda_y < 0$, a surface described by the AKPZ equation has the
same scaling properties as the EW equation. However, when $ \lambda_x \cdot 
\lambda_y > 0$ the scaling properties are described by the isotropic KPZ
equation (\ref{eq5}). Thus, changing the sign of $\lambda_{x}$ or $\lambda_y$
can induce morphological phase transitions from power law scaling 
$(w \sim t^{\beta}; w(L) \sim L^{\alpha})$ to logarithmic scaling $(w \sim
\log t; w(L) \sim \log L )$.  

\subsubsection{\bf Kuramoto-Sivashinsky (KS) equation}
\label{ta:A:3}

  The Kuramoto-Sivashinsky (KS) equation, originally proposed to describe
chemical waves and flame fronts \cite{ks}, is a deterministic equation of the
form:
\begin{equation}
\frac{\partial h}{\partial t} =  -|\nu| \nabla^2 h - K \nabla^4
h + \frac{\lambda}{2} (\nabla h)^2 \\ 
~~~~~[KS].
\label{eq7} 
\end{equation}
While it is deterministic, its unstable and
highly nonlinear character gives rise to 
chaotic solutions. The analysis of the KS equation for one dimensional
surfaces shows
\cite{ks1+1,yakhot,zaleski,hayot,sneppen,l'vov,chow} that in the limit of 
long time and length scales, the surface described by the KS equation is 
similar to that described by the KPZ equation, i.e.\ obeys self-affine 
scaling with exponents $z=3/2$ and $\beta=1/3$. The short time scale
solution of KS equation reveals an unstable pattern-forming behavior,
with a morphology reminiscent of ripples \cite{ks1+1}. 
For two dimensional surfaces, however, the results
are not clear. Computer simulations are somewhat contradictory, providing
evidence for both EW and KPZ scaling \cite{proc,ks2+1}. 

  The anisotropic KS equation was studied in Ref. \cite{Rost&Krug}, 
indicating that for some parameter values the nonlinearities cancel each other, 
and lead to unstable modes dominating the asymptotic morphology. At 
early times the surface displays a chaotic pattern, with stable domains
that nucleate and grow linearly in time until ripple domains of two different
orientations are formed. The pattern of domains of perpendicularly oriented
ripples coarsen with time until one orientation takes over the system. 

  There are various physical systems, including ion sputtering, in which 
the relevant equation for the surface height is a noisy version of the KS
equation (\ref{eq7}) \cite{GB,KM}. Dynamical renormalization 
group analysis \cite{CL} for the surface
dimensions $d=1$ and $2$ indicate that the large distance and long 
time behavior of such 
noisy generalization of Eq.\ (\ref{eq7}) is the same as that of the 
KPZ equation, the $d=2$ result being only quantitative.

\subsection{\bf Bradley and Harper theory of ripple formation}
\label{ta:B}

\begin{figure}[thb]
\begin{center}
%\hskip -1.0 cm
\epsfig{figure=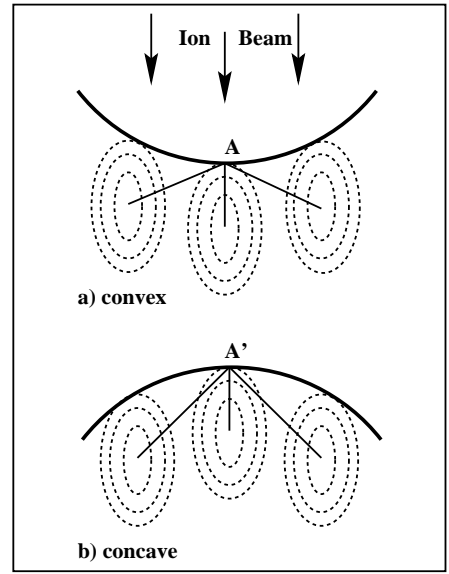,width=5.8 cm,height=7.6 cm}
\end{center}
\vskip 0.5 cm
\caption{Schematic illustration of the physical origin of the instability
during ion erosion of nonplanar surfaces. A surface element with convex 
geometry (a) is eroded faster than that with a concave geometry (b), due
to the smaller distances (solid lines) the energy has to travel from
the impact point to the surface (A or A' points).} 
\label{fig1}
\end{figure}

  A rather successful theoretical model, capturing many features of ripple
formation, was developed by Bradley and Harper (BH)\cite{bh}. They used
Sigmund's theory of sputtering \cite{sig,sigmun2} (see Sect.\ \ref{ta:G}) to 
relate the sputter yield to the energy deposited onto the surface by the 
incoming ions. This work has demonstrated for the first time that the yield 
variation with the local surface curvature induces an instability, which 
leads to the formation of periodically modulated structures. This instability
is caused by the different erosion rates for troughs and crests, the former
being eroded faster than the latter (see Fig.\ \ref{fig1}). Consequently, 
any surface perturbation increases exponentially in time. Viewing the surface
profile as a smooth analytical function of coordinates, BH assumed that 
the surface can be locally  approximated by a quadratic form. Due to the 
erosion mechanism, described in Fig.\ \ref{fig1}, the erosion rate depends on
the local curvature. Combining the curvature dependent erosion velocity with
the surface smoothing mechanism due to surface diffusion 
(see \cite{herr,mullins} and next section), BH derived a linear equation 
for surface morphology evolution
  
\begin{equation}
\frac{\partial h}{\partial t} =  -v(\theta)+
\nu_x(\theta) \partial_x^2 h + 
\nu_y(\theta) \partial_y^2 h - 
K \nabla^4 h.
\label{eq8} 
\end{equation}
Here $\nu_x(\theta),\nu_y(\theta)$ are the effective surface tensions
generated by the erosion process, dependent on the angle of incidence of the
ions, $\theta$, $K$ is the relaxation rate due to surface diffusion 
($K =D_{s}\gamma \Omega^2 n /k_{B}T \exp\left\{ \frac{-\Delta E}{k_{B} T}
\right\}$, where $\Delta E$ is the activation energy for surface diffusion,
$\gamma$ is the surface free energy per unit area, $T$ is temperature,
$D_{s}$ is the surface diffusion constant, $\Omega$ is the atomic volume 
and $n$ is the number of molecules per unit area on the surface).
The physical instability illustrated in Fig.\ 1 leads to the {\em negative}
signs of the $\nu_x, \nu_y$ coefficients in Eq.\ (\ref{eq8}).
Eq.\ (\ref{eq8}) is linearly unstable, with a Fourier mode $k_c$ whose 
amplitude exponentially dominates all the others. This mode is observed
as the periodic ripple structure. Using linear stability analysis, BH derived
from Eq.\ (\ref{eq8}) the ripple wavelength as 
\begin{equation}
\ell_c = 2\pi/k_{c} =2\pi \sqrt{2 K \over | \nu |} \sim (JT)^{-1/2}
\exp\left\{ \frac{-\Delta E}{k_{B} T} \right\},
\label{eq9}
\end{equation}
where $\nu$ is the largest in absolute value of the two negative surface 
tension coefficients, $\nu_x$ and $\nu_y$, and $J$ is the ion flux.
The calculation also predicts that the ripple direction is a function of 
the angle of incidence: for small $\theta$ the ripples are parallel to the 
ion direction, while for large $\theta$ they are perpendicular to it. As\
subsequent experiments have demonstrated \cite{chason94,vajo}, the BH model
predicts well the ripple wavelength and orientation. On the other hand, 
the BH equation (\ref{eq8}) is linear, predicting unbounded exponential
growth of the ripple amplitude, thus it cannot account for the stabilization
of the ripples and  for kinetic roughening, both phenomena being strongly
supported by experiments (see Sect.\ \ref{er:A}-\ref{er:B}).
Furthermore, the BH model cannot account for low temperature ripple 
formation since the only smoothing mechanism it considers is of thermal 
origin. At low temperatures the ion energy and flux dependence of the 
ripple wavelength also disagree with the BH predictions. Despite these
shortcomings, the BH theory represents a major step in understanding the
mechanism of surface evolution in ion sputtering since for the first time 
it uncovered the origin of the ion induced surface instability. Recently 
a generalization of BH linear theory has been successfully introduced
\cite{rusponi3} to account for the thermally activated {\it anisotropic} 
surface diffusion present in metallic substrates such as Cu(110).
     
\subsection{\bf Surface diffusion and deposition noise}
\label{ta:E}

  At high temperatures surface diffusion and fluctuations in the 
ion beam flux are relevant physical mechanisms 
taking place on the surface \cite{viscous}. In this section, 
we discuss the standard approach to include these phenomena in 
continuum models. Let us consider
the simplest scenario: atoms are deposited on a surface, whereupon they
diffuse. If we assume that surface diffusion is the only relaxation
mechanism present, the height $h$ obeys a continuity equation of the form
\begin{equation}
\frac{\partial h}{\partial t} + \tilde{\nabla} \cdot
\mbox{\boldmath $j$}= 0 ,
\label{eq10}
\end{equation}
where $\mbox{\boldmath $j$}$ is a surface current density tangent to the
surface, and $\tilde{\nabla}$ is calculated in a frame with axes parallel
to the surface \cite{marsili}. In general, $\mbox{\boldmath$j$}$ is given
by the gradient of a chemical potential $\mu$,
\begin{equation}
\mbox{\boldmath $j$} \propto -\tilde{\nabla} \mu(\mbox{\boldmath$r$}, t)
\equiv -\tilde{\nabla}^2 \frac{\delta {\cal F}[h]}{\delta h},
\label{eq11}
\end{equation}
where $\mu$ minimizes the free energy functional of the surface ${\cal F}[h]$
and $\tilde{\nabla}^2$ is the surface Laplacian or the Laplace--Beltrami
operator. Taking the latter to be proportional to the total surface area
\begin{equation}
{\cal F}[h] = \int d\mbox{\boldmath $r$} \sqrt{g},
\label{eq12}
\end{equation}
with $g$ as defined in Appendix \ref{AP:a}, and neglecting third or higher
powers of derivatives of $h$ one arrives \cite{mullins} at
\begin{equation}
\frac{\partial h}{\partial t} = - K \nabla^2 (\nabla^2 h) \equiv
- K \nabla^4 h,
\label{eq13}
\end{equation}
where $K$ is a positive constant. Eq.\ (\ref{eq13}) is the so-called linear
MBE equation \cite{revrough}. For an amorphous solid in equilibrium with
its vapor Eq.\ (\ref{eq13}) was obtained in \cite{herr,mullins}, together
with the expression for the coefficient $K$ as in Eq.\ (\ref{eq8}).

  In addition to the deterministic processes, there is considerable
randomness in sputter erosion due to 
fluctuations in the intensity of the ion beam. The ion flux is defined as
the number of particles arriving on the unit surface (or per lattice site)
in unit time. At large length scales the beam flux is homogeneous with an 
average intensity $J$, but there are local random fluctuations, $\eta({\bf x},t)
\equiv \delta J({\bf x},t)$, uncorrelated in space and time.  We can include
fluctuations in Eq.\ (\ref{eq13}) by considering the ion flux to be the
sum of the average flux $J$ and the noise $\eta$, which has zero average,

\begin{equation}
\langle\eta({\bf x},t)\rangle=0
\label{eq15}
\end{equation}
and is uncorrelated,
\begin{equation}
\langle\eta({\bf x},t)\eta({\bf x}',t')\rangle=
J \delta({\bf x}-{\bf x}')\delta(t-t'),
\label{eq16}
\end{equation}
where we have assumed a Poisson distribution for the shot noise.
Consequently, the stochastic growth equation describing surface diffusion
and fluctuations in an erosion process has the form

\begin{equation}
{\partial h \over \partial t } = - K \nabla^4 h - J  + \eta({\bf x},t).
\label{eq17}
\end{equation}
 This variant of Eq.\ (\ref{eq13}) was introduced independently by Wolf and
Villain \cite{Wolf90b}, and by Das Sarma and Tamborenea \cite{DasSarma91},
and played a leading role in developing our understanding of MBE. We will use
the methods leading to (\ref{eq17}) to incorporate the smoothing  by
surface diffusion in our model of ion erosion. Note, however, that as numerous
experimental studies \cite{r1,r2,r3,r4,rossn,rossn2} indicate, ion bombardment
leads to an enhancement of the surface adatom mobility and thus may
drastically change the relaxation mechanism, as compared to regular
surface diffusion. 

\subsection{\bf Microscopic models of ripple formation and roughening}
\label{ta:D}

  Computer simulations provide invaluable insight into microscopic processes
taking place in physical systems. Consequently, a number of recent studies
have focused on modeling ripple formation at the microscopic level. These
studies have proven useful in resolving issues related to low temperature 
ripple formation and provided important ideas regarding the physical 
mechanism governing ripple formation 
\cite{protsenko,koponen1,koponen4,koponen5,koponen,koponen3,koponen2,cuerno1,TMC,blee}.
Here we shortly discuss the conclusions reached in some of the most
representative numerical work. 

Monte Carlo simulations of sputter-induced roughening were reported by
Koponen {\it et al.} 
\cite{koponen1,koponen4,koponen5,koponen,koponen3,koponen2}. 
Roughening of amorphous carbon 
surfaces bombarded by $5$ keV Ar$^{+}$ ions was studied in 
\cite{koponen1,koponen4,koponen5} for incidence angles between 
$0^\circ$ and $60^\circ$. It
was found that ion bombardment induces self-affine topography on the
submicrometer scale, the roughness exponent being $\alpha \simeq 0.25-0.47$
depending on the angle of incidence \cite{koponen4,koponen5}. 
The growth exponent 
$\beta$ was found to be strongly dependent on the relaxation mechanism
used and changed from $\beta \simeq 0.3$ in the model without relaxation 
to $\beta \simeq 0.2 - 0.14$ when different relaxation rules were used in
the simulations. At the same time the roughness exponent $\alpha$ was 
found to be relatively insensitive to the relaxation process on the
nanometer scales. Analogous results were obtained for C ions 
\cite{koponen3}. In this Reference, the ripple wavelength
was found to be relatively independent of the
ion energy or the magnitude of surface diffusion. Ripple
formation was observed even at zero temperature, when surface diffusion
was switched off, indicating the presence of ion induced smoothing.
Furthermore, these simulations led to the observation of  traveling
ripples, as predicted by continuum theories (see section \ref{age:A:1}).
Similarly, for $5$ keV Ar$^{+}$ bombardment of amorphous carbon substrates,
the ripple wave vector is seen \cite{koponen2}
to change from parallel to normal to the 
beam direction
as the incidence angle is increased, in agreement with BH linear theory,
(see Sect.\ \ref{ta:M}). 
The ripple structure was again observed even when no explicit relaxation 
mechanism was 
incorporated in the simulations, and ripple travelling also occurs.
For length scales
comparable to the cascade dimensions, self-affine topography is observed.

 A discrete stochastic model was introduced in Ref.\cite{cuerno1,TMC}
to study the morphological evolution of amorphous one dimensional surfaces
under ion-bombardment. This is a solid-on-solid model incorporating the erosion 
rate dependence of surface curvature, the local slope dependence of the 
sputtering yield, and thermally activated surface diffusion. Up to 
four different dynamical regimes have been identified. Initially the surface 
relaxes by surface diffusion with a growth exponent $\beta \simeq 0.38$,
until the onset of the linear BH instability. The instability induces rapid 
growth ($\beta > 0.5$). In this regime the local slopes increase rapidly,
which triggers non-linear effects eventually stabilizing the 
surface, $\beta$ taking up the EW value $\beta \simeq 1/4$, which indicates
that an effective positive value of the surface tension has been generated.
Finally, in the asymptotic time limit $\beta$ reaches the KPZ value $\beta
\simeq 1/3$. This behavior is consistent with that displayed by the noisy 
KS equation \cite{CL}. Furthermore, the analytical study \cite{LCM} using 
the master equation approach to interface models \cite{VZLW} has shown that 
the noisy KS equation indeed provides the continuum limit of the discrete
stochastic model of Ref. \cite{cuerno1}. Conversely, the results of the 
simulations in \cite{cuerno1} support the theoretical conclusions 
of Ref.\  \cite{CL} that the
asymptotic behavior of the noisy KS equation is the same as that of the KPZ
equation for one and two dimensional surfaces.

  In summary, Monte Carlo simulations of the sputtering process 
of amorphous materials have shown that intermediate and high energy ion 
bombardment may lead to ripple formation in a wide parameter range. 
Furthermore, ripple formation was observed even at {\it zero temperature}.
Computer simulations have also confirmed  the linear dependence of the ripple
wavelength on the incident ion penetration depth and the fact that ripple
formation is a process fully determined by the incident ion characteristics
and not caused by any defects, irregularities or surface chemistry. The 
same simulations have confirmed that under some bombarding conditions the
surface is rough, and obeys scaling.               

\subsection{\bf Sigmund's theory of sputtering}
\label{ta:G}

 The erosion rate of ion bombarded surfaces is characterized by the 
sputtering yield, $Y$, defined as the average number of atoms leaving the
surface of a solid per incident particle. In order to calculate the yield 
and to predict the surface morphology generated by ion bombardment, we first
need to understand the mechanism of sputtering, resulting from the 
interaction of the incident ions and the substrate \cite{blue,book}. In the
process of sputtering the incoming ions penetrate the surface and transfer 
their kinetic energy to the atoms of the substrate by inducing cascades of
collisions among the substrate atoms, or through other processes such as
electronic excitations. Whereas most of the sputtered atoms are located at 
the surface, the scattering events that might lead to sputtering take place
within a certain layer of average depth $a$, which is the average penetration
depth of the incident ion. A qualitative picture of
the sputtering process is as follows: an incoming ion penetrates into the 
bulk of the material and undergoes a series of collisions with the atoms of
the substrate. Some of the atoms undergo secondary collisions, thereby 
generating another generation of recoiling atoms. A vast majority of atoms 
will not gain enough energy to leave their lattice positions permanently.
However, some of them will be permanently removed from their sites.
\begin{figure}[thb]
\centerline{
\epsfxsize=8cm
\epsfbox{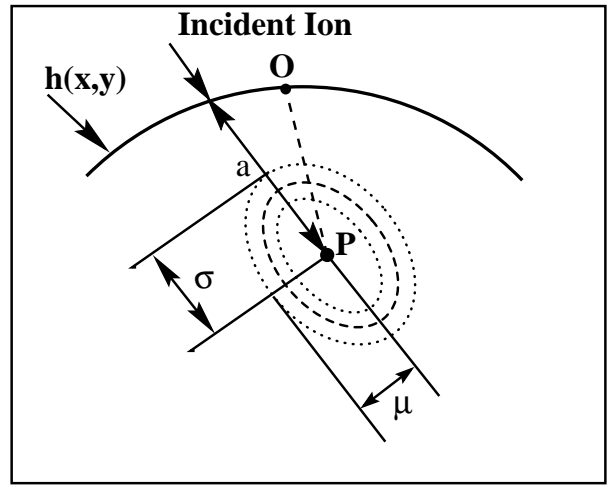}}
\vskip 1.0 cm
\caption{Schematic illustration of the energy distributed by an incident
ion. While the collision process induced by a ion is rather complex, 
according to Sigmund it can be reduced to the following effective process:
The ion penetrates the bulk of the material and stops at point $P$, where 
all its kinetic energy is released and spread out to the neighboring sites
following a Gaussian form with widths $\sigma$ and $\mu$.}
\label{fig2}
\end{figure}
The atoms located in the close vicinity of the surface, which can gain 
enough energy to break their bonds, will be sputtered. Usually the number 
of sputtered atoms is orders of magnitude smaller than the total number of
atoms participating in the collision cascade. 

  A quantitative description of the process of ion sputtering was developed 
by Sigmund \cite{sig}. Assuming an amorphous target, Sigmund derived a set 
of transport equations describing the energy transfer during the sputtering
process. A practically important result of Sigmund's theory is the 
prediction of the deposited energy distribution: the ion deposited at a 
point $P$ inside the bulk of the material spreads its kinetic energy 
according to the Gaussian distribution
\begin{equation}
E(\mbox{\boldmath $r'$})  =  \frac{\epsilon}{(2\pi)^{3/2} \sigma
\mu^2} \exp \left\{ - \frac{Z'^2}{2 \sigma^2} -
\frac{X'^2+Y'^2}{2 \mu^2} \right\}.
\label{eq20}
\end{equation}
In (\ref{eq20}) $Z'$ is the distance from point $P$ to point $O$ measured
along the ion trajectory, and $X'$, $Y'$ are measured in the plane 
perpendicular to it (see Fig.\ \ref{fig2} and the inset of Fig.\ \ref{fig5});
$\epsilon$ denotes the total energy carried by the ion  and $\sigma$ and 
$\mu$ are the widths of the distribution in directions parallel and
perpendicular to the incoming beam respectively. Deviations of the energy
distribution from Gaussian (\ref{eq20}) occur mainly when $M_{1} > M_{2}$,
where $M_{1}$ is the mass of the projectile and $M_{2}$ is the mass of the
target atom. As shown by Sigmund \cite{sig,sigmun2} and Winterbom \cite{win},
electronic stopping doesn't affect much the shape of deposited-energy 
distribution. Subsequently, Monte Carlo simulations of the sputtering
process have demonstrated that the deposited-energy distribution and 
damage distribution can be well approximated by Gaussian for intermediate 
and high energies. In general, comparison of Sigmund's theory with
experimental results has shown that it describes well the qualitative 
behavior of the backsputtering yield, and in many cases there is good
quantitative agreement as well \cite{blue,revsput,book}. 

  A quantity of central importance is the mean path length of an ion 
traveling inside the bulk of the material (see Fig.\ \ref{fig2}), often 
called penetration depth, given by 

\begin{equation}
a(\epsilon) = \frac{1-m}{2m } {\gamma^{m-1}} \frac{\epsilon^{2m }}{NC_m},
\label{eq21}
\end{equation}
where $N$ is the target atom density, $C_{m}$ is a constant dependent on the
parameters of the interatomic potential \cite{sigmun2} and the exponent 
$m = m(\epsilon)$ varies slowly from $m=1$ at high energies to $m \simeq 0$
at very low energies. In the region of intermediate energies, i.e.\ for
$\epsilon$ between $10$ and $100$ keV, $m \simeq 1/2$ and we can approximate
the penetration depth as $a(\epsilon) \sim \epsilon$.

  Eq.\ (\ref{eq20}) describes the effect of a single incident ion. 
Actually, the sample is subject to an uniform flux $J$ of bombarding ions, 
penetrating the solid at different points simultaneously, such that the 
erosion velocity at an arbitrary point $O$ depends  on the total power 
${\cal E}_O$ contributed by all the ions deposited within the range of the
distribution (\ref{eq20}). If we ignore shadowing effects and redeposition 
of the eroded material, the normal erosion velocity at $O$ is given by

\begin{equation}
V_O = p \; \int_{{\cal R}} d\mbox{\boldmath $r$} \;
\Phi(\mbox{\boldmath $r$})
\; E(\mbox{\boldmath $r$}),
\label{eq22}
\end{equation}
where the integral is taken over the region ${\cal R}$ of all points at
which the deposited energy contributes to ${\cal E}_O$, $\Phi(\mbox{\boldmath
$r$})$ is a local correction to the uniform flux $J$ due to variation 
of the local slopes, and the material 
constant $p$ depends on the surface binding energy and scattering 
cross-section \cite{sig,sigmun2} as 
\begin{equation}
p = \frac {3}{4 \pi^{2}} \frac{1}{N U_{o}C_{o}},
\label{eq23} 
\end{equation}
where $U_{o}$ is the surface binding energy and $C_{o}$ is a constant
proportional to the square of the effective radius of the 
interatomic interaction potential.

   While the predictions of Sigmund's theory have been checked on many 
occasions, it also has well known limitations. Next we list two,
that will limit our theory on the surface morphology as well: (a) the 
assumption of random slowing down and arbitrary collisions works
satisfactorily only at intermediate and high energies, i.e. when 
$\epsilon \sim 1-100$ keV, but may break down at low energies; (b) the
assumption of a planar surface may influence the magnitude of the
yield, since surface roughness has a tendency to increase the yield
\cite{mb2,mb3}.

\section{\bf Continuum equation for the surface height}
\label{ConE:M}

  Sigmund's theory, while offering a detailed description of ion bombardment,
is not able to provide direct information about the morphology of ion-sputtered
surfaces. While Eq.\ (\ref{eq22}) provides the erosion velocity, 
in the present form it cannot be used to make analytical predictions regarding
the dynamical properties of surface evolution. To achieve such a predictive
power, we have to eliminate the nonlocality contained in the integral 
(\ref{eq22}) and derive a continuum equation describing the surface 
evolution depending only on the local surface morphology. The main goal of 
this section is to provide a detailed derivation of such an equation starting
from Eq.\ (\ref{eq22}). The properties of the obtained equation will be 
discussed in the following sections. 

\begin{figure}[thb]
\begin{center}
%\hskip -1.0 cm
\epsfig{figure=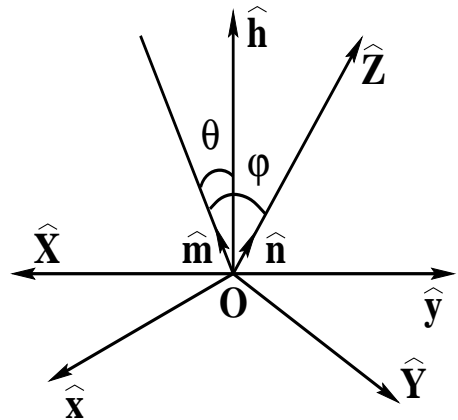,width=6 cm,height=5.5 cm}
\end{center}
\vskip 0.5 cm
\caption{Illustration the local reference frame $(\hat{X},\hat{Y},\hat{Z})$. 
The $\hat{Z}$ axis is parallel to the local normal to the surface $\hat{n}$.
The ions arrive to the surface along $-\hat{m}$. The $\hat{X}$ axis is in 
the plane defined by $\hat{Z}$ and $\hat{m}$, while the $\hat{Y}$ axis is
perpendicular to this plane. The laboratory coordinate frame 
$(\hat{x},\hat{y},\hat{h})$ has its $\hat{h}$ axis perpendicular to the 
flat substrate, $\hat{h}$ and $\hat{m}$ define 
the $(\hat{x},\hat{h})$ plane and $\hat{y}$ is perpendicular to it. 
The incidence angle measured in the local reference frame is $\varphi$, 
and $\theta$ in the laboratory frame.}
\label{fig3}
\end{figure}

 We start by summarizing the main steps that we  follow in the derivation of 
the equation for the surface morphology evolution: 

{\bf (i)} Using Eq.\ (\ref{eq22}), we calculate the normal component of
the erosion velocity $V_O$ at a generic point $O$ of the surface. This
calculation can be performed in a local frame of reference $(\hat{X},\hat{Y},
\hat{Z})$, defined as follows: the $\hat{Z}$ axis is chosen to be parallel 
to the local normal to the surface at point $O$. The $\hat{Z}$ axis forms a
plane with the trajectory of an ion penetrating the surface at $O$. We choose
the $\hat{X}$ axis to lie in that plane and be perpendicular to  $\hat{Z}$.
Finally, $\hat{Y}$ is perpendicular to the $(\hat{X}, \hat{Z})$ plane and 
completes the local reference frame, as shown in Fig.\ \ref{fig3}.

    {\bf (ii)}  Next we relate the quantities measured in the local frame
 $(\hat{X},\hat{Y},\hat{Z})$ to those measured in the laboratory frame
$(\hat{x},\hat{y},\hat{h})$. The latter is defined by the experimental 
configuration as follows:
$\hat{h}$ is the direction normal to the uneroded flat surface. 
The ion direction together with the $\hat{h}$ axis define the 
$(\hat{x},\hat{h})$ plane. Finally, the $\hat{y}$ axis is
perpendicular to the $(\hat{x},\hat{h})$ plane (see Fig.\ \ref{fig3} 
and Appendix \ref{AP:a}).
Furthermore, we have to take into account the fact that the local angle of
incidence $\varphi$, which is the angle between the ion trajectory and the 
{\it local} normal to the surface, changes from point to point along 
the surface. Consequently, $\varphi$ is a function of the local value of 
the slope at $O$ (as measured in the laboratory frame), and the angle 
$\theta$ between the ion trajectory and the normal $\hat{n}$ to the uneroded
surface.

{\bf (iii)} Finally, to obtain the equation of motion for the surface profile
$h(x,y,t)$, we have to project the normal component of the velocity of erosion
onto the global $\hat{h}$ axis. 
The time derivative of $h(x,y,t)$ at any point $O$
on the surface is proportional to the surface erosion velocity $V_{O}$ at 
that point and the local normal is defined by the gradient of the surface
profile $h(x,y,t)$ at $O$. 

  Having defined our objectives and outlined the strategy, we move on to the
description of the calculations. We consider point $O$ to be the origin of the
local system of coordinates $(\hat{X},\hat{Y},\hat{Z})$. To describe the 
surface profile in a neighborhood of $O$ we assume that the surface can 
be described in terms of a smooth analytical infinitely differentiable 
function, i.e.\ there are no singularities and overhangs, and thus we can 
approximate the surface profile at an arbitrary point $(X,Y,Z)$ by \cite{NOTE}
%----------------------------------------------------------------------------
\begin{eqnarray}
Z(X,Y) & \simeq & \frac{\Delta_{20} X^2}{a} + 
\frac{\Delta_{02} Y^2}{a}+ \nonumber \\
& + & \sum_{m,n=0, n+m=3,4}^{4} \frac{\Delta_{nm}}{a^{m+n-1}} X^{n} Y^{m} ,
\label{eq24}
\end{eqnarray}
%-----------------------------------------------------------------------------
where, for later convenience, we introduced the following notations:
\begin{eqnarray}
\Delta_{nm} =  \frac{a^{n+m-1}}{n!m!}
\frac{\partial^{n+m} Z(X,Y)}{\partial^{n} X \partial^{m} Y}. 
\label{eq25}
\end{eqnarray}
Here $\Delta_{20}$ and $\Delta_{02}$ are proportional to the principal
curvatures of the surface, i.e., to the inverses of the 
principal radii of curvature, $R_{X}$ and $R_{Y}$. It must be noted that,
in our approximation, $\hat{X}$ and $\hat{Y}$  (see Fig.\
\ref{fig3}) are the two principal directions of the surface at $O$, 
along which the curvatures are extremal. This implies the absence
in Eq.\ (\ref{eq24}) of cross-terms of the type $\sim X Y$, i.e., we 
neglected the term $\partial^2 Z(X,Y)/\partial X \partial Y$ at $O$. 

  Due to its exponential nature, the deposited energy distribution 
(\ref{eq20}) decays very fast and, consequently, only particles striking 
the surface at a point $(X,Y,Z)$ such that $X/a$, $Y/a$ are of order unity,
contribute non-negligibly to the energy reaching the surface at $O$. 
We further assume that the surface varies slowly enough so that $R_X$, $R_Y$
and the inverses of the higher order derivatives are much larger than 
the penetration depth $a$, i.e. the surface is smooth on length scales close
to $a$ (this fact is supported by nearly all experimental results). Now 
we can calculate the various factors appearing in the integral 
(\ref{eq22}).

\begin{figure}[thb]
\centerline{
\epsfxsize=6cm
\epsfbox{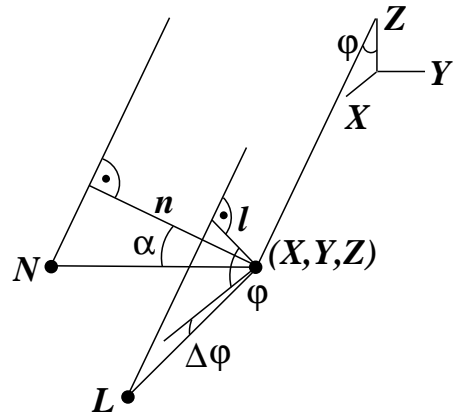}}
\vskip 1.0 cm
\caption{Illustration of the calculation of the local correction to the
average flux $J$ due to the surface curvature.}
\label{fig4}
\end{figure}

To proceed with Eq.\ (\ref{eq22}) we note that, 
with respect to local surface orientation,  only the normal component of the
incident flux contributes to ion erosion. Figure \ref{fig4} illustrates
the calculation of the normal component of the flux. In the figure we
consider a point at the surface $(X,Y,Z)$, and two other points also on 
the surface, at infinitesimal distances $L$ and $N$ away from the former.
We can estimate the correction to the average flux $J$ due to the surface 
slopes by projecting a square perpendicular to the ion beam with area 
$n\times l$ onto the surface area element intersected by the ion 
trajectories. The result is
%----------------------------------------------------------------------
\begin{equation}
\Phi(\mbox{\boldmath $r$}) \simeq J \frac{l \; n}{L \; N},
\label{eq27}
\end{equation}
where $J$ is the average flux. From Fig.\ \ref{fig4},
\begin{equation}
\Delta \varphi = \tan^{-1} \left( \frac{\partial Z}{\partial X} \right)
\simeq \frac{\partial Z}{\partial X},
\label{eq27a} 
\end{equation}
and 
\begin{equation}
\frac{\ell}{L} = \cos(\varphi + \Delta \varphi) \simeq \cos \varphi-
\frac{\partial Z}{\partial X} \sin \varphi.
\label{eq28}
\end{equation}
On the other hand, we also have (see Fig.\ \ref{fig4})
\begin{equation}
\frac{n}{N} = \cos \alpha \simeq 1 - \frac{1}{2}
\left(\frac{\partial Z}{\partial Y}\right)^2 \simeq 1,
\label{eq29}
\end{equation}
so that, combining Eqs.\ (\ref{eq27})-(\ref{eq29}),  and neglecting powers 
of derivatives of the height, we obtain the correction
to the flux
\begin{equation}
\Phi(\mbox{\boldmath $r$}) \simeq J (\cos \varphi + (\partial_{X} Z)
\sin \varphi).
\label{eq30} 
\end{equation}
%----------------------------------------------------------------------
%----------------------------------------------------------------------
\begin{figure}[thb]
\centerline{
\epsfxsize=7cm
\epsfbox{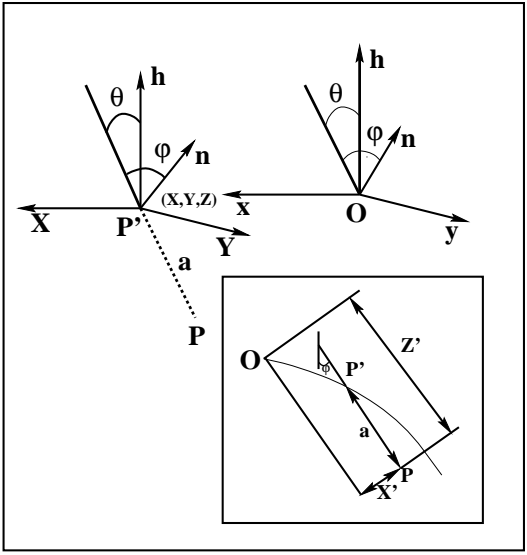}}
\vskip 1.0cm
\caption{Reference frame for the calculation of the erosion velocity at
point $O$. Following a straight trajectory (thick solid line) the ion
penetrates an average distance $a$ inside the solid (dotted line) after 
which it completely spreads out its kinetic energy. The energy released at
point $P$ contributes to erosion velocity at $O$.  Inset shows the lateral
view for $Y'=0$.}
\label{fig5}
\end{figure}
%------------------------------------------------------------------------

  Within the same approximation, the surface element $d\mbox{\bf r}$ in
Eq.\ (\ref{eq22}) can be obtained in the local coordinate system $(\hat{X},
\hat{Y}, \hat{Z})$ as

\begin{eqnarray}
d\mbox{\bf r} \simeq d X d Y.
\label{eq31}
\end{eqnarray}

 Next we determine the distances $X'$, $Y'$, $Z'$ appearing in the exponential
distribution (\ref{eq20}), evaluating them in the local coordinate system. 
Using Fig.\ \ref{fig5}, we have 
\begin{eqnarray}
X' & = & X \cos \varphi + Z \sin \varphi, 
\nonumber\\
Y' & = & Y , 
\nonumber \\
Z' & = & a + X \sin \varphi - Z \cos \varphi. 
\nonumber \\
\label{eq32}
\end{eqnarray}
%---------------------------------------------------------------------------
 Using these expressions, the correction to the ion flux (\ref{eq30}),
the deposited energy distribution (\ref{eq20}) and the expression for the
surface area element $d\mbox{\bf r}$ (\ref{eq31}), we can calculate the
integral (\ref{eq22}) providing the erosion velocity $V_O$. 
Introducing the dimensionless
variables $\zeta_{X}=X/a$, $\zeta_{Y} = Y/a$, and $\zeta_{Z} = Z/a$,
and extending the integration
limits to infinity, we obtain the following expression for the erosion 
velocity in the laboratory coordinate frame
%----------------------------------------------------------------------------
\begin{eqnarray}
V_O & = & \frac{\epsilon p J a^2}{\sigma \mu^2 (2 \pi)^{3/2} }
\exp{(-a_{\sigma}^2/2)} \nonumber \\
& \times & \int_{-\infty}^{\infty}\!\! \int_{-\infty}^{\infty}\,
d{\zeta_X}\,d{\zeta_Y} (\cos\varphi+
\frac{\partial \zeta_Z}{\partial \zeta_X} \sin\varphi)
\nonumber\\
& \times & \exp(-{\zeta_Y^2} L^2) \exp(-{\zeta_X} A) \exp(- \frac{1}{2} B_1
{\zeta_X^2} ) \nonumber \\
& \times & \exp(-4 D {\zeta_Z^2}) \exp(-2 C {\zeta_X}{\zeta_Z})
\exp( B_2 {\zeta_Z}),
\label{eq33}
\end{eqnarray}
%----------------------------------------------------------------------------
where we used the following notations
\begin{mathletters}
\begin{eqnarray} 
A & = & a_{\sigma}^2 \sin\varphi , \label{eq34a} \\
B_1 & = & a_{\sigma}^2 \sin^2\varphi + a_{\mu}^2 \cos^2\varphi,
\label{eq34b} \\
B_2 & = & a_{\sigma}^2 \cos \varphi , \label{eq34c} \\
C & = & \frac{1}{2} (a_{\mu}^2 - a_{\sigma}^2) \sin \varphi \cos \varphi,
\label{eq34d} \\
D & = & \frac{1}{8} (a_{\mu}^2 \sin^2\varphi + a_{\sigma}^2
\cos^2\varphi), \label{eq34e} \\
L & = & \frac{a_{\mu}}{\sqrt{2}}.
\label{eq34f}
\end{eqnarray}
\end{mathletters}
%-----------------------------------------------------------------------------
It must be noted that Eq.\ (\ref{eq33}) coincides with the two-dimensional
version of the local erosion velocity derived in Ref. \cite{bh}. Now 
we use the approximation for the surface profile given by Eq.\ (\ref{eq24}).
Taking a small $\Delta_{nm}$ (see Eq.\ (\ref{eq25})) expansion of the $C$ and
$B_2$ coefficients in Eq.\ (\ref{eq33}) and evaluating the Gaussian integrals
over $\zeta_{X}$ and $\zeta_{Y}$, we get 
%----------------------------------------------------------------------------
%----------------------------------------------------------------------------
\begin{eqnarray}
V_O & = & \frac{\epsilon p J a^2} {\sigma \mu \sqrt{2 \pi} }
\exp{(-a_{\sigma}^2/2)}\exp\left\{\frac{A^2}{2B_1}\right\}
\frac{1}{\sqrt{B_{1}}} \nonumber \\
& \times & \left[ \cos \varphi+\Gamma_{20} \Delta_{20}+\Gamma_{02}
\Delta_{02}+\Gamma_{30} \Delta_{30} \right. \nonumber \\
& + & \left. \Gamma_{21} \Delta_{21}  \Gamma_{40}\Delta_{40}
+\Gamma_{22} \Delta_{22}+\Gamma_{04}\Delta_{04}\right].
\label{eq35}
\end{eqnarray}
%----------------------------------------------------------------------------
 The expressions for the coefficients $\Gamma_{nm}$ can be found
in Appendix \ref{AP:b}. 

  Next we have to rewrite  $V_O$ in terms of the laboratory coordinates 
$(x,y,h)$, which we perform in two steps. First, we write the angle 
$\varphi$ as a function of $\theta$ and the slopes of the surface at $O$
as measured in the laboratory frame. Second, we perform the transformation
between the local and the laboratory coordinates. For both steps we will
have to make expansions in powers of derivatives of $h(x,y,t)$. In line 
with our earlier assumption on the smoothness of the surface, we  will 
assume that $h$ varies smoothly enough so that we can neglect products 
of derivatives of $h$ for third and higher orders. In the laboratory frame,
the neglect of overhangs allows us to describe a generic point at the 
surface, such as $O$, with coordinates $(x,y,h(x,y))$. Considering now the
unit vectors $\hat{n}$, $\hat{m}$ shown in Fig.\ \ref{fig3}, the angle 
$\varphi$ is given by

\begin{mathletters}
\begin{eqnarray}
\cos \varphi & = & \hat{m} \cdot \hat{n} = \frac{ \cos \theta -
(\partial_x h) \sin \theta}{\sqrt{1+ (\partial_x h)^2 +
(\partial_y h)^2}},
\label{eq36a} \\
\sin \varphi & = & \left(\sin^2\theta + 2 (\partial_x h)
\sin\theta \cos\theta (\partial_x h)^2 \cos^2\theta \right. 
\nonumber\\ & + & \left. (\partial_y h)^2 \right)^{1/2} 
\times \left(1 + (\partial_x h)^2 + (\partial_y h)^2\right)^{-1/2}.
\label{eq36b}
\end{eqnarray}
\end{mathletters}
Thus far, expressions (\ref{eq36a})-(\ref{eq36b}) are exact, and the values 
of $\partial_x h$ and $\partial_y h$ are already evaluated in 
the laboratory frame of reference. To implement our approximations, in 
principle we have to separate the cases for normal ($\theta = 0$) and
off--normal ($\theta \neq 0$) incidence. Nevertheless, it can be shown that
the former case can be obtained as a smooth limit of the latter. Therefore 
in the following we give the expressions pertaining to the off--normal 
incidence and refer the reader to Appendix \ref{AP:c} for details on the 
$\theta = 0$ limit. Expanding (\ref{eq36a}) and (\ref{eq36b}) in powers of 
the surface height derivatives, we obtain
\begin{mathletters}
\begin{eqnarray}
\cos \varphi & \simeq & \cos\theta - (\partial_x h) \sin\theta 
\nonumber\\
& & -\frac{1}{2} ((\partial_x h)^2 + (\partial_y h)^2) \cos\theta,
\label{eq38a} \\
\sin \varphi & \simeq & \sin\theta + (\partial_x h) \cos\theta -
\frac{1}{2} (\partial_x h)^2 \sin\theta 
\nonumber\\
& & + \frac{1}{2} (\partial_y h)^2
\frac{\cos^2\theta}{\sin\theta}. 
\label{eq38b}
\end{eqnarray}
\end{mathletters}
Note that these expressions are invariant under the coordinate
transformation $y \rightarrow -y$, but {\em not} under $x \rightarrow -x$,
a consequence of $\theta$ being non-zero and of our choice of 
coordinates. Naturally, the $\theta \rightarrow 0$ limit restores the
symmetry in the $x$ direction.

  Having obtained the expressions (\ref{eq38a}) and (\ref{eq38b}), we can 
return to Eq.\ (\ref{eq35}) to calculate the dependence of $V_O$ on the 
slopes at $O$. Finally, all derivatives in (\ref{eq25}) have to be 
expressed in terms of the laboratory coordinates. This can be accomplished
given the relation between the base vectors of the local frame
$(\hat{X},\hat{Y},\hat{Z})$ and those of the laboratory frame $(\hat{x},
\hat{y},\hat{h})$, derived in Appendix \ref{AP:a}. If the coordinates of
a generic vector $\mbox{\boldmath $r$}$ are given by 
\begin{eqnarray}
\mbox{\boldmath $r$} & = & X \hat{X} + Y \hat{Y} + Z \hat{Z} 
\;\;\;\; {\rm local} \;\; {\rm frame} , \nonumber \\
\mbox{\boldmath $r$} & = & x \hat{x} + y \hat{y} + z \hat{z}
\;\;\;\;\;\;\;\;\;\; {\rm laboratory} \;\; {\rm frame},
\label{eq39}
\end{eqnarray}
then these quantities are related to each other through
\begin{equation}
\left( \begin{array}{c} 
x \\
y \\
z \\
\end{array}
\right) = {\cal M} \left( \begin{array}{c}
X \\
Y \\
Z \\
\end{array}
\right) , 
\label{eq40}
\end{equation}
where ${\cal M}$ is a matrix which has as columns the components of the 
$(\hat{X},\hat{Y},\hat{Z})$ set of vectors in terms of the $(\hat{x}, 
\hat{y}, \hat{z})$ (see Appendix \ref{AP:a}). 
To obtain the expression for the erosion velocity, analogous to (\ref{eq35}), 
in the laboratory frame, we use Eqs.\ (\ref{eq38a}), (\ref{eq38b}), and
${\cal M}$ along with the chain rule for differentiation, and perform
expansions in powers of derivatives of $h(x,y,t)$. After some algebra we
obtain in the laboratory frame
\begin{eqnarray}
\frac{\partial^{n+m} h}{\partial^{n} X \partial^{m} Y} \simeq
\frac{\partial^{n+m} h}{\partial^{n} x \partial^{m} y},
\label{eq41}
\end{eqnarray}
up to fourth order in products of derivatives of $h(x,y)$. 
To summarize our results
thus far, within the approximations leading to (\ref{eq35})
and neglecting nonlinearities of cubic and higher orders in derivatives of 
$h$ in the laboratory frame, we obtained Eq.\ (\ref{eq41}), providing the
relation between the derivatives in the two reference frames, and relations
(\ref{eq38a}) and (\ref{eq38b}) for the angle of incidence measured
in the local frame as a function of the angle of incidence $\theta$
and of the surface slopes. 

  Finally, to relate the velocity of erosion $V_O$, which is normal to the
surface at $O$, to the velocity of erosion of the surface along the $h$ 
axis, $\partial h/\partial t$, we have to project the former onto the latter,
obtaining
\begin{equation}
\frac{\partial h}{\partial t} = - V_O \sqrt{g},
\label{eq42}
\end{equation}
where the negative sign accounts for the fact that $V_O$ is the rate at which 
the surface is eroded, i.e.\ the average height decreases. Furthermore, taking
into account surface diffusion effects, together with the fluctuations (shot
noise) in the flux of the bombarding particles, as discussed in Section
\ref{ta:E}, we complete (\ref{eq42}) by adding these physical effects 
\begin{equation}
\frac{\partial h}{\partial t} = -V_O \sqrt{g} - K \nabla^4 h 
+ \eta(\mbox{\boldmath $r$},t), 
\label{eq43}
\end{equation}

  Finally, we have to write down the contribution of the $-V_O \sqrt{g}$
term to the evolution equation (\ref{eq43}). Performing a small slope
expansion and using Eqs.\ (\ref{eq35}), (\ref{eq38a}), and (\ref{eq38b}), 
we obtain
%-----------------------------------------------------------------------

\bleq
\begin{eqnarray}
\frac{\partial h}{\partial t} & = & -v_0 + \gamma
\frac{\partial h}{\partial x}+ 
\xi_x \left(\frac{\partial h}{\partial x}
\right) \left(\frac{\partial^2 h}{\partial x^2} \right) 
+\xi_y \left(\frac{\partial h}{\partial x} \right) 
\left(\frac{\partial^2 h}{\partial y^2} \right)
%\nonumber \\
%& & 
+ \nu_x \frac{\partial^2 h}{\partial x^2} 
+\nu_y \frac{\partial^2 h}{\partial y^2} 
+\Omega_1 \frac{\partial^3 h}{\partial x^3}+
\Omega_2 \frac{\partial^3 h}{\partial x \partial y^2} \nonumber \\
& & 
-D_{xy} \frac{\partial^4 h}{\partial x^2 \partial y^2}
-D_{xx} \frac{\partial^4 h}{\partial x^4}
-D_{yy} \frac{\partial^4 h}{\partial y^4}
-K \nabla^4 h 
%\nonumber \\
%& & 
+\frac{\lambda_x}{2}
\left(\frac{\partial h}{\partial x} \right)^2 + 
\frac{\lambda_y}{2}\left(\frac{\partial h}{\partial y} \right)^2
+\eta(x,y,t),
\label{eq44}
\end{eqnarray}
where the coefficients are given by the expressions
\begin{eqnarray}
v_0 & = & F c, 
\label{eq45}
\end{eqnarray}

\begin{eqnarray}
\gamma & = & F \frac{s}{f^2} \left\{ a_{\sigma}^2 a_{\mu}^2 c^2
(a_{\sigma}^2-1) -a_{\sigma}^4 s^2 \right\}, 
\label{eq46}
\end{eqnarray}

\begin{eqnarray}
\nu_x & = & F a \frac{a_{\sigma}^2}{2 f^3} \left\{ 2a_{\sigma}^4
s^4 - a_{\sigma}^4 a_{\mu}^2 s^2 c^2 + a_{\sigma}^2 a_{\mu}^2 s^2 c^2- 
a_{\mu}^4 c^4 \right\}, 
\label{eq47}
\end{eqnarray}

\begin{eqnarray}
\lambda_x & = & F \frac{c}{2 f^4} \left\{ a_{\sigma}^8 a_{\mu}^2
s^4 (3+2c^2) + 4 a_{\sigma}^6 a_{\mu}^4 s^2 c^4 
%\right. \nonumber \\ & & 
- a_{\sigma}^4 a_{\mu}^6 c^4 (1+2s^2) 
\right. \nonumber \\ &-& \left. 
f^2 (2 a_{\sigma}^4
s^2 - a_{\sigma}^2 a_{\mu}^2 (1+2s^2))
%\nonumber \\ & & \left.
-a_{\sigma}^8 a_{\mu}^4 s^2 c^2 - f^4 \right\},
\label{eq48}
\end{eqnarray}

\begin{eqnarray}
\nu_y & = & -F a \frac{c^2 a_{\sigma}^2}{2f},
\label{eq49}
\end{eqnarray}

\begin{eqnarray}
\lambda_y & = & F \frac{c}{2f^2} \left\{ a_{\sigma}^4 s^2 +
a_{\sigma}^2 a_{\mu}^2
c^2 - a_{\sigma}^4 a_{\mu}^2 c^2 - f^2 \right\},
\label{eq50}
\end{eqnarray}

\begin{eqnarray}
\xi_x & = & F a \frac{a_{\sigma}^2 s c}{2f^5} \left\{ 
-6 s^6 a_{\sigma}^{8} + a_{\sigma}^8 a_{\mu}^2 s^4 (4+3c^2)
-a_{\sigma}^8 a_{\mu}^4 c^2 s^2
%\right. \nonumber \\ & & 
+ a_{\sigma}^6 a_{\mu}^4 c^2 s^2 (4-6 s^2)
+ a_{\sigma}^6 a_{\mu}^2 s^4 (-3 +15 s^2) 
\right. \nonumber \\ &+& \left. 
  a_{\sigma}^4 a_{\mu}^4 3 c^2 s^2 (4+3s^2)
- a_{\sigma}^4 a_{\mu}^6 3 c^4 (1+s^2) 
%\nonumber\\ & & \left.
+ a_{\sigma}^2 a_{\mu}^6 c^4 (9-3s^2)
- 3 a_{\mu}^8 c^6 \right\},
\label{eq51}
\end{eqnarray}

\begin{eqnarray}
\xi_y & = & F a \frac{a_{\sigma}^2 s c}{2 f^3} \left\{
-a_{\sigma}^4 a_{\mu}^2 c^2 + a_{\sigma}^4 s^2 (2+c^2) 
%\right. \nonumber\\ & & \left.
- a_{\mu}^4 c^4 + a_{\sigma}^2 a_{\mu}^2 c^2 (3 -2 s^2) \right\}, 
\label{eq52}
\end{eqnarray}

\begin{eqnarray}
\Omega_{1} & = & - F a^2 \frac{3}{6}  \frac{1}{f^2} \frac{s}{a_{\mu}^2}  
\left\{  f^2 - f a_{\sigma}^4 c^2 
%\right. \nonumber\\ & & \left. 
- (a_{\mu}^2-a_{\sigma}^2) c^2 (f+a_{\sigma}^4 s^2) \right\}, 
\label{eq53}
\end{eqnarray}

\begin{eqnarray}
\Omega_{2} & = & F a^2 \frac{1}{6} \frac{1}{f^4} \left\{ -3 s f^2 
( f+a_{\sigma}^4 s^2)
%\right. \nonumber\\ & & \left.
+ a_{\sigma}^2 c^2  (3 a_{\sigma}^2 s f+a_{\sigma}^6 s^3) f 
%\right. \nonumber\\ & & \left. 
+ 2 (a_{\mu}^2-a_{\sigma}^2) c^2 (3 f^2 s  + 
6 a_{\sigma}^4 s^3+ a_{\sigma}^8 s^5)\right\}, 
\label{eq54}
\end{eqnarray}

\begin{eqnarray}
D_{xx} & = & F \frac{a^3}{24} \frac{1}{f^5} \left\{
-4 (3 a_{\sigma}^2 s^2 f+a_{\sigma}^6 s^4 ) f^2
%\right. \nonumber\\ & & \left. 
+a_{\sigma}^2 c^2 (3 f^2+6 a_{\sigma}^4 s^2 f +a_{\sigma}^8 s^4) f 
\right. \nonumber \\ &+& \left. 
 2 (a_{\mu}^2-a_{\sigma}^2) c^2 (15 a_{\sigma}^2 s^2 f^2
+10 a_{\sigma}^6 s^4 f+ a_{\sigma}^{10} s^6)
\right\}, 
\label{eq55}
\end{eqnarray}

\begin{eqnarray}
D_{yy} & = & F \frac{a^3}{24} \frac{1}{f^5} 
\frac{3 a_{\sigma}^2}{a_{\mu}^2}  
\left\{ f^4 c^2 \right\}, 
\label{eq56}
\end{eqnarray}

\begin{eqnarray}
D_{xy} & = & F \frac{6 a^3}{24} \frac{1}{f^5} 
\frac{f^2}{a_{\mu}^2}
%\nonumber \\ & & \times 
\left\{ -2 (a_{\sigma}^2 s^2) f^2 
+a_{\sigma}^2 c^2 (f^2+a_{\sigma}^4 s^2 f) 
%\right. \nonumber \\ & &  
%\left.
+2 (a_{\mu}^2-a_{\sigma}^2) c^2 ( 3 a_{\sigma}^2 s^2 f + 
a_{\sigma}^6 s^4) \right\}.
\label{eq57} 
\end{eqnarray}
\eleq
\narrowtext
\noindent
%-----------------------------------------------------------------------
In the above expressions, we have defined
\begin{equation}
F \equiv \frac{J \epsilon p a}{ \sigma \mu \sqrt{2 \pi f}}
e^{-a_{\sigma}^2 a_{\mu}^2 c^2/2f}.
\label{eq58} 
\end{equation}
and, as introduced in Appendix \ref{AP:b},
\begin{eqnarray}
a_{\sigma} \equiv \frac{a}{\sigma}, %\;\;,\;\;
a_{\mu} \equiv \frac{a}{\mu}, %\;\;,\;\;
\nonumber\\
s \equiv \sin \theta, %\;\;,\;\;
c \equiv \cos \theta, %\;\;,\;\;
\nonumber\\
f \equiv a_{\sigma}^2 s^2 + a_{\mu}^2 c^2.
\label{eq59}
\end{eqnarray}
Equation (\ref{eq44}) with the coefficients (\ref{eq45})-(\ref{eq59}), 
fully describe the nonlinear time evolution of sputter eroded surfaces, 
provided that the leading relaxation mechanisms are 
thermally activated surface diffusion and ion-induced effective 
smoothing. Due to its highly
nonlinear character, Eq.\ (\ref{eq44}) can predict rather complex morphologies
and dynamical behaviors. In the remainder of the paper we will focus on 
the physical interpretation of the coefficients (\ref{eq45})-(\ref{eq59}),
uncovering their dependence on the experimental parameters, and we discuss 
the morphologies described by Eq.\ (\ref{eq44}). Consistent with the 
symmetries imposed by the geometry of the problem, the coefficients in 
Eq.\ (\ref{eq44}) are symmetric under the transformation $y\rightarrow -y$ 
but not under $x \rightarrow -x$, while for $\theta \rightarrow 0$ the 
system is isotropic in the $x$ and $y$ directions, specifically $\gamma = 
\xi_x = \xi_y = \Omega_1 = \Omega_2=0$, $\lambda_x =\lambda_y$, $\nu_x 
= \nu_y$, and $D_{xx}=D_{yy}= \frac{1}{2} D_{xy}$.

\section{\bf Analysis of the growth equations}
\label{age:M}

 This section is devoted to the study of the morphological properties  
predicted by Eq.\ (\ref{eq44}). This is not a simple task, due to large number
of linear and nonlinear terms, each of which influence the surface morphology.  
The complexity of the problem is illustrated by some special cases of Eq.
(\ref{eq44}), for which the behavior is better understood. For example, when
nonlinear terms and the noise are neglected $(\xi_{x}=\xi_{y}=\lambda_x=
\lambda_y=0, \eta=0)$, Eq.\ (\ref{eq44}) reduces to a linear generalization
of BH theory, which 
predicts ripple formation. It is also known that the isotropic KS equation,
obtained by taking  $\nu_x=\nu_y, D_{xx}=D_{yy}=D_{xy}/2$, and $\lambda_x =
\lambda_y$, asymptotically predicts kinetic roughening, with morphology and
exponents similar to those seen experimentally in ion sputtering 
\cite{eklund,eklund2}. 
For positive $\nu_x$ and $\nu_y$, Eq.\ (\ref{eq44}) reduces to
the anisotropic KPZ equation, whose scaling behavior is controlled by the 
sign of the product $\lambda_x \cdot \lambda_y$ \cite{wolf}. Finally, recent
integration by Rost and Krug \cite{Rost&Krug} of the noiseless anisotropic 
KS equation (i.e., when $\eta=0$) showed that when $\lambda_x 
\cdot \lambda_y < 0$, ripples unaccounted for by the linear theory appear,
their direction being rotated with respect to the ion direction. 

  To predict the morphology of ion-sputtered surfaces, we need to gain a
full understanding of the behavior predicted by (\ref{eq44}) in the 
physically  relevant two dimensional case, going beyond the special 
cases. Help is provided by the recent numerical integration of (\ref{eq44})
by Park {\it et al.} \cite{kahng} that indicates a clear separation in time 
of the linear and nonlinear behaviors. The results show that 
before a characteristic time $t_{c}$ has been reached, the morphology is 
fully described by the linear theory, as if nonlinear terms were not
present. However, after $t_{c}$ the nonlinear terms completely determine 
the surface morphology.  These results offer a natural layout for our
discussion. In section \ref{age:A} we will limit our discussion to the 
linear theory. However, even in this case we have to distinguish four 
different cases, depending on whether the surface diffusion in the system is 
thermally generated or of the effective type associated with the ion erosion
process. Consequently, in Sections \ref{age:A:1}
- \ref{age:A:2}, we discuss the high temperature case, when relaxation is 
by thermal surface diffusion, treating separately the symmetric 
($\sigma = \mu$), 
and asymmetric ($\sigma \neq \mu$) cases. Next we turn our attention to
low temperature ripple formation,  when surface relaxation is dominated by
erosion, and we again distinguish the symmetric and asymmetric cases 
(Sects. \ref{age:A:3} and \ref{age:A:4}). Finally, Sections \ref{age:B:1}
- \ref{age:B:4} are devoted to the effect of the nonlinear terms, addressing
such important features as ripple stabilization, rotated ripples and kinetic
roughening.

\subsection{\bf Linear theory}
\label{age:A}

\subsubsection{\bf Ripple formation at high temperatures: Symmetric case}
\label{age:A:1}

  In this section we discuss the process of ripple formation in the symmetric
case $\sigma=\mu$, when the relaxation is by thermally activated 
surface diffusion.
Thus we assume that the magnitude of the thermally activated surface diffusion
coefficient, $K$, is much larger than $D_{xx}$, $D_{xy}$, $D_{yy}$,
generated by the ion bombardment process. This is always the case for high
temperatures since $K$ increases as $(1/T) \exp(-\Delta E/k_{B}T)$ with $T$, 
while ion induced effective smoothing terms are independent of $T$. 
Dropping the nonlinear terms in Eq.\ (\ref{eq44}), we obtain 
\begin{eqnarray}
\frac{\partial h}{\partial t} & = & -v_0 + \gamma
\frac{\partial h}{\partial x} + 
\nu_x \frac{\partial^2 h}{\partial x^2} 
+\nu_y \frac{\partial^2 h}{\partial y^2} 
\nonumber\\ 
& & +\Omega_{1} \frac{\partial^3 h}{\partial x^3} 
+\Omega_{2} \frac{\partial^3 h}{\partial x \partial^2 y}
-K \nabla^4 h+\eta(x,y,t),
\label{eq60}
\end{eqnarray}
where the coefficients can be obtained from Eqs.\ (\ref{eq45}) -(\ref{eq58}) 
by taking $\sigma=\mu$: 

\begin{eqnarray}
v_0 & = & F c, \nonumber\\ 
\gamma& = & F s (a_{\sigma}^2 c^2 - 1), \nonumber\\
\nu_x & = & \frac{Fa}{2} \left\{ 2 s^2- c^2 -
a_{\sigma}^2 s^2 c^2 \right\}, \nonumber\\
\nu_y & = & -\frac{Fa}{2} c^2, \nonumber\\
\Omega_1 & = & -\frac{F a^2 s}{ 2 a_{\sigma}^2} (1- a_{\sigma}^2 c^2) ,
\nonumber\\
\Omega_2 & = & \frac{F a^2}{6 a_{\sigma}^2} s \left\{a_{\sigma}^2
(3 c^2-3 s^2) + a_{\sigma}^4 c^2 s^2 -3\right\}. 
\label{eq61} 
\end{eqnarray}
  
  Since the surface morphology depends on the signs and absolute values of the 
coefficients in Eq.\ (\ref{eq60}), in the following we discuss in detail 
their behavior as functions of the angle of incidence $\theta$ and the 
reduced penetration depth $a_{\sigma}$. 

\begin{figure}[thb]
\begin{center}
\hskip -0.5 cm
\epsfig{figure=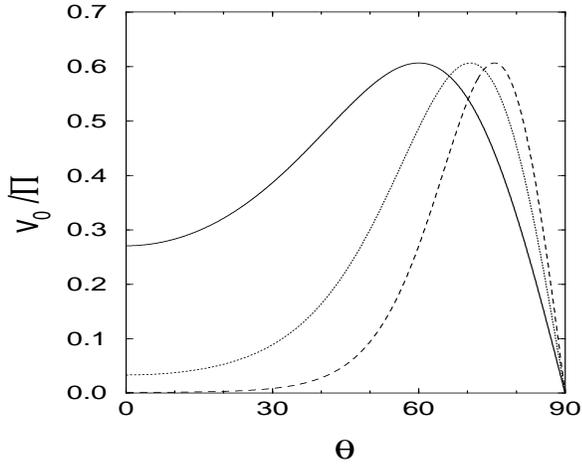,width=3.0in,height=6.4 cm}
\end{center}
%\vskip 0.1 cm
\caption{The erosion velocity, $v_{0}/\Pi$ as a function of $\theta$. The 
three curves correspond to the reduced penetration depth $a_\sigma=1$ (solid
line), $a_\sigma=2$ (dotted line), $a_\sigma=3$ (dashed line). The velocity
has been normalized by a factor $\Pi=p \epsilon J/(\sqrt{2\pi} a)$, 
independent of $\theta$.}
\label{fig6}
\end{figure}

 {\it Erosion velocity, $v_{0}$:}
The $v_{0}$ term describes the erosion velocity of a flat surface. This term
does not affect the ripple characteristics, such as ripple wavelength and 
ripple amplitude, and can be eliminated from the surface evolution equation
by the coordinate transformation $\tilde{h} = h + v_{0}t$. This corresponds
to a transformation to the coordinate frame moving with the eroded surface.
However, since $v_{0}$ is the largest contribution to the erosion rate and 
is the only one that contributes in the linear theory, it is worthwhile to
investigate its dependence on $\theta$ and $a_{\sigma}$. Fig.\ \ref{fig6} 
shows the $v_{0}$ dependence on the angle of incidence $\theta$ for three
different values of the reduced penetration depth $a_{\sigma}$. From Eq.
(\ref{eq61}), $v_{0}$ is positive for all $\theta$ and $a_{\sigma}$. In
experiments $v_{0}(\theta)$ corresponds to the secondary ion yield variation
with the incidence angle $\theta$, i.e. $v_{0}(\theta) = J Y_{flat}(\theta)/n$,
where $n$ is the density of target atoms. 
Note that $v_{0}$ has the 
characteristic increasing part for small $\theta$, followed by saturation 
and decrease for large $\theta$, similar to the measured yield 
\cite{revsput}.

{\it Traveling ripples, $\gamma$, $\Omega_{1}$, $\Omega_{2}$:}
If we consider a periodic perturbation with wave vector $(q_x,q_y)$
in the form 
\begin{eqnarray}
h & = & -v_{0}t +A \exp\left[ i(q_x x+q_y y -\omega t)+rt \right],
\label{eq62}
\end{eqnarray}
from Eq.\ (\ref{eq60}) we obtain the mode velocity
\begin{eqnarray}
\omega  & = & -\gamma q_x + \Omega_{1} q_x^3 + \Omega_2 q_x q_y^2,
\label{eq63}
\end{eqnarray}
and the growth rate
\begin{eqnarray}
r  & = & -(\nu_x q_x^2 +\nu_y q_y^2 + K (q_x^2+q_y^2)^2).
\label{eq64}
\end{eqnarray}
Thus the coefficients $\gamma, \Omega_1, \Omega_2$ contribute to the Fourier 
mode velocity $\omega$ in an anisotropic way that reflects the asymmetry
of the $x$ and $y$ directions for oblique ($\theta \neq 0$) ion incidence. 
The coefficients $\nu_x, \nu_y, K$, on the other hand, contribute to 
the growth rate of the mode amplitude. Carter \cite{NOTE,carter3} pointed 
out that dispersive terms,
such as $\Omega_{1}$ and $\Omega_{2}$, destroy the translational invariance
of the periodic morphology because the different modes travel with different
velocities. Note, however, that the existence of a ripple structure means
that there is essentially only one Fourier mode describing the surface
morphology, which will thus move across the surface with velocity $\omega$.
The coefficient $\gamma$ contributes only to the velocity of the ripples 
along the $x$ direction, leaving unaffected the $y$ component of the 
ripple velocity. Thus, as expected, $\gamma=0$ for normal incidence 
($\theta = 0$). Similarly to the $v_{0}$ term, $\gamma$ does not affect 
the ripple characteristics and can actually be eliminated using the
transformation $\tilde{h}= h(x-\gamma t, t)$.

\begin{figure}[thb]
\begin{center}
\hskip -0.5 cm
%\vskip  0.1 cm
\epsfig{figure=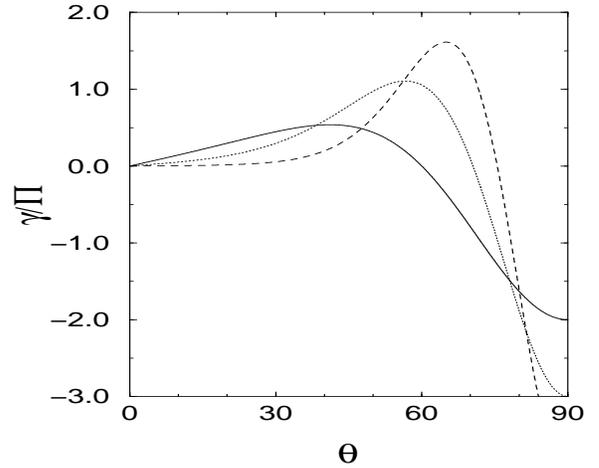,width=3.0in,height=6.4 cm}
\end{center}
%\vskip 0.1 cm
\caption{The coefficient $\gamma/\Pi$ as a function of the angle of 
incidence $\theta$ for three reduced penetration depths: $a_{\sigma}=1$ 
(solid line); $a_\sigma=2$ (dotted line); $a_\sigma=3$ (dashed line).}
\label{fig7}
\end{figure}
As can be seen in Fig.\ \ref{fig7}, $\gamma$ can change sign with $\theta$,
indicating that ripples travel in both positive and negative directions 
along the $x$ coordinate, depending on the angle of incidence and the
penetration depth: ripples travel in the positive $x$ direction for small
$\theta$ and in the negative $x$ direction for larger $\theta$. Travelling
ripples were observed in numerical simulations of Koponen {\it et al.}
\cite{koponen2}.
%----------------------------------------------------------------------------
%----------------------------------------------------------------------------

\begin{figure}[thb]
\begin{center}
%\hskip -1.0 cm
\vskip 0.5 cm
\epsfig{figure=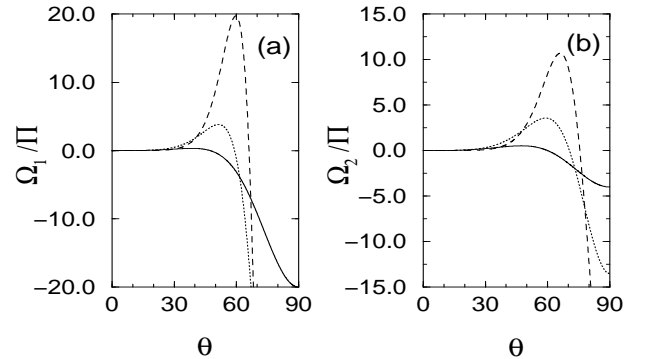,width=3.2in,height=5 cm}
\end{center}
\vskip 0.5 cm
\caption{The reduced third order coefficients $\Omega_1/\Pi$ (a) and
$\Omega_2/\Pi$ (b) as functions of the angle of incidence $\theta$ for
three reduced penetration depths: $a_{\sigma}=2$ (solid line); 
$a_\sigma = 3$ (dotted line); $a_\sigma=4$ (dashed line).} 
\label{fig8}
\end{figure}

As discussed above, the terms $\Omega_{1}$, $\Omega_{2}$ also contribute 
to the travelling of the ripples, and thus have no further effect on the 
surface morphology. 
Fig.\ \ref{fig8} shows the coefficients $\Omega_{1}$ and $\Omega_{2}$ as
functions of the angle of incidence $\theta$. We find that the absolute value
of these coefficients at small angles is small compared to $\gamma$ (see 
Fig.\ \ref{fig7}), thus the main contribution to the ripple velocity comes 
from the ($\gamma \partial h/ \partial x$) term.
On the other hand, for angles $\theta \ge 60^\circ$, these terms are 
comparable to or larger than $\gamma$. 

%----------------------------------------------------------------------------
%----------------------------------------------------------------------------
\begin{figure}[thb]
\begin{center}
%\hskip -1.0 cm
\vskip  0.1 cm
\epsfig{figure=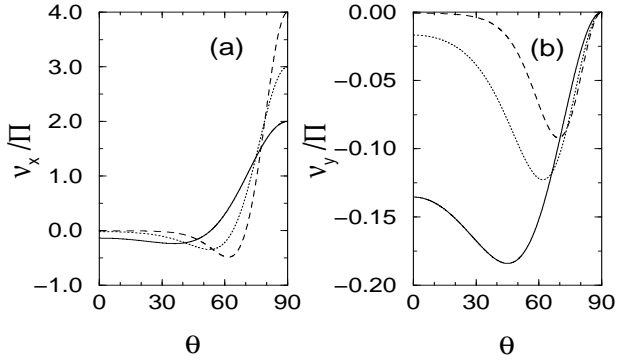,width=3.2in,height=5.0 cm}
\end{center}
\vskip 0.5 cm
\caption{The surface tension coefficients $\nu_x/\Pi$ (a) and
$\nu_y/\Pi$ (b) as functions of the angle of incidence $\theta$ for
three reduced penetration depths: $a_{\sigma}=2$ (solid line); 
$a_\sigma = 3$ (dotted line); $a_\sigma=4$ (dashed line).}
\label{fig9}
\end{figure}
{\it The coefficients $\nu_x$ and $\nu_y$:}
As we discussed above (see Sect.\ \ref{ta:B}) the negative surface tension
coefficients are the origin of the instability responsible for ripple 
formation. Consequently, they play a particularly important role in 
determining the surface morphology. The coefficients $\nu_x$ and $\nu_y$ 
are not equal due to the fact that the direction of the ion beam breaks 
the symmetry along the surface. As seen in Eq.\ (\ref{eq61}), $\nu_y$ is 
always negative, while $\nu_{x}$ can change sign as $\theta$ and 
$a_{\sigma}$ vary, as shown in Fig.\ \ref{fig9}. The sign and the 
magnitude of $\nu_{x}$ and $\nu_{y}$ determine both the wavelength and 
the orientation of the ripples.

{\it Ripple wavelength and orientation:}
The experimental studies on ripple formation have mainly focused on the 
measurement of the ripple characteristics, such as the ripple wavelength 
and amplitude. Thus, a successful theory must address and predict these
quantities. In the following we outline the method for calculating the 
ripple wavelengths $\ell_x$ and $\ell_y$. Taking into account the noisy
character of Eq.\ (\ref{eq60}), the experimentally observed ripple wavelength
corresponds to the unstable Fourier mode which yields the maximum value 
of the structure factor. The structure factor, $S({\bf q},t)$, is calculated 
from the Fourier transform $h({\bf q},t)$ of the instantaneous surface 
profile and is defined as
\begin{eqnarray}
S({\bf q},t) = \langle h(-{ \bf q},t) h({\bf q},t) \rangle,
\label{eq65} 
\end{eqnarray}
where
\begin{eqnarray}
h({\bf q},t) = \int\!\! \frac{d{\bf r}}{(2\pi)^2} \exp(i{\bf q}{\bf r})
h({\bf r}, t).
\label{eq19}
\end{eqnarray}

\begin{figure}[thb]
\begin{center}
%\hskip -1.0 cm
\epsfig{figure=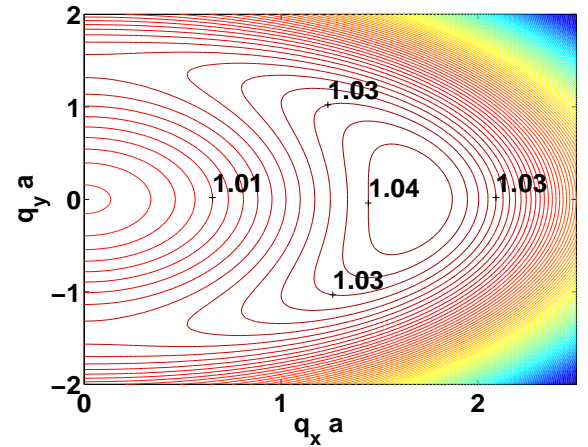,width=3.0in,height=6.0 cm}
\end{center}
\vskip 0.1 cm
\caption{Contour plot of the structure factor $2 S(q_{x},q_{y})/J$ as a 
function of the two dimensionless wavevectors $q_{x}a$ and $q_{y}a$ 
calculated for the angle of incidence $\theta =30^\circ$. The reduced 
coefficients $\nu_{x}/\Pi$ and $\nu_{y}/\Pi$ are obtained using Eq.\
(\ref{eq61}), their values being $\nu_{x}/\Pi=-0.057$, and $\nu_{y}
/\Pi=-0.0418$, while $K/\Pi$ is taken to be 0.01. These parameter values 
correspond to Region I in Fig.\ \ref{fig12}.}
\label{fig10}
\end{figure}

  Fourier transforming Eq.\ (\ref{eq60}) and inserting the expression for the
Fourier transforms of $h({\bf r},t)$ into (\ref{eq65}), we obtain 

\begin{eqnarray}
S({\bf q},t) = \langle h(-{   \bf q},t) 
h ({\bf q},t) \rangle = -\frac{J}{2}  \frac{1-\exp(r({\bf q}) t)}{r({\bf q})},
\label{eq66} 
\end{eqnarray}
where $r$ is the growth rate of the mode ${\bf q}$ given by Eq.\ 
(\ref{eq64}) and is {\em positive} for all unstable Fourier modes 
in the system. We find that, depending on the sign of $\nu_x$ and 
the relative magnitude of $\nu_{x}$ and $\nu_{y}$, 
we can distinguish two cases:

  {\bf(i)} For $\nu_x < \nu_y < 0$, which, according to Eq.\ 
(\ref{eq61}), holds when
\begin{eqnarray}
a_{\sigma} > \sqrt{\frac{2}{c^2}},
\label{eq67}
\end{eqnarray}
the ripple structure is oriented in the $x$ direction, with ripple wavelength

\begin{eqnarray}
\ell_x= 2\pi \sqrt\frac{2 K}{|\nu_x|}.  
\label{eq68}
\end{eqnarray}
This means that the maximum of $S({\bf q},t)$ is at 
$(\sqrt{\frac{|\nu_x|}{2 K}},0)$. 
To illustrate this, in Fig.\ \ref{fig10} we show the dependence of the 
structure factor on the wavevectors $q_{x}$ and $q_{y}$. The contour plot
indicates the existence of a global maximum at $(\sqrt{\frac{|\nu_x|}{2 K}},
0)$, indicating that the ripples are oriented along the $x$ direction. 

(ii) For $\nu_x > \nu_y$, which holds when
 
\begin{eqnarray}
a_{\sigma} < \sqrt{\frac{2}{c^2}},
\label{eq69}
\end{eqnarray}
the ripple structure is oriented along the $y$-direction, with
ripple wavelength 
\begin{eqnarray}
\ell_y=2 \pi \sqrt\frac{2 K}{|\nu_y|}.
\label{eq70}
\end{eqnarray}

%------------------------------------------------------------------------
\begin{figure}[thb]
\begin{center}
%\hskip -1.0 cm
\epsfig{figure=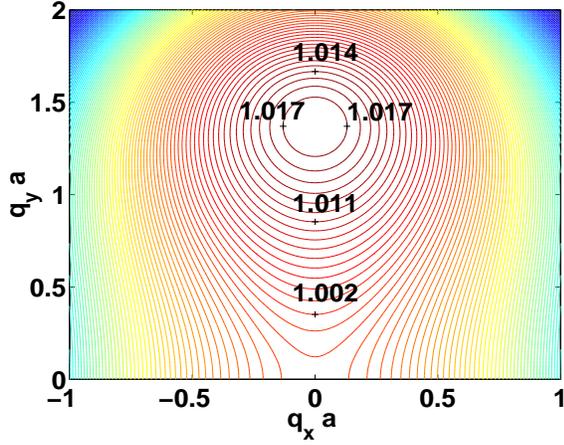,width=3.0in,height=6.0 cm}
\end{center}
\vskip 0.1 cm
\caption{Contour plot of the structure factor $2 S(q_{x},q_{y})/J$ as a
function of the two dimensionless wavevectors $q_{x}a$ and $q_{y} a$ 
calculated for the angle of incidence $\theta=60^\circ$. The reduced
coefficients $\nu_{x}/\Pi$ and $\nu_{y}/\Pi$ are obtained using Eq.\
(\ref{eq61}), their values being $\nu_{x}/\Pi=0.0758$, and $\nu_{y}
/\Pi=-0.0379$, while $K/\Pi$ is taken to be 0.01.These parameter values
correspond to Region II in Fig.\ \ref{fig12}.}
\label{fig11}
\end{figure}

  Figure \ref{fig11} shows an example of this regime, indicating the
existence of a global maximum at point $(0, \sqrt{\frac{|\nu_y|}{2 K}})$, 
corresponding to the ripple structure oriented along the $y$ direction.

\begin{figure}[thb]
\begin{center}
%\hspace*{5cm}
\vskip   0.5 cm
\epsfig{figure=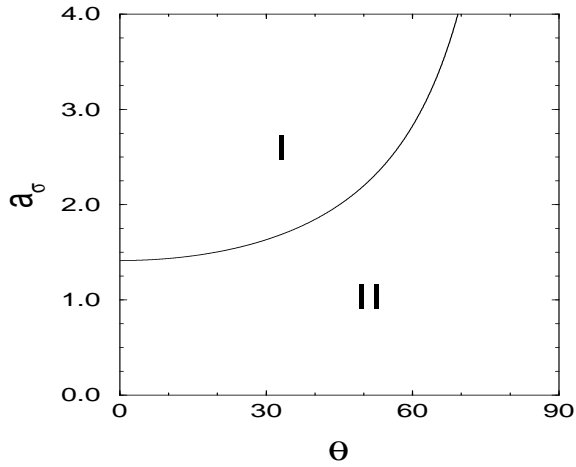,width=3.0in,height=6.4 cm}
\end{center}
\vskip 0.5 cm
\caption{Ripple orientation phase diagram for the isotropic case $\sigma =
\mu=1$. Region I: $\nu_x < \nu_y <0$; Region II: $\nu_x > \nu_y$.}
\label{fig12}
\end{figure}

{\it  Phase diagram for ripple orientation---}
The results obtained on ripple formation can be summarized in a $(\theta,
a_\sigma)$ morphological phase diagram, shown in Fig.\ \ref{fig12}, which 
has the following regions:

 {\it Region I---} For small $\theta$ both $\nu_x$ and $\nu_y$ are negative 
such that $\nu_x < \nu_y$, thus the ripples are oriented along the 
$x$ direction. Their wavelength is
$\ell_x = 2 \pi \sqrt{2 K / |\nu_x|} $. 
  
%-----------------------------------------------------------------------

\begin{figure}[thb]
\begin{center}
%\hskip -1.0 cm
\vskip  0.1 cm
\epsfig{figure=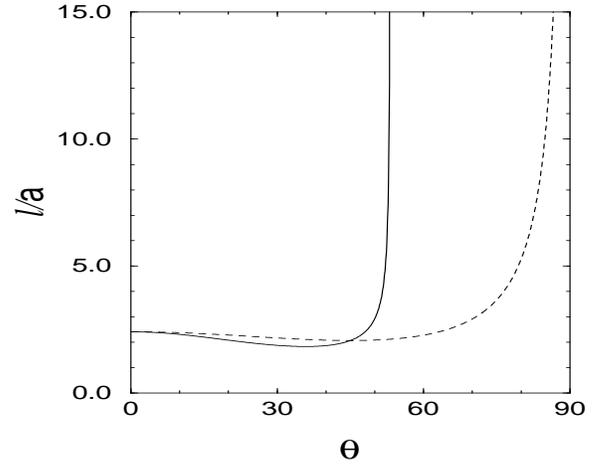,width=3.0in,height=6.4 cm}
\end{center}
\vskip 0.5 cm
\caption{Ripple wavelengths $\ell_x$ (solid line) and $\ell_y$ (dotted
line) as functions of the angle of incidence $\theta$ for $K p \epsilon J/
(\sqrt{2\pi} a) = 0.01$. The reduced penetration depth is taken as
$a_\sigma=2$.}
\label{fig13}
\end{figure}
The amplitude of the ripples is expected to be weakly modulated by the 
larger wavelength $\ell_y = 2 \pi \sqrt{2 K / |\nu_y}|$. The ripple 
amplitude grows as $h_{0} \sim \exp(r_{x} t)$, where $r_{x} =
r(\sqrt{\frac{|\nu_x|}{2K}},0)$   (see Eq.\ (\ref{eq62})). 
The boundary of this region is defined by $\nu_{x}(a_{\sigma},\theta)
=\nu_{y}(a_{\sigma},\theta)$, i.e.
 
\begin{eqnarray}
a_{\sigma} = \sqrt{\frac{2}{c^2}}.
\label{eq71}
\end{eqnarray}

 {\it Region II---} This region is characterized by $\nu_y < \nu_x$. The
ripples are directed along the $y$ direction and have wavelength 
$\ell_y = 2 \pi \sqrt{2 K / |\nu_y}|$. Note that Region II extends down
to small values of the incidence angle $\theta$ for small enough 
reduced penetration depth $a_{\sigma}$. This somewhat unphysical result 
is a consequence of the assumption of a symmetric ($\sigma = \mu$)
distribution of deposited energy. We will see in the next section that
the more physical asymmetric case with $\sigma > \mu$ leads in most cases 
to ripples only oriented along the $x$ direction for small enough
angles of incidence, as generally observed.

  Figure \ref{fig13} shows the $\theta$ dependence of the ripple wavelengths
along the $x$ and $y$ directions. In the framework of this model, where 
thermal surface diffusion is the only smoothing mechanism, the observed
ripple orientation corresponds to the direction featuring the smallest value 
of $\ell$, and changes when $\ell_{x}=\ell_{y}$. The
prediction for the ripple wavelength close to $90^\circ$ is questionable
since reflection \cite{sig,sigmun2} and shadowing \cite{bales}, 
not incorporated in the model, start to play an important role during 
ion-bombardment at these high angles.

{\it Summary:} 
The dependence of $\ell$ on the main physical parameters characterizing 
the sputtering process is given by
\begin{eqnarray}
\ell = 2 \pi \sqrt \frac{2K}{|\nu|} \sim \sqrt\frac{2K}{Fa}
\label{eq72}
\end{eqnarray}
This prediction has a number of consequences, some of which have been verified
experimentally (see Section \ref{cE:M}):

{\bf(a)} Since the penetration depth, $a$, is proportional to $\epsilon^{2m}$
(see \ref{ta:G}), and $\sigma \sim \mu \sim a$, we have $a_{\sigma} \sim $
const, and $F \sim (\epsilon a)/(\sigma \mu) \sim \epsilon^{1-2m}$, which is
independent of $\epsilon$, when $m=1/2$. Consequently 
\begin{equation}
\ell \sim \epsilon^{-1/2}, 
\label{eq73}
\end{equation}
i.e.\ the ripple wavelength is expected to decrease with the ion energy.

{\bf(b)} 
Taking into account that $K$ is independent of the flux and $\nu \sim J$, 
we obtain that the ripple wavelength is also a decreasing function of the 
incident ion flux, given by
\begin{equation}
\ell \sim \frac{1}{J^{1/2}}.
\label{eq74} 
\end{equation}

{\bf(c)} As was mentioned above, the negative surface tension is the 
origin of the instability leading to ripple formation. When both $\nu_{x}$
and $\nu_{y}$ are negative the experimentally observed ripple structure
has the direction for which the growth rate $r$ is largest. However, 
in general, we expect a superposition of both wavelengths, where the long
wavelength will appear as a modulation of the ripple amplitude. Indeed, 
such modulations have been observed both experimentally and numerically 
\cite{kahng}.    

{\bf(d)} An important prediction of this model, illustrated in Fig.\
\ref{fig13}, is the existence of the critical angle $\theta_c$ where the 
ripple orientation changes. In the case when surface diffusion is thermally
activated, this transition coincides with the condition $\nu_{x}=\nu_{y}$.

\subsubsection{\bf Ripple formation at high temperatures: Asymmetric case}
\label{age:A:2}

  The results of the previous section were derived for the isotropic case,
$\sigma = \mu$. While this approximation considerably simplifies our 
discussion, most systems present some anisotropy in the deposited energy 
distribution. In this section we 
demonstrate that the existence of anisotropy does not modify the 
overall qualitative result on the existence of the two parameter regions
corresponding to ripples oriented along the $x$ or $y$ directions. 
However, anisotropy does change the numerical value of the ripple 
wavelength and the exact boundary between the two morphological regions:
we demonstrate that, for large enough anisotropies, if the incidence 
angle is small only ripples oriented along the $x$ direction are possible. 

\begin{figure}[thb]
\begin{center}
\hspace*{1cm}
\vskip 0.1 cm
\epsfig{figure=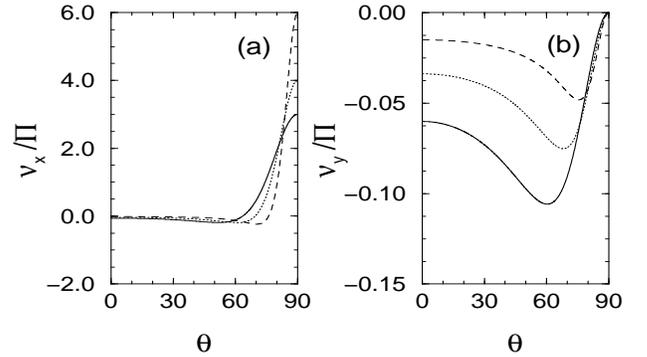,width=3.2in,height=5.0 cm}
\end{center}
\vskip 0.5 cm
\caption{The effective surface tensions $\nu_x/\Pi$ (a) and $\nu_y/\Pi$ 
(b) plotted as functions of the incidence angle $\theta$ in the asymmetric 
case. The curves correspond to values of the asymmetry parameter $\tau=1.5$
(solid line), $\tau=2$ (dotted line), $\tau=3$ (dashed line), with
$a_{\sigma}=2$.} 
\label{fig14}
\end{figure}

Fig.\ \ref{fig14} shows the coefficients $\nu_x$ and $\nu_y$, given by Eq.
(\ref{eq61}), as functions of the angle of incidence $\theta$, for three
different degrees of asymmetry $\tau =\sigma/\mu$ in the 
physical \cite{bh,sig,sigmun2} $\tau > 1$ range. As one can observe the
qualitative behavior of $\nu_x$ and $\nu_y$ is similar to that observed in 
the symmetric case. One interesting feature, however, must be emphasized: 
the increasing asymmetry leads to larger ripple wavelength, since the 
absolute values of $\nu_x$ and $\nu_y$ decrease.
%---------------------------------------------------------------------------
With respect to the the third order linear terms $\Omega_1$ and $\Omega_2$,
their behavior as functions of the angle of incidence can be also seen to 
be qualitatively analogous to the symmetric case. Thus the asymmetry does 
not change our conclusions regarding the travelling ripples.
%---------------------------------------------------------------------------
{\it Ripple wavelength:}
 The calculation of ripple characteristics in the asymmetric case is identical
to the one used in the symmetric case. Therefore, we limit ourselves to the
presentation of the results. Again, there are two possible ripple
directions, and the dominant one can be found from the  maximum of the 
structure factor (\ref{eq66}) or, as can be seen to be equivalent, 
from the maximum of the growth rate (\ref{eq64}): 

 {\bf(i)} When $\nu_x < \nu_y < 0$, i.e.,

\begin{eqnarray}
a_{\sigma} > \sqrt{\frac{ s^2 (2+c^2) + \tau^2 c^2 (1 +2 c^2) - 
\tau^4 c^4}{\tau^2 c^2}}, 
\label{eq75}
\end{eqnarray}
the ripple structure is oriented along the $x$-direction with ripple 
wavelength $\ell_x= 2\pi \sqrt\frac{2 K}{|\nu_{x}|}$. 

\begin{figure}[thb]
\begin{center}
\hskip  -0.5 cm
\vskip   0.1 cm
\epsfig{figure=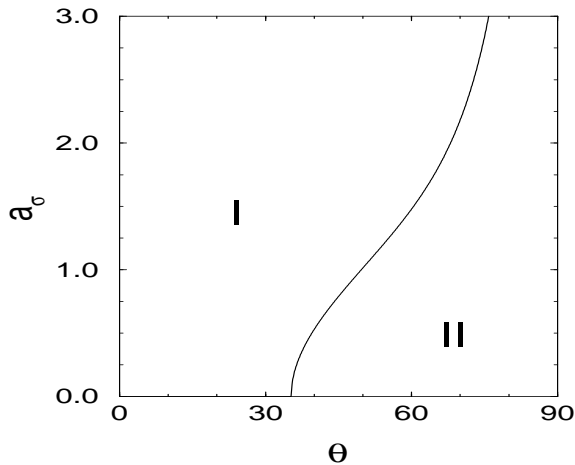,width=3.0 in,height=6.4 cm}
\end{center}
\vskip 0.5 cm
\caption{Ripple orientation phase diagram for the asymmetric case with
asymmetry parameter $\tau=2$ and $a_{\sigma} =2$. 
Region I: $\nu_{x} < \nu_{y} < 0$; 
Region II: $\nu_{x} > \nu_{y}$.}
\label{fig16}
\end{figure}

{\bf(ii)} When $\nu_x > \nu_y$, i.e.,
 
\begin{eqnarray}
a_{\sigma} < \sqrt{\frac{ s^2 (2+c^2) + \tau^2 c^2 (1 +2 c^2) - 
\tau^4 c^4}{\tau^2 c^2}}, 
\label{eq76}
\end{eqnarray}
the ripples are oriented along the $y$-direction, with ripple wavelength
$\ell_y=2 \pi \sqrt\frac{2 K}{|\nu_{x}|}$. 
%------------------------------------------------------------------------
%--------------------------------------------------------------------------
{\it  Phase diagram for ripple orientation:}
  To consider the effect of asymmetry on the different regimes in ion 
sputtering, we have studied the ripple orientation phase diagram for 
different values of $\tau$. As $\tau$ changes, we find a smooth
evolution which does not uncover any new morphological regime. However, 
the topology of the phase diagram does change as $\tau$ increases.
For $\tau < \sqrt{3}$ the topology of the phase diagram is 
similar to the symmetric case (see Fig.\ \ref{fig12}). As Fig.\ \ref{fig16}
illustrates, for $\tau \ge \sqrt{3}$ the ripples oriented along the $y$
direction, predicted by the linear theory for small enough $\theta$ and 
$a_{\sigma}$, are absent, which is consistent with most experimental
observations. The characteristics of Region $I$ and Region $II$ of the 
phase diagram are the same as in the symmetric case. 
  
\subsubsection{\bf Ripple formation at low temperatures: Symmetric case}
\label{age:A:3}

  In the previous two sections we discussed the process of ripple formation 
when the origin of surface smoothing is surface diffusion,
described by the $-K \nabla^4 h$ term. However, in the series expansion
of the erosion velocity we found linear fourth order terms of the form 
$-D_{xx} \partial_x^4 h$, $-D_{xy} \partial_x^2 \partial_y^2 h$, and $-D_{yy} 
\partial_y^4 h$, which are formally equivalent to the thermally induced 
surface diffusion terms. These terms 
arise as a higher order correction to the local surface curvature,
being fully determined by the process of surface erosion. Consequently, 
these terms do not imply actual mass transport along the surface, as 
thermal surface diffusion does. In this section we show that, in some 
parameter regions, these terms have a smoothing effect that counteracts 
the erosion instability, in such a way that they can also lead to ripple 
formation. We believe this explains the ripples observed at low temperatures 
both experimentally \cite{maclaren} and in computer simulations 
\cite{koponen2}. 

  Neglecting the thermally induced relaxation terms (i.e, taking 
$K = 0$), nonlinear terms and the terms $v_{0}$, $\gamma$, $\Omega_1$ 
and $\Omega_2$, that do not affect the ripple characteristics, from
Eq.\ (\ref{eq44}) we obtain the linear equation

\begin{eqnarray} 
\frac{\partial h}{\partial t} & = & \nu_x
\frac{\partial^2 h}{\partial x^2} +
\nu_y \frac{\partial^2 h}{\partial y^2} 
-D_{xx} \frac{\partial^4 h}{\partial x^4}
\nonumber\\ & & -
D_{xy} \frac{\partial^4 h}{\partial x^2 \partial y^2}-
D_{yy} \frac{\partial^4 h}{\partial y^4}+
\eta(x,y,t).
\label{eq78}
\end{eqnarray}
 The expressions for the coefficients of the ion-induced effective smoothing 
terms can be obtained from Eqs.\ (\ref{eq55})-(\ref{eq57}) using $\sigma=\mu$:

\begin{eqnarray} 
D_{xx} & = & \frac{F a^{3}}{24 a_{\sigma}^2 } \left\{a_{\sigma}^4 s^4 c^2 +
 a_{\sigma}^2 ( 6 c^2 s^2 -4 s^4 ) + 3 c^2 -12 s^2 \right\},\nonumber \\
D_{xy} & = & \frac{F a^{3} }{24 a_{\sigma}^2} 6 \left\{a_{\sigma}^2 s^2 c^2+
 c^2-2 s^2 \right\},\nonumber \\
D_{yy} & = & \frac{F a^{3}}{24 a_{\sigma}^2} 3 c^2.
\label{eq79}
\end{eqnarray}
              
From Eq.\ (\ref{eq79}), $D_{yy}$ is always positive, 
while $D_{xy}$ and $D_{xx}$ change sign with $\theta$. 
As we discuss below, the positive $D_{xx}$ and $D_{yy}$ coefficients play a 
role similar to thermally activated surface diffusion.
Furthermore, the absolute value of $D_{xx}$ is comparable with the thermally
activated surface diffusion coefficient even at high temperatures (see Sect.\
\ref{cE:M}). 
%----------------------------------------------------------------------------
%----------------------------------------------------------------------------

\begin{figure}[thb]
\begin{center}
\hskip -0.5 cm
\vskip 0.1 cm
\epsfig{figure=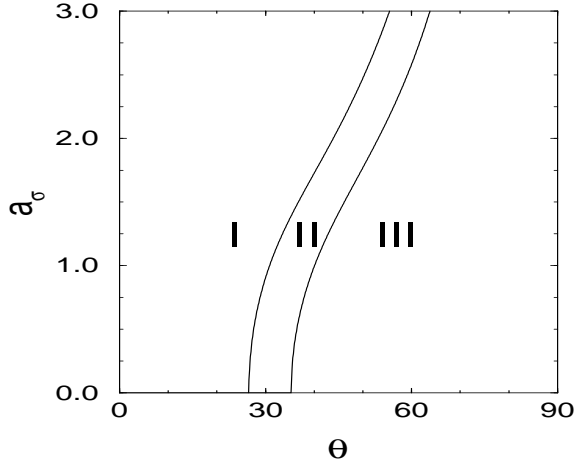,width=3.0 in,height=6.4 cm}
\end{center}
\caption{Morphological phase diagram in the symmetric case $\sigma =
\mu=1$. Different regions correspond to:
Region I:
$\nu_x<0$, $\nu_y < 0$, $D_{xx} > 0$, $D_{yy} >0$ and  $r_x > r_y$;
Region II:
$\nu_x<0$, $\nu_y < 0$, $D_{xx} < 0$, $D_{yy} >0$;
Region III:
$\nu_x>0$, $\nu_y<0$, $D_{xx} <0$ and $D_{yy} >0$. }
\label{fig21}
\end{figure}

\begin{figure}[thb]
\begin{center}
%\hskip -1.0 cm
\epsfig{figure=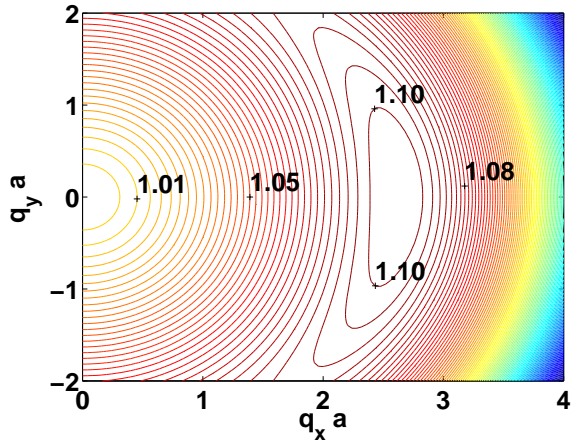,width=3.0 in,height=6.0 cm}
\end{center}
\vskip 0.5 cm
\caption{Contour plot of the structure factor $2 S(q_{x},q_{y})/J$ shown
as a function of the two dimensionless wavevectors $q_{x}a$ and $q_{y}a$
calculated for the angle of incidence $\theta =30^\circ$. The reduced
coefficients $\nu_{x}/\Pi$ and $\nu_{y}/\Pi$ are obtained using  Eq.\
(\ref{eq61}), their values being $\nu_{x}/\Pi=-0.0578$ and $\nu_{y}
/\Pi =-0.0418$. These parameter values correspond to Region I in Fig.\
\ref{fig21}.}
\label{fig18}
\end{figure}

{\it Ripple wavelength and orientation:}
The ripple wavelength and orientation can be calculated following the 
arguments presented in Section \ref{age:A:1}, being determined by the maxima
of the structure factor $S({\bf q},t)$. The growth rate $r$ is now given 
by
\begin{eqnarray}
r(q_{x},q_{y}) & = & -(\nu_x q_x^2 +\nu_y q_y^2 + D_{xx} q_x^4
\nonumber\\ & & +D_{xy} q_x^2 q_y^2+ D_{yy} q_y^4).
\label{eq80}
\end{eqnarray}
%---------------------------------------------------------------------------
In principle the asymmetry of the $D_{ij}$ coefficients may lead to maxima
of $S({\bf q},t)$ at nonzero $q_{x}$ and $q_{y}$ values, which correspond 
to ripples forming a nonzero angle with both the $x$ and $y$ directions.
However, straightforward calculations indicate that the following condition
holds
\begin{eqnarray}
D_{xy} \nu_{y}=2 \nu_{x} D_{yy},
\label{eq75a} 
\end{eqnarray}
for all values of $a_{\sigma}$ and $\theta$ in (\ref{eq61}) and (\ref{eq79}).
This identity implies that no extrema $(q_{x}^{*}, q_{y}^{*})$ of $S({\bf
q},t)$ exist other than of the form $(q_{x}^{*},0)$ or $(0,q_{y}^{*})$. 
Of these two 
solutions the one with the largest positive value of $r(q_{x},q_{y})$ 
corresponds to the observed ripple structure. For small angles of incidence
so that $D_{xx} \ge 0$ (Region I in Fig.\ \ref{fig21}), 
\noindent it can be easily seen that $\nu_{x} < 0$, and the absolute 
maximum of $S({\bf q},t)$ is at $(q_{x}^{*},0)$
with $q_{x}^{*}=\sqrt{|\nu_{x}|/2 D_{xx}}$, thus the ripple structure is
aligned along the $x$ direction (see for example Fig.\ \ref{fig18}).
%---------------------------------------------------------------------------
%---------------------------------------------------------------------------
\begin{figure}[thb]
\begin{center}
%\hskip -1.0 cm
\epsfig{figure=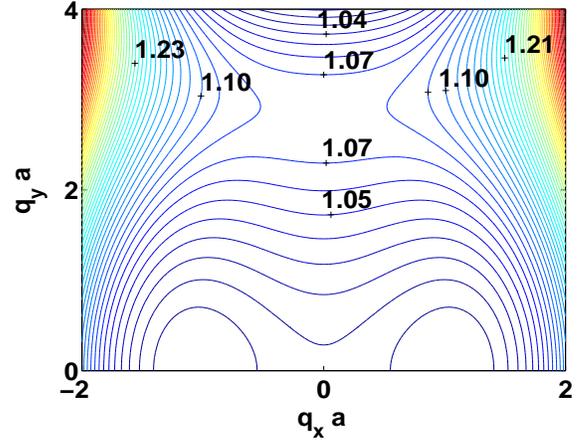,width=3.0 in,height=6.0 cm}
\end{center}
\vskip 0.5 cm
\caption{Contour plot of the structure factor $2 S(q_{x},q_{y})/J$ shown
as a function of the two dimensionless wavevectors $q_{x}a$ and $q_{y}a$
calculated for the angle of incidence $\theta =30^\circ$. The reduced
coefficients $\nu_{x}/\Pi$ and $\nu_{y}/\Pi$ are obtained using  Eq.\ 
(\ref{eq61}), their values being $\nu_{x}/\Pi=-0.0758$ and $\nu_{y}
/\Pi =-0.0379$. These parameters values correspond to Region II in Fig.\
\ref{fig21}.}
\label{fig19}
\end{figure}
%---------------------------------------------------------------------------
Crossing the $D_{xx}=0$ line into Region II in Fig.\ \ref{fig21}
the $(q_{x}^{*},0)$ solution disappears, the structure factor having an
extremum at $(0,q_{y}^{*})$, with $q_{y}^{*} = \sqrt{|\nu_{y}|/2 D_{yy}}$.
However, this extremum is not a {\em global} maximum,
see for example Fig.\ \ref{fig19}. The
lower boundary of Region II is provided by the condition $\nu_{x}=0$.
When we cross it (entering Region III in Fig.\ \ref{fig21}), we have $D_{xx}
< 0$ and $\nu_{x} > 0$. Under this condition, again, there exists an extremum
of $S({\bf q},t)$ at $(q_{x}^{*},0)$, with $q_{x}^{*} =\sqrt{\nu_{x}/2
|D_{xx}|}$. However, the structure factor takes its absolute maximum at
$(0,q_{y}^{*})$.

{\it Phase diagram for ripple orientation:}
In summary, three different regions can be determined in 
the morphological phase diagram shown in Fig.\ \ref{fig21} for the 
case of ion-induced effective smoothing in the symmetric $\sigma =\mu$ 
case.

   {\it Region I---} In this region, the surface tension coefficients
$\nu_x$ and $\nu_y$ are negative, while $D_{xx}$ and $D_{yy}$ are both 
positive. The observed ripple structure corresponds to the maximum of 
$S({\bf q},t)$, which indicates that the ripple structure is oriented along 
the $x$ direction. The lower boundary of this region, separating it from Region
II, is given by the $D_{xx}(a_{\sigma},\theta)=0$ line or, equivalently, by

\begin{eqnarray}
a_{\sigma} =\sqrt{\frac{(2 s^2 -3 c^2) + \sqrt{6 c^4 + 4 s^4}}{s^2 c^2}}. 
\label{eq82}
\end{eqnarray}

   {\it Region II---} In this region both $D_{xx}$ and $\nu_x$ are negative.
This region is bounded below by the $\nu_{x}(a_{\sigma},\theta) = 0$ line, 
given by 

\begin{eqnarray}
a_{\sigma} =\sqrt{\frac{(2 s^2 -c^2)}{s^2 c^2}}. 
\label{eq82X}
\end{eqnarray}

In a continuum description, the maximum of $r({\bf q})$ is at infinite
${\bf q}$, thus our theory possibly breaks down in the sense that not even 
non-linear effects can be expected to stabilize the surface under such 
conditions. In such a case, a higher order Taylor expansion should be 
carried out in Sec.\ \ref{ConE:M} in order to be able to describe our system.
Additional effects, such as shadowing, could also start to play a role
under such conditions. 

  {\it Region III ---} In this region $D_{xx}$ is negative and $\nu_x$ is
positive. Thus the instability given by the negative $D_{xx}$ is smoothed 
out by the positive $\nu_{x}$. Since the structure factor takes on its 
absolute maximum at the finite wave vector $(0,q_y^*)$, in principle there 
is still a ripple structure oriented along the $y$ direction. However, 
remarks similar to those made in Region II might apply here, since we still
have $D_{xx} < 0$.

{\it Summary:}
In the presence of ion-induced effective smoothing the dependence of the ripple
wavelength on the physical parameters is different from the case of
thermal surface diffusion (see Sect.\ \ref{age:A:1}). 
Here we summarize some of the differences. 

  {\bf(a)} The dependence of $\ell$ on the ion energy is given by 
\begin{eqnarray}
\ell = \sqrt{\frac{2D}{|\nu|}} \sim \sqrt{\frac{Fa^3}{Fa}} \sim a 
\sim \epsilon^{2m},
\label{eq83}
\end{eqnarray}
indicating that the {\it ripple wavelength depends linearly on the 
penetration depth $a$}. This is very different from the behavior predicted 
by Eq. (\ref{eq60}), derived for thermal surface diffusion, 
and indicates that monitoring 
the ripple wavelength dependence on $\epsilon$ can be used to identify the
dominant smoothing mechanism. Such a linear behavior of $\ell$ on $\epsilon$
has indeed been seen experimentally (see Section \ref{er:A}). 

{\bf(b)} From  Eq.\ (\ref{eq83}) it also follows that the ripple wavelength
is {\it independent of the incident ion flux}. This prediction is again 
quite different from the case dominated by thermal surface diffusion, 
given by Eq.\
(\ref{eq74}). Such a flux independent behavior has been observed 
experimentally (see Section \ref{er:A}).
 
  Finally, analogues of characteristics (c) and (d), discussed in 
Sect.\ \ref{age:A:1}, apply here as well.

\subsubsection{\bf Ripple formation at low temperatures: Asymmetric case}
\label{age:A:4}

  In this section we discuss the effect of asymmetry ($\sigma \neq \mu$) 
of the energy distribution on the morphological regimes predicted by Eq.
(\ref{eq78}). We find that the coefficients of Eq.\ (\ref{eq78}) vary slowly 
with the asymmetry, but this does not change the qualitative picture presented
in the previous section, regarding the ripple wavelength and orientation, or 
the major morphological regimes found in the isotropic case,
including the conditions of validity of our continuum theory.
Specifically, we find that the
asymmetry enlarges the region in $\theta$ where $D_{xx}$ and $D_{yy}$ are
positive, thus shifting Region $II$ to larger values of $\theta$.

%---------------------------------------------------------------------------
 
{\it Phase diagram for ripple orientation---}
  The topology of the morphological phase diagram and the characteristics 
of the three main regions are not changed by the asymmetry. We find that the
only effect of the asymmetry is to move the boundaries smoothly to
larger values of $\theta$ as $\tau$ increases. The condition 
$D(a_{\sigma},\theta)=0$ (see Eq.\ (\ref{eq82})) now takes the form 

\begin{eqnarray}
a_{\sigma} =\sqrt{(s^2+\tau^2 c^2) \frac{(2 s^2 -3 \tau^2 c^2) + 
\sqrt{6 \tau^4 c^4 + 4 s^4}}{\tau^2 s^2 c^2}}, 
\label{eq84}
\end{eqnarray}
and the condition $\nu_x(a_{\sigma},\theta)=0$ (see Eq.\ (\ref{eq82X}))
becomes 

\begin{eqnarray}
a_{\sigma} =\sqrt{\frac{2 s^4 +\tau^2 c^2 s^2 -\tau^4 c^4}{\tau^2 s^2 c^2}}. 
\label{eq84X}
\end{eqnarray}

\subsection{\bf Nonlinear theory}
\label{age:B}

 As we demonstrated in the previous section, linear theory can predict 
many features of ripple formation, such as the ripple wavelength
and orientation, both at high and low temperatures. However, a 
number of important experimental features are incorrectly predicted 
by linear theory. They include the stabilization of the ripple amplitude 
(according to the linear theory the amplitude increases indefinitely 
at an exponential rate) or the presence of kinetic 
roughening (completely unaccounted for by the linear approach). To account 
for these features, we have to consider the effect of the nonlinear
terms. Consequently, this section is devoted to the effect of 
the nonlinear terms on the surface morphology. There is an important difference
between the linear and nonlinear theories: while all predictions of the linear
theory can be calculated analytically (as we demonstrated in the previous
section), the discussion of the nonlinear effects requires a combination of
analytical and numerical tools. Even with these tools, the understanding of 
the nonlinear effects is far less complete than that of the linear theory.

\subsubsection{\bf High temperature morphology: Symmetric case}
\label{age:B:1}
 
In the high temperature regime, where thermal surface diffusion dominates over 
ion-induced effective smoothing, the nonlinear equation of the 
interface evolution has the form

\begin{eqnarray}
\frac{\partial h}{\partial t} & = & -v_0 + \gamma
\frac{\partial h}{\partial x} + \xi_x 
\left(\frac{\partial h}{\partial x}\right) 
\left(\frac{\partial^2 h}{\partial x^2} \right)
\nonumber\\ & &+ 
\xi_y \left(\frac{\partial h}{\partial x} \right) 
\left(\frac{\partial^2 h}{\partial y^2} \right) 
+ \nu_x \frac{\partial^2 h}{\partial x^2} 
\nonumber\\ & & +
\nu_y \frac{\partial^2 h}{\partial y^2} + \frac{\lambda_x}{2}
\left(\frac{\partial h}{\partial x} \right)^2+ 
\frac{\lambda_y}{2} \left(\frac{\partial h}{\partial y} \right)^2
\nonumber\\
 & & + \Omega_{1} \frac{\partial^3 h}{\partial x^3}+ 
\Omega_{2} \frac{\partial^3 h}{\partial x \partial y^2}
-K \nabla^4 h+\eta(x,y,t),
\label{eq85}
\end{eqnarray}
where the coefficients of the linear terms, $v_{0}$, $\gamma$, $\nu_x$, 
$\nu_y$, $\Omega_1$, $\Omega_2$, and $K$ have been discussed in Sections 
\ref{age:A:1} and \ref{age:A:2}. The coefficients of the nonlinear terms 
in the symmetric case ($\sigma = \mu$) are

\begin{eqnarray}
\lambda_x & = & \frac{F}{2} c \left\{ a_{\sigma}^2
(3s^2-c^2) - a_{\sigma}^4 s^2 c^2 \right\},\nonumber \\
\lambda_y & = & - \frac{F}{2} c \{ a_{\sigma}^2 c^2 \},
\nonumber\\
\xi_x & = &  \frac{F a}{2} s c \left\{
-a_{\sigma}^4 c^2 s^2 + a_{\sigma}^2 (4 s^2 -3c^2) +6\right\} ,
\nonumber \\ 
\xi_y & = &  \frac{F a}{2} s c \left\{2- a_{\sigma}^2 c^2 \right\} .
\label{eq86} 
\end{eqnarray}

  Next we discuss the physical interpretation and the behavior of these
coefficients as functions of $\theta$ and $a_{\sigma}$.
%----------------------------------------------------------------------------

\begin{figure}[thb]
\begin{center}
%\hskip -1.0 cm
\vskip 0.1 cm
\epsfig{figure=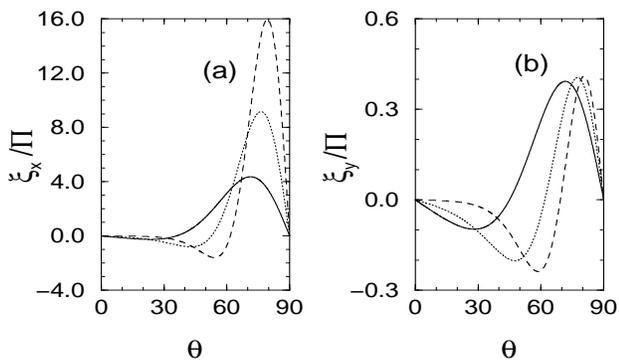,width=3.2 in,height=5.0 cm}
\end{center}
\vskip 0.5 cm
\caption{The reduced coefficients $\xi_x/\Pi$ (a) and $\xi_y/\Pi$ (b) shown 
as functions of $\theta$ for different values of the reduced penetration
depths: $a_{\sigma}=2$ (solid line); $a_{\sigma}=3$ (dotted line);
$a_{\sigma}=4$ (dashed line).}
\label{fig24}
\end{figure}

{\it The coefficients $\xi_{x}$ and $\xi_{y}$:}
Fig.\ \ref{fig24} shows the nonlinear coefficients $\xi_{x}$ 
and $\xi_{y}$ as functions of the angle of incidence $\theta$. 
As the numerical analysis of Eq.\ (\ref{eq85}) shows, 
these nonlinearities are responsible for the development of
overhangs on the surface \cite{CUN}. 
Even though the $\xi_{x}$ and $\xi_{y}$ terms are expected not to
determine the asymptotic scaling behavior, they can play a 
relevant role at intermediate time scales after the development of the
ripple structure, particularly in the region of large $\theta$.
The precise contribution of these nonlinearities to the surface
dynamics is currently under investigation \cite{CUN}.

{\it The coefficients $\lambda_x$ and $\lambda_y$:}
As we discussed in Sec. \ref{ta:A:2}, the morphology of the surface 
described by Eq.\ (\ref{eq85}) depends on the relative signs of the the 
nonlinear 
terms $\lambda_x$ and $\lambda_y$. As it is evident from Eq.\ (\ref{eq86}), 
$\lambda_y$ is negative for all angles of incidence and penetration depths.

%-----------------------------------------------------------------------------

\begin{figure}[thb]
\begin{center}
%\hskip -1.0 cm
\vskip 0.1 cm
\epsfig{figure=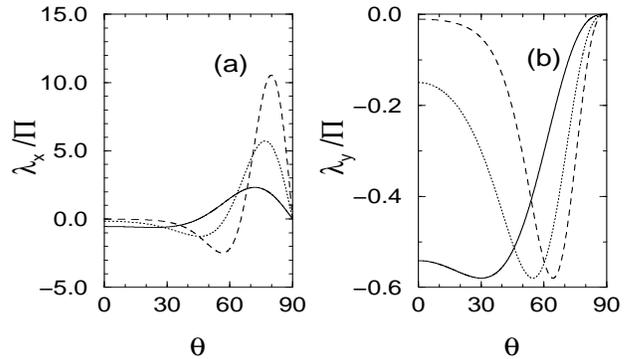,width=3.2 in,height=5.0 cm}
\end{center}
\vskip 0.5 cm
\caption{The reduced coefficients $\lambda_x/\Pi$ (a) and
$\lambda_y/\Pi$ (b) shown as functions of $\theta$. The different curves 
correspond to different values of the reduced penetration depth: 
$a_\sigma=2$ (solid line); $a_\sigma=3$ (dotted line); $a_\sigma=4$ 
(dashed line).}
\label{fig25}
\end{figure}
However, as shown in  Fig.\ \ref{fig25}, the sign of $\lambda_x$ depends 
on $\theta$ and $a_{\sigma}$: $\lambda_x$ is negative for small $\theta$ 
and changes sign for larger angles of incidence.
In principle the nonlinear terms completely determine the surface
morphology. Since the nonlinear terms are always present, an important 
question is whether the linear regimes are relevant at all. Recent results 
by Park {\it et al.} \cite{kahng} indicate that, while the nonlinear effects
indeed change the surface morphology, the regime described by the linear 
terms is still visible for a wide range of parameters. By numerically
integrating Eq.\
(\ref{eq85}) they have shown that there is a clear separation of the linear
and nonlinear regimes in time: for times up to a crossover time $t_{c}$ the
surface erodes as if the nonlinear terms would be completely absent, 
following the predictions of the linear theory. After $t_{c}$, however, the
nonlinear terms take over and completely determine the surface morphology. 
The transition from the linear to the nonlinear regime can be seen either 
by monitoring the surface width (which is proportional to the ripple 
amplitude) or the erosion velocity. The simulations indicate that the width
increases exponentially with time, as predicted by the linear theory, until
$t_{c}$, after which the width growths at a much slower rate \cite{kahng}. 
This transition is typically accompanied by the disappearance of ripples
predicted by the linear theory and the appearance of either kinetic 
roughening or of a new rotated ripple structure. The erosion velocity is
constant in the linear regime (before $t_{c}$), while it increases or 
decreases after $t_{c}$, depending on the relative signs of the nonlinear 
terms. 

{\it Crossover time:}
The crossover time $t_{c}$ from the linear to the nonlinear behavior can
be estimated \cite{kahng} by comparing the strength of the linear terms with 
that of the nonlinear terms. 
Let the typical surface width at the crossover time $t_{c}$
be $W_{o}=\sqrt{W^2(L,t_{c})}$. Then from the linear equation we obtain
%----------------------------------------------------------------------
\begin{figure}[thb]
\begin{center}
\hskip -0.5 cm
\epsfig{figure=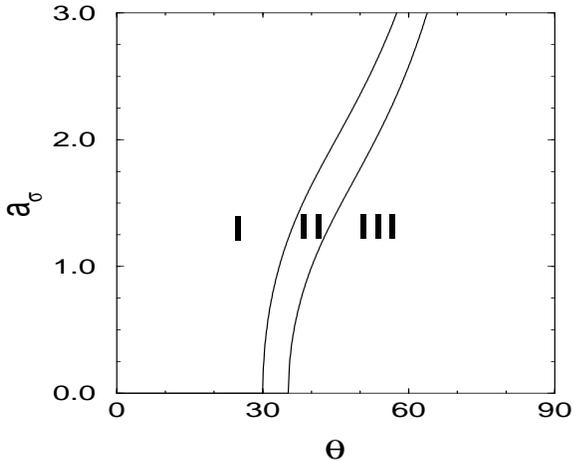,width=3.0 in,height=6.4 cm}
\end{center}
\vskip 0.5 cm
%\hskip 0.1 cm
\caption{Phase diagram for the isotropic case $\sigma=\mu=1$. 
Region I: 
$\nu_x < 0$, $\nu_y < 0$, $\lambda_x <0$, $\lambda_y < 0$; 
Region II:
$\nu_x < 0$, $\nu_y < 0$, $\lambda_x >0$, $\lambda_y < 0$;
Region III:
$\nu_x>0$, $\nu_y <0$, $\lambda_x >0$, $\lambda_y < 0$.}
\label{fig26}
\end{figure}
%----------------------------------------------------------------------
\begin{equation}
W_{o} \sim \exp(\nu t_{c}/\ell^2),
\label{eq87}
\end{equation}
while from $\partial_t h \sim \lambda (\nabla h)^2$ we estimate 
\begin{equation}
W_{o}/t_{c} \sim \lambda W_{o}^2/\ell^2.
\label{eq88}
\end{equation}
Combining these relations we obtain
\begin{equation}
t_{c} \sim (K/\nu^2)\ln(\nu/\lambda). 
\label{eq89}
\end{equation}
In this expression, $\nu$, $K$, and $\lambda$ correspond to the direction
parallel to the ripple orientation. The predicted $\lambda$ dependence
of $t_{c}$ has been confirmed by numerical simulations \cite{kahng}.

{\it Surface morphology:}
  The surface morphology in the nonlinear regime depends on the relative
signs of $\nu_x$, $\nu_y$, $\lambda_x$ and $\lambda_y$. The different
morphological regimes can be summarized in a phase diagram, shown
in Fig.\ \ref{fig26}. Next we discuss each of the phases separately.  

  {\it Region I---} For small $\theta$ the nonlinear terms 
$\lambda_x$ and $\lambda_y$ have the same (negative) sign, 
the boundary of this region being given by the condition that
$\lambda_x(a_{\sigma},\theta)=0$, or equivalently
\begin{eqnarray}
a_{\sigma} = \sqrt{\frac{3s^2-c^2}{c^2s^2}}.
\label{eq90}
\end{eqnarray}
In this region {\em both} $\nu_x$ and $\nu_y$ are 
negative, thus at short time scales $(t \le t_{c})$, 
the linear theory (see Section \ref{age:A:1}) predicts ripples 
oriented along the direction ($x$ for large $a_{\sigma}$ and $y$ 
for small $a_{\sigma}$) for which the absolute value 
of $\nu$ is largest, with ripple wavelength
$\ell = 2 \pi \sqrt{\frac{2K}{|\nu|}}$. On the other hand, at long times
$(t \gg t_{c})$, the ripple structure disappears and the surface
undergoes kinetic roughening \cite{kahng}. Since $\lambda_{x} \cdot
\lambda_{y} > 0$, we expect the dynamics of the kinetic roughening
regime to be described by the KPZ equation, i.e. the surface width follows
$W \sim L^{\alpha}$, $W \sim t^{\beta}$, where the scaling exponents are
$\alpha \simeq 0.38$ and $\beta \simeq 0.25$ (see Sect.\ \ref{ta:A:1}). 

{\it Region II---}
 In this region both the $\nu_x$ and the $\nu_y$ coefficients are 
still negative, but in contrast with Region I the product 
$\lambda_x \cdot \lambda_y$ is negative.
Recent studies by Park {\it et al.} \cite{kahng} have shown that in time 
three morphological regimes can be distinguished. 
For short times, $t \le t_{I}$, the ripple structure predicted by the 
linear theory (see Section \ref{ta:A:1}) 
is observed, with ripples oriented along the direction which 
has the largest value of $|\nu|$. For intermediate
times $t_{I} \le t \le t_{II}$, the surface is rough. If this roughness
would follow kinetic roughening, one would expect logarithmic scaling, 
described by the Edwards-Wilkinson equation (see Sections
\ref{age:B:2}-\ref{age:B:4}), since $\lambda_x \cdot \lambda_y < 0$. 
However, this transient regime is somewhat different from what we expect 
during kinetic roughening. The numerical simulations often show the 
development of individual ripples, which soon disappear, 
and no long-range order is present in the system. 
However, at a second crossover time, $t_{II}$, a new
ripple structure suddenly forms, in which the ripples are stable and rotated
an angle $\theta_{c}$ with respect to the $x$ direction. 
Rost and Krug \cite{Rost&Krug} 
have shown [for the deterministic limit of Eq.\ (\ref{eq85})] that,
by defining $\alpha_{\nu}=\nu_x/\nu_y$ and $\beta_{\lambda}=
\lambda_x/\lambda_y$, the fact that $\alpha_{\nu} > 0$ and $\beta_{\lambda} <
0$ throughout Region II implies the existence of ``cancellation modes''
which dominate the dynamics and lead to this rotated ripple morphology.
(Note the parameter ratios $\alpha_{\nu}$ and $\beta_{\lambda}$ are not to be 
confused with the roughness and growth exponents $\alpha$ and $\beta$ 
introduced in Section \ref{st:2}.) The angle $\theta_{c}$ 
calculated by Rost and Krug has the value $\theta_{c}
= \tan^{-1}{\sqrt{-\lambda_x/\lambda_y}}$ (see also Appendix \ref{AP:d}),
and can be obtained by moving to a rotated frame of coordinates that
cancels the nonlinear terms in the transverse direction. 
The boundary of 
Region $II$ is given by the condition $\nu_x(a_{\sigma},\theta)=0$, 
Eq.\ (\ref{eq82X}).   

  {\it Region III ---} This region is characterized by a positive $\nu_x$ 
and a negative $\nu_y$. At short time scales, $t \le t_{c}$, the periodic 
structure associated with the instability is oriented along the $y$ direction
and has wavelength $\ell_y = 2 \pi \sqrt{ \frac{2K}{|\nu_y|}}$. Much less
is known, however, about the nonlinear regime. Such an anisotropic and
linearly unstable equation is unexplored in the context of growth equations.
The analysis by Rost and Krug \cite{Rost&Krug} for the corresponding 
deterministic equation
predicts that, given that $\beta_{\lambda} < \alpha_{\nu} < 0$ does hold 
all over Region III, again cancellation modes induce a rotated ripple 
morphology.

{\it Summary:}
  Even though several aspects of the scaling behavior predicted by Eq.
(\ref{eq85}) remain to be clarified, we believe that this equation
contains the relevant ingredients for understanding roughening by ion
bombardment. To summarize, at short time scales the morphology consists of
a periodic structure oriented along the direction determined by the largest
in absolute value of the negative surface tension coefficients \cite{bh}.
Modifying the values of $a_{\sigma}$ or $\theta$ changes the orientation of
the ripples \cite{bh}. At long time scales we expect two different
morphological regimes. One is characterized by the KPZ exponents, which 
{\it might} be observed in Region $I$ in Fig.\ \ref{fig26}. Indeed, the
values of the exponents reported by Eklund {\em et al.} 
\cite{eklund,eklund2} are consistent within the experimental errors with 
the KPZ exponents. In Region $II$  kinetic roughening is not expected. Rather,
nonlinear terms lead to a new ripple structure that is rotated with respect
to the ion direction. Region III is less understood; analysis of the 
deterministic equation \cite{Rost&Krug} again predicts a rotated ripple
structure. By tuning the values of 
$\theta$ and/or $a_{\sigma}$ one may induce transitions among these
morphological regimes.

\subsubsection{\bf High temperature morphology: Asymmetric case}
\label{age:B:2}

  In this section we discuss the effect of asymmetry on the scaling regimes
of Eq.\ (\ref{eq85}). Here again we obtain that the effect of asymmetry
does not bring in new qualitative features. 
Specifically, we find that the qualitative behavior of $\xi_x$ and $\xi_y$
is not affected by the asymmetry. As the asymmetry grows, the absolute value
of the coefficients in the region of small angles decreases and the peaks at 
large $\theta$ increase. 
Similarly, while the numerical values of $\lambda_{x}$ and $\lambda_{y}$ 
are affected by the asymmetry $\tau$, their qualitative features are not.

  Finally, we find that the morphological diagram is topologically 
equivalent to the phase diagram obtained in the symmetric case (see Fig.
\ref{fig26}), the only difference being in the position of the boundaries.
As $\tau$ changes, we find a smooth evolution of the boundaries, which does 
not uncover any new regimes. Since the morphological properties of the 
system in the three regimes are the same as those discussed in the symmetric
case, we will not discuss them again.

\subsubsection{\bf Low temperature morphology: Symmetric case}
\label{age:B:3}

  In this section we discuss the effect of the effective surface smoothing 
on the surface morphology in the nonlinear regime. In the absence of
thermally activated surface diffusion, from  Eq.\ (\ref{eq44}) we obtain the 
following equation governing the morphology evolution 
 \begin{eqnarray} 
\frac{\partial h}{\partial t} & = & -v_0 + 
\gamma \frac{\partial h}{\partial x} 
+ \xi_x \left(\frac{\partial h}{\partial x}\right)
\left(\frac{\partial^2 h}{\partial x^2} \right)
\nonumber\\ & & +
\xi_y \left(\frac{\partial h}{\partial x} \right)
\left(\frac{\partial^2 h}{\partial y^2} \right) 
+\nu_x \frac{\partial^2 h}{\partial x^2} 
\nonumber\\ & & +
\nu_y \frac{\partial^2 h}{\partial y^2} + 
\frac{\lambda_x}{2} \left(\frac{\partial h}{\partial x} \right)^2+
\frac{\lambda_y}{2} \left(\frac{\partial h}{\partial y} \right)^2
\nonumber \\
& & + \Omega_1 \frac{\partial^3 h}{\partial x^3}
+ \Omega_2 \frac{\partial^3 h}{\partial x \partial y^2} 
- D_{xx} \frac{\partial^4 h}{\partial x^4}\nonumber\\
& &- D_{xy} \frac{\partial^4 h}{\partial x^2 \partial y^2}-
D_{yy} \frac{\partial^4 h}{\partial y^4}
+
\eta(x,y,t).
\label{eq91}
\end{eqnarray}
%----------------------------------------------------------------------------
\begin{figure}[thb]
\begin{center}
\hskip -0.5 cm
\vskip  0.1 cm
\epsfig{figure=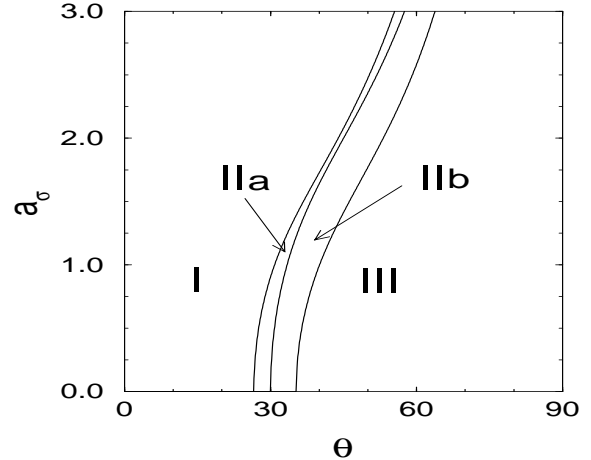,width=3.0in,height=6.4cm}
\end{center}
\caption{Nonlinear phase diagram for the isotropic case $\sigma=\mu=1$ 
at low temperatures. 
Region  I: 
$\nu_x< 0$, $\nu_y < 0$, $D_{xx} >0$, $D_{yy} > 0$ and $\lambda_x < 0$,
$\lambda_y < 0$; 
Region  IIa:
$\nu_x< 0$, $\nu_y < 0$, $D_{xx} < 0$, $D_{yy} > 0$ and $\lambda_x < 0$,
$\lambda_y < 0$; 
Region IIb:
$\nu_x< 0$, $\nu_y < 0$, $D_{xx} < 0$, $D_{yy} > 0$ and $\lambda_x > 0$,
$\lambda_y < 0$; 
Region III:
$\nu_x > 0$, $\nu_y < 0$, $D_{xx} < 0$, $D_{yy} > 0$ and $\lambda_x > 0$,
$\lambda_y < 0$.}
\label{fig30}
\end{figure}
The terms $\gamma$, $\nu_{x}$, $\nu_y$, $\Omega_1$, $\Omega_2$,
$\xi_x$, $\xi_y$, $\lambda_x$, $\lambda_y$, as well 
as the ion-induced effective smoothing coefficients $D_{xx}$, 
$D_{xy}$ and $D_{yy}$ have been discussed in the previous sections. 

  In the following we discuss the morphological phase diagram predicted 
by Eq.\ (\ref{eq91}) and shown in Fig.\ \ref{fig30}. The different regimes 
have the following characteristics:
%---------------------------------------------------------------------------

   {\it Region I:} The surface tensions $\nu_x$ and $\nu_y$ are negative
while $D_{xx}$ and $D_{yy}$ are positive, and $\lambda_x$, $\lambda_y$ are both
negative. This regime has been previously described in Section 
\ref{age:B:1} (Regime I in Fig.\ \ref{fig26}), the only difference being 
that here the ion-induced effective surface smoothing plays the role of $K$. 
The boundary of this region is given by $D_{xx}(a_{\sigma},\theta) = 0$,
Eq.\ (\ref{eq82}). 

 {\it Region IIa:} Here $\nu_x$, $\nu_y$, $\lambda_x$, and
$\lambda_y$ are still negative, $D_{yy}$ is positive, while $D_{xx}
< 0$. Consequently, along the $x$ direction the surface is unstable, all
modes growing exponentially. However, the nonlinear terms $\lambda_x$ and
$\lambda_y$ have the same sign. The nonlinear regime in this parameter
region has not yet been explored numerically, thus its morphology is unknown. 
The boundary of this region is given by $\lambda_{x}(a_{\sigma},\theta) =0$, 
Eq.\ (\ref{eq90}).  

 {\it Region IIb:} In this region $D_{xx}$ is negative, $D_{yy}$ is 
positive, $\nu_x < 0$, $\nu_y < 0$, and $\lambda_x > 0$, $\lambda_y < 0$. 
Thus, the only difference of this region with respect to Region IIa
is that the nonlinear terms have different signs.
Similarly to Region IIa, nothing is known about the nonlinear behavior. 
The boundary of this region is given by $\nu_{x} (a_{\sigma}, \theta) =0$, 
Eq.\ (\ref{eq82X}).  

 {\it Region III:}  Here we have $\nu_x > 0$, $\nu_y < 0$, $D_{xx}< 0$, 
$D_{yy} > 0$, $\lambda_x > 0$ and $\lambda_y < 0$. This region has
similar features to Region III in the phase diagram of Fig.\ 
\ref{fig26}, except for the negative value of the $D_{xx}$ coefficient.
%-------------------------------------------------------------------------
\begin{figure}[thb]
\begin{center}
\hskip -0.5 cm
\epsfig{figure=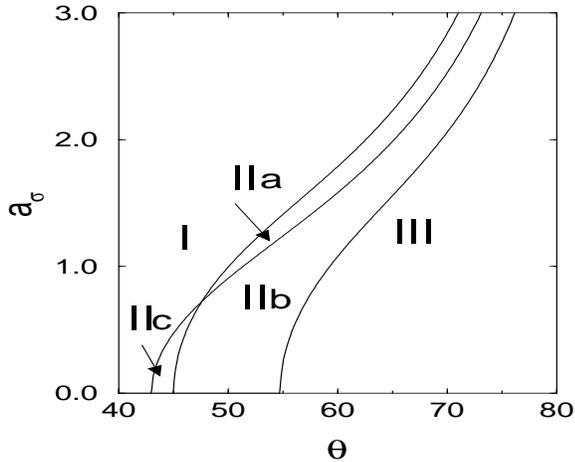,width=3.0 in,height=6.4 cm} 
\end{center}
%\vskip 0.5 cm
\caption{Nonlinear phase diagram for the anisotropic case with $\tau=2$
and $a_{\sigma}=2$. Different regions in the diagram correspond to: 
Region  I: 
$\nu_x < 0$, $\nu_y < 0$, $D_{xx} > 0$, $D_{yy} > 0$ and $\lambda_x < 0$,
$\lambda_y < 0$; 
Region  IIa:
$\nu_x < 0$, $\nu_y < 0$, $D_{xx} < 0$, $D_{yy} > 0$ and $\lambda_x < 0$,
$\lambda_y < 0$; 
Region IIb:
$\nu_x < 0$, $\nu_y < 0$, $D_{xx} < 0$, $D_{yy} > 0$ and $\lambda_x > 0$,
$\lambda_y < 0$; 
Region IIc:
$\nu_x < 0$, $\nu_y < 0$, $D_{xx} > 0$, $D_{yy} > 0$ and $\lambda_x > 0$,
$\lambda_y < 0$;
Region III:
$\nu_x > 0$, $\nu_y < 0$, $D_{xx} < 0$, $D_{yy} > 0$ and $\lambda_x > 0$,
$\lambda_y < 0$.}
\label{fig31}
\end{figure}
%------------------------------------------------------------------------

\subsubsection{\bf Low temperature morphology: Asymmetric case}
\label{age:B:4}

  In this section we discuss the effect of asymmetry on the long distance
properties of Eq.\ (\ref{eq91}). The effect of the
asymmetry on the coefficients appearing in the equation
were discussed earlier. Therefore we concentrate here merely on the 
morphological phase diagram predicted by Eq.\ (\ref{eq91}) for the
asymmetric case, which is displayed in Fig.\ \ref{fig31}. As before, 
asymmetry ($\tau \neq 1$) leads to a smooth shift of the boundaries of 
the regions provided by the lines where the coefficients
$D_{xx}(a_{\sigma}, \theta)$, $\nu_x(a_{\sigma}, \theta)$, and 
$\lambda_x(a_{\sigma}, \theta)$ change sign. In the presence of
effective smoothing, 
however, asymmetry in the deposited energy distribution
induces the appearance of a fifth morphological regime. 
This is caused by the smooth motion of the
boundary determined by $\lambda_x(a_{\sigma}, \theta) =0 $, which intersects 
for some value of $\tau$ the boundary defined by $D_{xx}(a_{\sigma}, 
\theta)=0$. Comparison of Fig.\ \ref{fig31} and Fig.\ \ref{fig30} illustrates
how the boundaries move with $\tau$. Regions I, IIa, IIb and 
III are analogous of those shown in Fig.\ \ref{fig30}. Region 
IIc in the phase diagram, on the other hand, is analogous of Region 
II of the high temperature phase diagram, shown in Fig.\ \ref{fig26}, and
all the conclusions obtained in that section regarding the morphological 
properties in this regime also apply here.
  
\section{\bf Comparison with experiments}
\label{cE:M}

  In this section we compare the predictions of the theory presented in
this paper with experimental results on ripple formation and surface
roughening. For a better presentation, we choose to structure the material
around well known features of the morphological evolution, present the 
theoretical predictions and discuss to which extent are they supported 
by the available experimental data. We also discuss predictions that have 
not been tested in sufficient detail but could offer future tests of the 
theory.

{\it Ripple amplitude:}
A key quantity in ripple formation is the time evolution of the 
ripple amplitude. As we have shown in Section \ref{age:A:1}, at early 
times ($t \le t_{c}$) the ripple amplitude grows exponentially, following
$h \sim \exp{(r(q_x^{*}, q_y^{*})t)}$, where $r$ is the growth rate of the
most unstable mode $(q_x^{*}, q_y^{*})$. According to the linear theory,  
this growth continues indefinitely. In contrast, the nonlinear theory 
predicts that the amplitude should stabilize after time $t_{c}$, where 
$t_{c}$ is given by Eq.\ (\ref{eq89}). This is consistent with the 
experimental investigations \cite{vajo,aziz}. On the other hand,
recently Erlebacher {\it et al.} \cite{erleb} also found that at initial
stages the ripple morphology is growing exponentially. Furthermore, they
observed that at some time $t_c$ the exponential growth stops and the ripple 
amplitude saturates. Measuring the temperature dependence of the saturation
curves, they found that rescaling the time $t$ with a factor $\nu^2/4K$ and
the amplitude $h$ with $\sqrt{\nu/2K}$, the different curves representing
the amplitude as a function of time collapse onto a single one. This result 
is in excellent agreement with our prediction that suggest that plotting 
the result in terms of the rescaled parameters, $t/t_{c}$ and $h 
\sqrt{\nu/2K}$, the different curves should collapse \cite{kahng}. They 
also offer direct proof that the nonlinear terms play a major role in
determining the amplitude of the ripples, indicating that the incorporation
of the nonlinear mechanisms in the theory of ripple formation is essential.

{\it Temperature dependence of the ripple wavelength:}
A key quantity that provides direct information about the nature of the
relaxation mechanism is the temperature dependence of the ripple wavelength. 
Our results indicate that there are two mechanisms contributing to surface
relaxation: thermally induced surface diffusion (Sects.\ 
\ref{age:A:1}-\ref{age:A:2}) and ion-induced smoothing 
(Sects.\ \ref{age:A:3}-\ref{age:A:4}). 
At high temperatures thermal surface diffusion is 
rather intensive, thus it is the main mechanism determining the relaxation
process, the ripple wavelength being given by (\ref{eq68}) or (\ref{eq70}).
Since the surface diffusion constant $K$ decreases with 
$T$ as $(1/T) \exp(-\Delta E/k_{B}T)$, the ripple wavelength is 
also expected to decrease exponentially 
with $T$. Indeed, such an exponential temperature dependence of $\ell$ has 
been observed by various groups \cite{maclaren,umbach}. However, in Section
\ref{age:A:3} we demonstrated the existence of ion induced smoothing, 
that is present at {\it any} temperature. Thus, up to some 
inessential numerical factors, the total surface diffusion constants 
have a form $D^{T}= K+D$, where $D$ is independent of temperature.
Since $K$ decreases exponentially with $T$, at low enough temperatures we
have $K \ll D$, indicating that the main relaxation mechanism is ion-induced.
Consequently, below a certain critical temperature
$T_{c}$, given by $K(T_{c})=D$, one expects the ripple wavelength to be
independent of $T$. Support for this scenario has been provided by the 
experiments in \cite{maclaren} and \cite{umbach}
and the molecular dynamics simulations of Koponen {\it et al.}
\cite{koponen2}.
Consequently, as the theoretical results in this paper indicate, the 
temperature dependence of $\ell$ can be used to identify the relaxation
mechanism: when $\ell$ increases exponentially with $T$, we are dealing with
relaxation by thermal surface diffusion, 
while a temperature independent wavelength 
is an indication of ion induced smoothing.      

{\it Ripple orientation:}
An important feature of ripple formation is that, as the linear theory 
predicts, the ripple orientation depends on the angle of incidence $\theta$.
The dependence of the ripple orientation on the experimental parameters
has been  summarized in the phase diagrams shown in Figs. \ref{fig12},
\ref{fig16}, \ref{fig21}, \ref{fig26}, \ref{fig30}, and \ref{fig31}. In
general, for physical values of the asymmetry parameter ($i.e.$ for $\tau>1$),
we find that for small angles the
ripples are oriented in the $x$ direction (along the incoming ions), and 
they change orientation to the $y$ direction for large $\theta$. The
boundary separating these two morphological regions depends on the parameters
characterizing the sputtering process, such as the ion penetration depth and
the geometry of the deposited energy distribution. Such transition in the 
ripple orientation has been found in the simulations of \cite{koponen}, 
where for $\theta 
\le 45^\circ$ the observed ripples were oriented along the $x$ direction, 
while for $\theta \ge 45^\circ$, they changed their orientation to $y$.
Furthermore, the nonlinear theory predicts that after the nonlinear terms
take over, new ripples, with orientation different from both $x$ and $y$
directions, might appear (see Sect.\ \ref{age:B:1}). To the best of our 
knowledge, such rotated ripples have not been observed experimentally as 
yet. Nevertheless, this morphology might also lead to additional effects,
such as shadowing, which have been neglected in our approach.

{\it Ripple wavelength dependence on the flux:}
Depending on the nature of the relaxation mechanism, the linear theory has 
two different predictions on the flux dependence of the ripple wavelength:
for high temperatures, when thermal surface
diffusion dominates, one expects $\ell
\sim J^{-1/2}$ [see Eq.\ (\ref{eq74})], while at low temperatures, 
characterized 
by ion induced smoothing, we expect the wavelength to be 
independent of flux (see Sect.\ \ref{age:A:3}). Consequently, due to its 
strong dependence on the relaxation mechanism, the flux dependence of the 
ripple wavelength can also be used to identify the relaxation mechanism.
Indeed, a number of experiments \cite{vajo,umbach} are compatible with 
the prediction of a flux independent wavelength. Other aspects of ripple
characteristics (such as energy or temperature dependence) also lead to the
same conclusion. On the other hand, we are not aware of results indicating
decreasing ripple wavelength with increasing flux. However, support for the
relevance of thermal surface diffusion comes from the experiments of Chason 
{\it et al.} \cite{chason94a}, who reported that the growth rate 
$r(q_x^{*}, q_y^{*})$ as a function of flux follows the predictions of 
the linear theory with thermal surface diffusion.    

{\it Ripple wavelength dependence on the ion energy:}
The linear theory indicates that the ion energy dependence of the ripple
wavelength can be used to distinguish between the two relaxation mechanisms:
at high temperatures we expect $\ell \sim \epsilon^{-1/2}$ [see Eq.\
(\ref{eq73})], i.e., the wavelength decreases with the energy, while at
low temperature we have $\ell \sim \epsilon^{2m} \sim \epsilon$ [see Eq.\
(\ref{eq83})], i.e., the wavelength should increase with energy, strikingly
different predictions. A number of experimental groups have found that the
ripple wavelength increases linearly with the ion energy
\cite{stevie,karen1,vajo}. However, while we are not aware of any {\it direct}
observation of a decreasing ripple wavelength with increasing ion energy, the
growth rate dependence on the ion energy measured by Chason {\it et al.}
\cite{ChasonMRS} provided results which are in
agreement with the predictions based on thermal surface diffusion.

{\it The magnitude of the effective surface diffusion constant:}
Since the transition between the low and high temperature regimes is 
determined by the relative magnitude of $D_{xx}, D_{xy}, D_{yy}$ 
(ion induced smoothing), 
and $K$ (thermal surface diffusion),
we need to estimate the magnitude of these constants. 
In the following we give an order of magnitude estimate for the effective 
surface diffusion constant $D_{xx}$ and compare it to $K$, using data 
from \cite{vajo,erleb} for Si(001). Taking $Y=2.6$, $J=670$ $\mu$A/cm$^2$, 
$\epsilon = 9$ keV, $a=100$ \AA, $a_{\sigma}=2$, $a_{\mu}=4$, 
and $\theta=40^{\circ}$, Eq.\ (\ref{eq55})
gives $D_{xx} \simeq 12 \times 10^{-28}$ cm$^4$/s. For comparison, 
at T = 550 C it is estimated \cite{erleb} that $K \simeq 34 \times 10^{-28}$
cm$^4$/s. Hence, since $K$ decreases exponentially with temperature, 
ion induced smoothing can be significant at low temperatures 
(including room temperature), in some cases being comparable or more relevant  
than thermal surface diffusion. 

{\it Kinetic roughening:}
An important feature of our theory is that it goes beyond the linear 
approach, handling systematically the nonlinear effects as well. As we
demonstrated in Sect.\ \ref{age:B}, the presence of the nonlinear terms can
affect both the dynamics and the morphology of the surface. The first and
the most dramatic consequence is the stabilization of the ripple amplitude,
discussed above. Furthermore, after the stabilization of the ripple 
amplitude the surface morphology is rather different from the morphology
predicted by the linear theory. In particular, depending on the signs of
$\lambda_x$ and $\lambda_y$, different morphological features can develop. 
When $\lambda_x \lambda_y$ is positive, at large times the surface 
undergoes kinetic roughening, following the predictions of the KPZ 
equation. 
This behavior has, indeed, been observed experimentally \cite{eklund,eklund2} 
and numerically \cite{cuerno1}, providing direct support for the predictions
of the nonlinear theory. When $\lambda_x \lambda_y$ is negative, 
direct numerical integration of the nonlinear theory \cite{kahng,Rost&Krug}
indicated the existence of a new, rotated ripple structure. 
The absence of experimental data on this phase might be due to the required 
large sputtering times: the simulations indicate
\cite{kahng} that between the linear regime and the formation of the rotated
ripple structure there is a rather long transient regime with an apparently 
rough surface morphology. The above predictions apply when the surface diffusion
terms, ion or thermally induced, act to smooth the surface (i.e., $D^T \ge 
0$). However, at low temperatures, when ion induced smoothing dominates, 
surface diffusion can generate an instability that can further modify this 
behavior (see Sect.\ \ref{age:B:3}). In general, while rather detailed 
experimental data are available describing the linear regime of ion
sputtering, explanation of the nonlinear regime is only at its 
beginning, hiding the possibility of new interesting phases and
behaviors.

\section{\bf Conclusions}
\label{CON}
  
  In this paper we investigated the morphological properties of surfaces
eroded by ion bombardment. Starting from the expression for the erosion 
velocity derived in the framework of Sigmund's theory of sputtering
of amorphous targets, we
derived a stochastic partial differential equation for the surface height, 
which involves up to fourth order derivatives of the height, and 
incorporates surface diffusion and the fluctuations arising in the 
erosion process due to the inhomogeneities in the ion flux. In some 
special cases the derived nonlinear theory reduces to the much studied
KS or the KPZ equations, well known descriptions of 
dynamically evolving surfaces. However, in contrast with these theories,
which have been derived using symmetry and conservation considerations
\cite{revrough}, here we derived
the continuum theory directly from a microscopic model of sputtering,
and thus all coefficients can be explicitly expressed in terms of the 
physical parameters (such as angle of incidence, ion penetration depth,
etc.) characterizing the ion bombardment process. An important feature of
the derived nonlinear continuum theory is that the linear and the nonlinear
regimes are separated in time. As a result, they can be discussed separately,
the former controlling the behavior at early times, the latter at late
times. Furthermore, an important result of our calculations is that 
higher order effects of the sputtering process can smooth the surface. 
This effective mechanism was necessary to explain ripple 
formation at low temperatures, when thermally induced surface diffusion is not
relevant. Consequently, based on these two ingredients (separation of
time scales between linear and the nonlinear regimes and the existence of
two different relaxation mechanisms) we have discussed four different 
cases. In the linear high temperature regime the equations reduce to the
linear theory of Bradley and Harper, predicting ripple formation, and 
explaining such experimentally observed phenomenon as ripple orientation
(and its change with $a_{\sigma}$ and $\theta$), exponential increase in
ripple amplitude (valid for short times), or flux and energy dependence
of the ripple wavelength. On the other hand, phenomena not explained by
this approach, such as the stabilization of the ripple amplitude, can be
explained by considering the nonlinear terms as well. We also show that,
depending on the sign of the coefficients of the nonlinear terms, the 
late time morphology of the surface is either rough, or dominated by new
rotated ripples. The rough phase is expected to be described by the KPZ
equation, which has its own significance: while the introduction of the
KPZ equation has catalyzed an explosion in the study of the morphological
properties of growing surfaces, there are very few actual surfaces that
are described by it (and not by one of its offsprings, such as the MBE
or related equations \cite{revrough}), and most notably none, as far as 
we know, that describe crystal surfaces. The sputtering problem provides 
one of the first systems 
that is convincingly described by this continuum theory. Many of the previous
mysteries of low temperature ripple formation have also been solved by the
present theory. By deriving the higher order ion-induced 
effective smoothing terms, we
can explain the existence of ripples at temperatures where thermally
induced surface diffusion is not active. We showed that the derived 
effective smoothing affects both the linear and the nonlinear regimes, 
governing the early time ripple formation, and the late time nonlinear 
behavior. The coexistence of thermal and ion induced smoothing can explain
the stabilization of the ripple wavelength at low temperatures, 
in contrast with its exponential $T$ dependence at higher temperatures. 
On the other hand, our theory has limitations, most of which can
be already identified in Sigmund's and Bradley and Harper's theories,
with which it is related. Namely, it is devised for amorphous substrates,
whereupon it neglects effects such as viscous relaxation \cite{viscous},
which might be the cause for the failure of the theory to predict the 
absence in many experiments of ripples at low (but non-zero) angles 
of incidence. This issue should constitute one of the most important
extensions of our present theory. Perhaps related with this, we have seen 
that there exist parameter regions at low temperatures within which our
theory breaks down, due to the unstable higher order derivative terms
that occur. A relevant issue is thus to determine the correct continuum 
description of the surface under these conditions.

  Most of the predictions offered by the presented continuum theory have
been already verified experimentally. However, many unexplained predictions
remain at low temperatures both in the linear and the nonlinear regimes, 
as well as regarding the nonlinear regime at high temperatures. We hope
that the rather precise derivations offered in this paper will guide
such future experimental work. Furthermore, some of the morphologies 
expected in the nonlinear regime need further theoretical understanding
as well, allowing for the continuation of this inquiry. With 
the dramatic advances in computer speed, the understanding of some of
these questions, either through numerical integration of the continuum 
theory or through discrete models, might be not too far.           

\acknowledgements

We would like to acknowledge discussions with E.\ Chason, B.\ Kahng,
H.\ Jeong, F.\ Ojeda, and L.\ V\'azquez. This research was supported by 
NSF-DMR CAREER and ONR-YI awards (A.-L.\ B.\ and M.\ M.) and DGES (Spain) 
grant PB96-0119 (R.\ C.).

\appendix

\section{}
\label{AP:a}

The algebraic relation between the coordinates of the laboratory frame and
the local frame, depicted in Fig.\ \ref{fig3}, follows from the definitions
given in point {\bf (i)} of Sect.\ \ref{ConE:M}. Accordingly, the unit vector 
along the $\hat{Z}$ axis is the normal at point $O$
\begin{equation}
\hat{Z} \equiv \hat{n} = \frac{(-\partial_x h, -\partial_y h,
1)}{\sqrt{g}} ,
\label{A1.1}
\end{equation}
where $g \equiv 1 + (\partial_x h)^2 + (\partial_y h)^2.$
The vector $\hat{m}$ drawn on Fig.\ \ref{fig3} has components
\begin{equation}
\hat{m} = (\sin\theta, 0, \cos\theta).
\label{A1.2}
\nonumber
\end{equation}
Therefore, the unit vector along the $\hat{Y}$ axis reads:
\begin{eqnarray}
\hat{Y} & \equiv & \frac{\hat{n} \times \hat{m}}{| \hat{n} \times
\hat{m} |}  =  
\frac{1}{\sqrt{g} \sin\varphi} (-(\partial_y h) \cos\theta, 
\sin\theta  
\nonumber\\
& & +(\partial_x h) \cos\theta, (\partial_y h) \sin\theta) ,
\label{A1.3}
\end{eqnarray}
and finally (\ref{A1.1}) and (\ref{A1.3}) yield for the unit vector
along the $\hat{X}$ axis:
\begin{eqnarray}
\hat{X} & = & \hat{Y} \times \hat{Z}  
\nonumber \\ &=& 
\frac{1}{g \sin\varphi} 
\left( \sin\theta + (\partial_x h) \cos\theta 
%\right. \nonumber\\ &+& \left. 
(\partial_y h)^2 \sin\theta,
\right. \nonumber\\ & & \left. 
(\partial_y h) \cos\theta - (\partial_x h) (\partial_y h)
\sin\theta,
\right. \nonumber \\ & & \left. 
(\partial_x h) \sin\theta  + ((\partial_x h)^2 + 
%\right. \nonumber\\ & & \left. 
(\partial_y h)^2 ) \cos\theta \right) .
\label{A1.4}
\end{eqnarray}
%---------------------------------------------------------------------------
The matrix $\cal M$ defined in Eq.\ (\ref{eq40}) and which relates the 
coordinates in the local and laboratory frames reads 
\begin{equation}
\begin{mathletters}
\cal M = \left( 
\matrix{
\frac{s+c\partial_y h + s(\partial_y h)^2}{\sqrt{g} r} & -
\frac{c \partial_y h}{r} & -\frac{\partial_x h}{\sqrt{g}} \cr
\frac{c\partial_y h -s \partial_x h \partial_y h }{\sqrt{g}r} & 
\frac{s+c \partial_x h}{r}  & -\frac{\partial_y h}{\sqrt{g}} \cr
\frac{s\partial_x h -c ((\partial_x h)^2+ (\partial_y h)^2) }
{\sqrt{g}r} & \frac{s \partial_y h}{r}  &  -\frac{1}{\sqrt{g}} \cr
}\right)\,,
\normalsize
\label{A1.5}
\end{mathletters}
\end{equation}
where $s = \sin\theta$, $c = \cos \theta$, and \newline 
$r \equiv \sqrt{ (s + c\partial_x h)^2 + (\partial_y h)^2}$.    
%--------------------------------------------------------------------

\section{}
\label{AP:b}
%--------------------------------------------------------------------

  If we perform a small $\Delta_{nm}$ expansion in Eq.\ (\ref{eq33})
we obtain
%\bleq
%\end{multicols}
%\widetext
\begin{eqnarray}
V_O & = & \frac{\epsilon p J a^2}{\sigma \mu (2 \pi)^{3/2}} 
\exp{(-a_{\sigma}^2/2)} 
\nonumber \\ &\times&
\int_{-\infty}^{\infty}\!\! \int_{-\infty}^{\infty}\,
\,d{\zeta_X}\,d{\zeta_Y} \exp(-\zeta_Y L^2)
\nonumber\\ &\times&
\exp(-{\zeta_X} A -\frac{1}{2}{\zeta_X^2} B_1 )  
\left[ \cos \varphi \;(1+B_2 {\Delta_{20}} {\zeta_X^2}
\right. \nonumber\\ &+& \left.
 B_2 {\Delta_{02}} {\zeta_Y^2}
+B_2 {\Delta_{30}} {\zeta_X^3} 
+B_2 {\Delta_{12}} {\zeta_X}{\zeta_Y^2}
\right. \nonumber\\ &+&
B_2 {\Delta_{22}} {\zeta_X^2}{\zeta_Y^2}+
B_2 {\Delta_{40}} {\zeta_X^4}+ 
B_2 {\Delta_{04}} {\zeta_Y^4} 
\nonumber \\ &-&
 2 C {\Delta_{20}} {\zeta_X^3}
-2 C {\Delta_{02}} {\zeta_Y^2}{\zeta_X}
-2 C {\Delta_{30}} {\zeta_X^4} 
\nonumber \\ &-&
 2 C {\Delta_{12}} {\zeta_X^2}{\zeta_Y^2}
-2 C {\Delta_{22}} {\zeta_X^3}{\zeta_Y^2}
-2 C {\Delta_{40}} {\zeta_X^5}
\nonumber \\ &-& 
2 C {\Delta_{04}} {\zeta_X}  {\zeta_Y^4}) 
+\sin \varphi \; (2 \Delta_{20} {\zeta_X}+3 \Delta_{30} {\zeta_X^2} 
\nonumber\\ &+& \left.
\Delta_{12} {\zeta_Y^2} 
+2 \Delta_{22} {\zeta_X}{\zeta_Y^2}
+4 \Delta_{40} {\zeta_X^3} ) \right]. 
\label{A2.1}
\end{eqnarray}
%--------------------------------------------------------------------------
Evaluating the Gaussian integrals in this formula we obtain Eq.\ (\ref{eq35}), 
where the coefficients $\Gamma_{nm}(\varphi)$ are given by
\begin{eqnarray}
\Gamma_{20}(\varphi) & = & -\frac{2A}{B_{1}} \sin\varphi +\frac{B_2}{B_1}
\left[1+\frac{A^{2}}{B_1}\right] \cos\varphi 
\nonumber \\ &+& 
\frac{2AC}{B_{1}^2} \left[3+\frac{A^2}{B_1} \right] \cos\varphi, 
\nonumber \\
\Gamma_{02}(\varphi) & = & 2 \frac{\mu^{2}}{a^2} \cos\varphi 
\left( \frac{B_2}{2}+\frac{AC}{B_1} \right),\nonumber\\
\Gamma_{30}(\varphi) & = &  \sin\varphi \left( \frac{1}{B_1}+
\frac{A^2}{B_1^2} \right) - B_2 \cos\varphi \left(\frac{3A}{B_1^2}+
\frac{A^3}{B_1^3}\right)
\nonumber \\ &-&
 2 C \cos\varphi \left(\frac{3}{B_1^2} + \frac{6A^2}{B_1^3}
+\frac{A^4}{B_1^4} \right), \nonumber\\
\Gamma_{12}(\varphi) & = & 2 \frac{\mu^{2}}{a^2} \left\{ \sin\varphi 
- B_2 \cos\varphi \frac{A}{B_1} 
\right. \nonumber \\ &-& \left. 
 2 C \cos\varphi
\left(\frac{1}{B_1}+\frac{A^2}{B_1^2} \right) \right\}, 
\nonumber\\
\Gamma_{40}(\varphi) & = & \left\{ 4 \sin\varphi \left( 
\frac{-3A}{B_1^2}-\frac{A^3}{B_1^3} \right) 
\right. \nonumber\\ &+& \left.
B_2 \cos\varphi \left( \frac{3}{B_1^2}+ 
\frac{6A^2}{B_1^3}+ \frac{A^4}{B_1^4} \right)
\right. \nonumber\\ &+& \left. 
2 C \cos\varphi \left( \frac{15 A}{B_1^3} 
%\right. \right. \nonumber \\
%& & \left. \left. 
+\frac{10A^3}{B_1^4}+\frac{A^5}
{B_1^5} \right) \right\}. \nonumber\\
\Gamma_{22}(\varphi) & = & 2 \frac{\mu^{2}}{a^2} \left\{ -2 \sin\varphi
\frac{A}{B_1} + B_2 \cos\varphi \left( \frac{1}{B_1} + \frac{A^2}{B_1^2}
\right) 
\right. \nonumber \\ &+& \left. 
2 C \cos\varphi \left( \frac{3A}{B_1^2} + \frac{A^3}{B_1^3} 
\right) \right\}. \nonumber\\
\Gamma_{04}(\varphi) & = & 3 \frac{\mu^{4}}{a^4} \left\{ B_2 \cos\varphi 
+ 2 C \cos\varphi \frac{A}{B_1} \right\}. 
\label{A2.2}      
\end{eqnarray}
%---------------------------------------------------------------------------
 Taking into account Eqs.\ (\ref{eq36a}), (\ref{eq36b}) relating 
the local ($\varphi$) and the global ($\theta$) angles of incidence 
through the surface slopes $\partial_x h$, $\partial_y h$, a small 
slope approximation leads to  
\begin{eqnarray}
e^{A^2/2B_1} & \simeq & e^{a_{\sigma}^4 s^2/2f} \left\{ 1 +
\frac{a_{\sigma}^4}{2f}
\left[ \frac{a_{\mu}^2}{f} s (\partial_x h) + \frac{a_{\mu}^2}{f}
c^2 (\partial_y h)^2 
\right. \right. \nonumber \\ &+&\left. \left. 
\frac{a_{\mu}^2}{f^2} (\partial_x h)^2 \left(
a_{\mu}^2 c^2 (1+2s^2) - a_{\sigma}^2 s^2 (1+2c^2) 
\right. \right. \right.\nonumber\\ &+& \left. \left. \left. 
\frac{a_{\sigma}^4
a_{\mu}^2}{f} s^2 c^2 \right) \right] \right\}, 
\nonumber \\
B_1^{-1/2} & \simeq & \frac{1}{\sqrt{f}} \left\{ 1 -
\frac{a_{\sigma}^2 - a_{\mu}^2}{f} s c (\partial_x h)
\right. \nonumber\\ &+& \left.
\frac{a_{\sigma}^2 - a_{\mu}^2}{2 f^2} (\partial_x h)^2
\left(a_{\sigma}^2 s^2 (1+c^2) 
\right. \right. \nonumber \\ &-& \left. \left. 
a_{\mu}^2 c^2 (1+s^2) \right) - 
\frac{a_{\sigma}^2 - a_{\mu}^2}{2f} c^2 (\partial_y h)^2 \right\}. 
\label{A2.4}
\end{eqnarray}
Also, we have
\begin{eqnarray}
\Gamma_{20}(\theta) & = & \Gamma_{20}^{(0)}(\theta) +
\Gamma_{20}^{(1)}(\theta)
(\partial_x h), \nonumber \\
\Gamma_{02}(\theta) & = & \Gamma_{02}^{(0)}(\theta) +
\Gamma_{02}^{(1)}(\theta)
(\partial_x h), \nonumber \\
%--------------------------------------------------------------------
\Gamma_{20}^{(0)}(\theta) & = & \frac{a_{\sigma}^2}{2 f^3} 
\left\{-2 a_{\sigma}^6 s^4 c^2 + 2 a_{\sigma}^4 s^4 (s^2 - 2 c^2) 
\right. \nonumber \\ &+& \left. 
a_{\sigma}^4 a_{\mu}^2 s^2 c^2 (s^2 - c^2) + a_{\sigma}^2 a_{\mu}^2 
\right. \nonumber\\ &\times& \left. 
s^2 c^2 (7s^2 - 5c^2)+a_{\mu}^4 c^4 (5s^2 - c^2) \right\} , 
\nonumber \\ 
%---------------------------------------------------------------------
\Gamma_{20}^{(1)}(\theta) & = & \frac{a_{\sigma}^2 s c}{f^4} 
\left\{-2 a_{\sigma}^6 s^4 - a_{\mu}^6 c^4 + a_{\sigma}^6 a_{\mu}^2 
\right. \nonumber\\ &\times& \left. (s^2+s^2 c^2) 
-a_{\sigma}^4 a_{\mu}^4 (c^2 + c^2 s^2) 
\right. \nonumber\\ &+& \left.
a_{\sigma}^4 a_{\mu}^2 (5 s^2 - 3s^2 c^2) 
+ 4 a_{\sigma}^2 a_{\mu}^4 c^2 \right\} , \nonumber \\
%-------------------------------------------------------------------------
\Gamma_{02}^{(0)}(\theta) & = & - \frac{c^2 a_{\sigma}^2}{2 f},
\nonumber \\
\Gamma_{02}^{(1)}(\theta) & = & \frac{a_{\sigma}^4 c s}{f^2}.
\label{A2.3}
\end{eqnarray}
%----------------------------------------------------------------------------
%\eleq
In the above expressions we used the notations 
\begin{eqnarray}
a_{\sigma}  \equiv  \frac{a}{\sigma},
% \;\;,\;\;
a_{\mu} \equiv \frac{a}{\mu},
% \;\;,\;\;
s \equiv \sin \theta,
% \;\;,\;\;
\nonumber\\
c \equiv \cos \theta,
% \;\;,\;\;
f \equiv a_{\sigma}^2 s^2 + a_{\mu}^2 c^2.
\label{A2.5}
\end{eqnarray}
%----------------------------------------------------------------------------

\section{}
\label{AP:c}

  Equations (\ref{eq38a})-(\ref{eq38b}) relating the incidence angle
as measured in the local and laboratory reference frames apply only 
in the off-normal incidence case ($\theta \neq 0$). In the following we 
derive the correct expressions for the normal incidence case ($\theta =0$).
Indeed, if $\theta=0$, the vectors $\hat{n}$ and $\hat{m}$ shown 
in Fig.\ \ref{fig3} are given by
\begin{eqnarray}
\hat{n} = \frac{(-\partial_x h, -\partial_y h, 1)}{\sqrt{g}}, \,\,\,
\hat{m} = (0,0,1).
\label{A3.1}
\end{eqnarray}
Proceeding now as in (\ref{eq36a})-(\ref{eq36b}), we obtain
\begin{eqnarray}
\cos \varphi=\frac{1}{\sqrt{g}}, \;\; 
%\nonumber\\
\sin \varphi=\sqrt{\frac{(\partial_x h)^2 +(\partial_y h)^2}{g}},
\label{A3.2}
\end{eqnarray}
which are the $\theta \rightarrow 0$ limit of Eqs.\ (\ref{eq36a}) and
(\ref{eq36b}). The small gradient expansion performed on Eq.\ (\ref{A3.2})
now gives
\begin{eqnarray}
\cos \varphi  & \simeq & 1- \frac{1}{2}((\partial_x h)^2 +(\partial_y h)^2),
\nonumber\\
\sin \varphi  & \simeq &  \sqrt{(\partial_x h)^2 +(\partial_y h)^2}.
\label{A3.3}
\end{eqnarray}
Using Eqs.\ (\ref{A3.3}) in the expansions leading to Eq.\ (\ref{eq44}),
it can be seen that the expressions obtained 
for the coefficients indeed are the $\theta \rightarrow 0$ limit of Eqs.\
(\ref{eq45})-(\ref{eq59}).

\section{}
\label{AP:d}

  The solution corresponding to a rotated ripple structure
follows from Eq.\ (\ref{eq85}). 
Indeed, in the absence of the $\xi_x$ and $\xi_y$ terms, if we consider a 
solution of
(\ref{eq85}) of the form $h(x,y,t) = g(x-vy,t)$ with $v$ an arbitrary 
constant, the surface morphology evolution equation takes the form
\begin{eqnarray}
\partial_t g & = & -v_{0}+\gamma \partial_l g + (\nu_x+v^2\nu_y)
\partial_l^2 g \nonumber\\
& &+\frac{1}{2}( \lambda_x + v^2 \lambda_y) (\partial_l g)^2 
+(\Omega_1 + v^2 \Omega_2) \partial_l^3 g 
\nonumber\\
& & -K(1+v^2)^2 \partial_l^4 g,
\label{A4.1}
\end{eqnarray}
where $g(l)=g(x-vy)$ is the steady wave solution \cite{SAMT}. From 
(\ref{A4.1}) it follows that the nonlinearity vanishes when $\lambda_x+ 
v^2 \lambda_y=0$, or $v=\sqrt{-\lambda_x/\lambda_y}$. In this case we obtain 
an exponentially growing ripple structure with ripples forming an angle
$\theta_c =\tan^{-1}(v)= \tan^{-1}(\sqrt{-\lambda_x/\lambda_y})$
with respect to the $x$ axis.
   
%----------------------------------------------------------------------------
%----------------------------------------------------------------------------

\end{multicols}

\end{document}